\documentclass[12pt,preprint]{aastex}
\usepackage{natbib}
\usepackage{mathrsfs}
\usepackage{amssymb}
\usepackage{rotating,graphicx}
\usepackage{color,soul}
\usepackage{placeins}

\shorttitle{The internal rotation profile of the B star KIC\,10526294}
\shortauthors{Triana et al.}

\begin{document}

\title{The internal rotation profile of the B-type star KIC\,10526294\\
	 from frequency inversion of its dipole gravity modes and statistical
         model comparison}

   \author{S.~A. Triana, E. Moravveji \altaffilmark{1}, P.~I. P\'apics \altaffilmark{2},}
   \affil{Institute of Astronomy, KU Leuven, Celestijnenlaan 200D, Leuven, Belgium}
   \author{C. Aerts,}
   \affil{Institute of Astronomy, KU Leuven, Celestijnenlaan 200D, Leuven, Belgium}
   \affil{Department of Astrophysics/IMAPP, Radboud University Nijmegen, 6500 GL Nijmegen, The Netherlands}
   \author{S. D. Kawaler,}
   \affil{Department of Physics and Astronomy, Iowa State University, Ames, IA 50011, USA}
   \and
   \author{J. Christensen-Dalsgaard}
   \affil{Stellar Astrophysics Centre, Department of Physics and Astronomy, Aarhus University, DK–8000 Aarhus C, Denmark}

   \altaffiltext{1}{Postdoctoral Fellow of the Federal Science Policy Office of Belgium (Belspo)}

   \altaffiltext{2}{Postdoctoral Fellow of the Fund for Scientific Research of Flanders (FWO), Belgium}

\begin{abstract}
  The internal angular momentum distribution of a star is key to determine its
  evolution. Fortunately, the stellar internal rotation can be probed through
  studies of rotationally-split nonradial oscillation modes.  In particular,
  detection of nonradial gravity modes (g modes) in massive young stars has
  become feasible recently thanks to the {\it Kepler\/} space mission. Our aim
  is to derive the internal rotation profile of the {\it Kepler\/}  B8V
  star KIC\,10526294 through asteroseismology. We interpret the observed
  rotational splittings of its dipole g modes using four different approaches
  based on the best seismic models of the star and their rotational kernels.
  We show that these kernels can resolve differential rotation within the 
  radiative envelope if a smooth rotational profile is assumed and the
  observational errors are small. Based on {\it Kepler\/} data, we find that the rotation rate
  near the core-envelope boundary is well constrained to $163\pm89$ nHz.
  The seismic data are consistent with rigid
  rotation but a profile with counter-rotation within the envelope has a statistical
  advantage over constant rotation. Our
    study should be repeated for other massive stars
    with a variety of stellar parameters in order to deduce the physical
    conditions that determine the internal rotation profile of young massive
    stars, with the aim to improve the input physics of their models. 
\end{abstract}


\section{Introduction}
One of the major important ingredients in the computation of stellar evolution
models, from the star formation process until the death of the star, is
rotation \citep[e.g.,][for a recent monograph in the subject]{Maeder2009}.  Even
in the case of slow rotation, the dynamical and mixing processes related to it
are not negligible as they substantially affect the structure of the
star. Unfortunately, direct measurements of the internal rotation profile of
stars are not possible. Hence the inclusion of rotational effects in models
rests on uncalibrated theoretical prescriptions.

It is a fortunate circumstance that the oscillation frequencies of a star are
affected by its rotational properties \citep{ledoux1951nonradial}. The frequency
splitting of the oscillation modes is well understood in the case of slowly
rotating stars for which a first-order perturbation method is sufficient to
model the frequencies 
\cite[e.g.][for an extensive description]{Aerts2010asteroseismology}.  In
this work, we assume that we are dealing with an unevolved star that does not
possess a magnetic field and whose central frequencies of the rotationally-split
multiplets are not affected by the slow rotation. Moreover, we assume that the
deformation from spherical symmetry due to the centrifugal forces can be
ignored. In that case the frequency splitting of the oscillation modes, as measured
in the observer's frame, is due
to a combination of mode advection and the Coriolis force
 and can be computed from the so-called rotational kernels
\citep[cf.\ Eq.(3.356) in][]{Aerts2010asteroseismology}.

Helioseismology delivered a very detailed view of the internal rotation profile
$\Omega (r,\theta)$ of the Sun for the radial range $r\in
[0.2\,R_{\odot},1.0\,R_{\odot}]$ ($R_\odot$ denotes the solar radius) 
and for all co-latitudes $\theta$, through
frequency inversion of the rotational splittings of its hundreds of detected
acoustic modes \citep[e.g.,][]{JCD2002,Thompson2003}, a technique that has found
its way even to laboratory experiments \citep{triana2014}.
Given that acoustic modes
do not have sufficient probing power in the very inner regions and that the Sun 
does not
reveal gravity modes, it is not possible to deduce the
rotational profile for $r\lesssim0.2\,R_{\odot}$.

It is not currently within reach to derive the rotation profile for distant
stars with similar precision as the Sun, but applications of asteroseismology
did allow to deduce averaged rotation rate ratios $\Omega_{\rm core}/\Omega_{\rm envelope}$
 from forward
modeling of the rotational splitting for a few main-sequence B stars from
ground-based monitoring campaigns \citep{Aerts2003,Pamyatnykh2004,Briquet2007},
as well as two $\delta\,$Sct-$\gamma\,$Dor-type hybrid 
main-sequence  pulsators from {\it Kepler\/} space-based
photometry \citep{Kurtz2014,Saio2015}. Moreover, $\Omega_{\rm core}$ was derived
from gravity-dominated mixed modes in hundreds of red giants observed with {\it
  Kepler} \citep{Beck2012,Mosser2012}, while the estimate of their   
$\Omega_{\rm envelope}$ is uncertain due to the remaining dominant 
influence of the core
regions on the measured splittings of the pressure-dominated mixed modes.
\citet{Deheuvels2012} and \citet{Deheuvels2014} performed frequency
inversions for seven selected subgiants in different evolutionary stages
relying on the splitting of their dipole mixed modes. They selected profiles
representing a linear decrease in rotation frequency in the core regions
followed by a constant rotation profile in the extended convective envelope,
with $\Omega_{\rm core}/\Omega_{\rm envelope}$ ranging from 2 to about 20. 
A similar result was obtained for the red giant KIC\,5006817, which is the
primary of an eccentric binary, by
\citet{Beck2014}. 
Only in some of those studies, {e.g. \citet{Deheuvels2014}},
 statistical model comparison was used to evaluate
  the likelihood of the optimal shape of the rotational frequency throughout the stars
and few continuous
  and discontinuous functions for $\Omega (r)$ were
  considered. Despite their frequent occurrence in e.g., geophysics, 
counter-rotating solutions were considered inappropriate for stars. Independently
of  that
restriction,
the {\it Kepler\/} results for $\Omega (r)$ so far deliver 
an important calibration for the improvement of evolutionary models for
single and binary 
low-mass stars, given that the theoretical predictions of angular-momentum transport result in
core rotation rates are higher by at least an order of magnitude than observed \citep[see e.g.][for the input physics in question]{Eggenberger2012,vanSanders2013,Cantiello2014}. 
The need for
redistribution and loss of angular momentum during evolution is also required on
the basis of the internal rotation properties of white dwarfs derived from both
forward modeling and inversion of their rotationally-split 
g mode oscillation frequencies
\citep{Charpinet2009,Corsico2012}.

In this work, we provide the first frequency inverted rotation profile of an
unevolved intermediate-mass B-type main-sequence star from its
rotationally-split dipole gravity modes detected in four years of {\it Kepler\/}
data. The paper is organized as follows: we summarize the observational
  data and the resuls of forward seismic modeling in Section\,2. In Sections\,3
  and 4 we examine in detail the rotational kernels and the splittings
  associated with linear rotation models.  In Section\,5 we present piece-wise
  two- and three-zone rotation models. Section\,6 is devoted to inversion
  methods, including theory and results, and in Section\,7 we present results
  for the rotation profile based on Monte Carlo simulations. We summarize and
  conclude in Section\,8.

\section{Observational input and results of forward 
seismic modeling}\label{sec}

A first detailed asteroseismic analysis of the main sequence B star KIC\,10526294,
was presented by \mbox{\cite{papics2014}}, including an estimation of the amount of
core overshooting following earlier approaches for main-sequence B stars with a
well-developed convective core.  They characterized KIC\,10526294 as a slowly
rotating SPB star (see e.g. \mbox{\cite{Aerts2010asteroseismology}} for a 
definition) exhibiting a series of 19 quasi-equally
spaced dipole modes.  { KIC\,10526294 is so far the only multiperiodic SPB star 
with unambiguous detection of rotationally-split triplets from
the \mbox{\it Kepler\/} light curve. For this reason, it allows us to probe its
interior structure to a deeper level than for any other SPB so far.}

\begin{deluxetable}{cccc}
  \tablecolumns{2} \tablewidth{0pc} \tablecaption{\label{data}
Symmetric components of the observed rotational splittings of KIC\,10526294}
  \tablehead{ \colhead{Central frequency [$\mu$Hz]} & \colhead{Splitting 
$\delta_{nlm}$ [nHz]} & \colhead{Error Set\,1 [nHz]} & \colhead{Error Set\,2 [nHz]}}
    \startdata

    5.4655 &   45.28   &   14.72  &  16.54  \\ 
    5.6272 &   29.49   &    0.69  &   8.70 \\ 
    5.7978 &   33.91   &    7.37  &   8.81 \\ 
    5.9873 &   32.64   &    7.47  &  10.33 \\ 
    6.1739 &   41.99   &    4.42  &   7.56 \\ 
    6.3959 &   35.74   &    0.36  &   4.91 \\ 
    6.6200 &   29.43   &    4.06  &   7.38 \\ 
    6.8703 &   30.42   &    0.55  &   8.55 \\ 
    7.1235 &   33.07   &    3.46  &   6.09 \\ 
    7.4213 &   29.99   &    1.00  &   9.41 \\ 
    7.7616 &   41.55   &   15.04  &  17.17 \\ 
    8.1163 &   28.73   &    1.03  &   6.42 \\ 
    8.5036 &   29.50   &    2.59  &   7.75 \\ 
    8.9398 &   28.18   &    3.18  &   7.23 \\ 
    9.4090 &   27.53   &    0.85  &   4.63 \\ 
    9.9115 &   26.11   &    1.67  &   5.82 \\ 
   10.4495 &   26.41   &    7.02  &  10.04 \\ 
   11.0429 &   25.74   &    0.49  &   5.21 \\ 
   11.7293 &   23.32   &    4.55  &   9.34 \\  
   
   \enddata
\end{deluxetable}

With the purpose of detailed seismic modeling, \citet{moravveji2015}
  elaborated on optimal frequency error estimation, taking into account the
  signal-to-noise ratio, sampling, and correlated nature of the {\it
    Kepler\/} data, following the method by \cite{degroote2009}.  This resulted in a
correction factor of 3.0 to be applied to the formal errors obtained from the
nonlinear least-squares fit.
We estimated the
splitting for each dipole mode as the average splitting between the measured $m
= +1$ and $m=-1$ peaks with respect to the central $m=0$ peak. This comes down
to considering only the symmetric component of the splittings. The total
variance was then estimated as the variance of the symmetric component plus the
\emph{inter}-variance:
\begin{equation}
\sigma^2_{\mathrm{Total}}=\frac{1}{2}\left(\sigma^2_{-1}+\sigma^2_{+1}\right)+\sigma^2(\delta_{-1},\delta_{+1}),
\end{equation}
where $\delta_{\pm1}$ denotes the $m=\pm1$ splittings and $\sigma_{\pm1}$ their individual
uncertainties. This results in
larger errors for those splittings with larger asymmetric components. 
Further, as explained by the authors, 
the error estimates for the triplet components as computed  from the
Rayleigh limit and taking into account mode crowding effects are too large
\citep[][their Fig.\,8]{papics2014}.
Nevertheless, we also used those overestimated values with the
argument that they deliver the most conservative upper limit to the true frequency
errors. We list both error sets in Table\,\ref{data} and show in this work that
our conclusions on $\Omega (r)$ are essentially independent on the choice of error set.
The best error estimates (Error Set\,1) are used
throughout the main text,  while all results relying on the 
too large errors (Error Set\,2) are treated in Appendix\,\ref{apx_A}.

\begin{figure}[ht!]
\centering
\includegraphics[width=\linewidth]{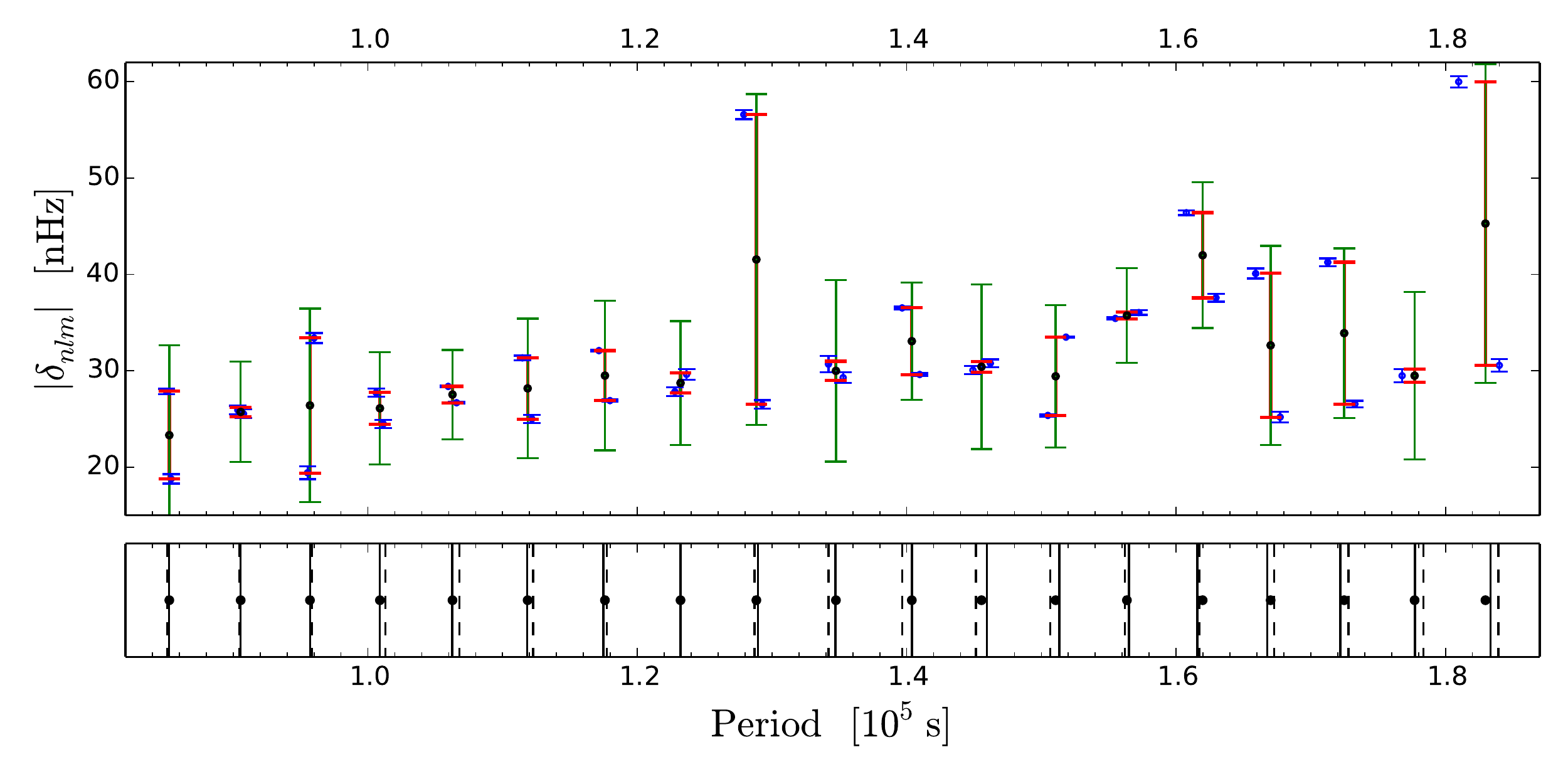}
\caption{Top: Observed rotational splittings (top panel, blue error bars), and
  the symmetric component used as inversion inputs (top panel, red error
  bars). The errors on the inversion inputs were taken from
    \citet{moravveji2015} and the green error bars are from \cite{papics2014},
    see text for details). We chose the abscissae of the inversion inputs so as
    to coincide with the central $m=0$ peak of the observed triplets. Bottom:
    the location of the central peaks as black dots, and the mode periods from
    models: vertical solid lines for the best model in \citet{moravveji2015}
    (Model\,1), vertical dashed lines for the best model in \citet{papics2014}
    (Model\,2).}
\label{fig:data}
\end{figure}

Following an essentially identical approach as in
\citet{papics2014}, \citet{moravveji2015} were able to find seismic models that
match more closely the observed frequencies thanks to the inclusion of extra
 diffusive
mixing in the stellar envelope,
in addition to core overshooting. Such additional mixing was already found necessary for
the B3V SPB star HD\,50230 \citep{Degroote2010}. The parameters of the two
models are given in Table\,\ref{modelparameters}.  We base the present work on the best matching Model\,1
from \citet{moravveji2015} but we checked that the main qualitative features of
the resulting rotation profiles from inversion do not depend on the choice of
the seismic model, as long as it is able to reproduce reasonably well the main
observed characteristics of the star. We return to this in
Section\,\ref{ProfilesInversion}.
\begin{figure}[ht!]
\centering
\includegraphics[width=\linewidth]{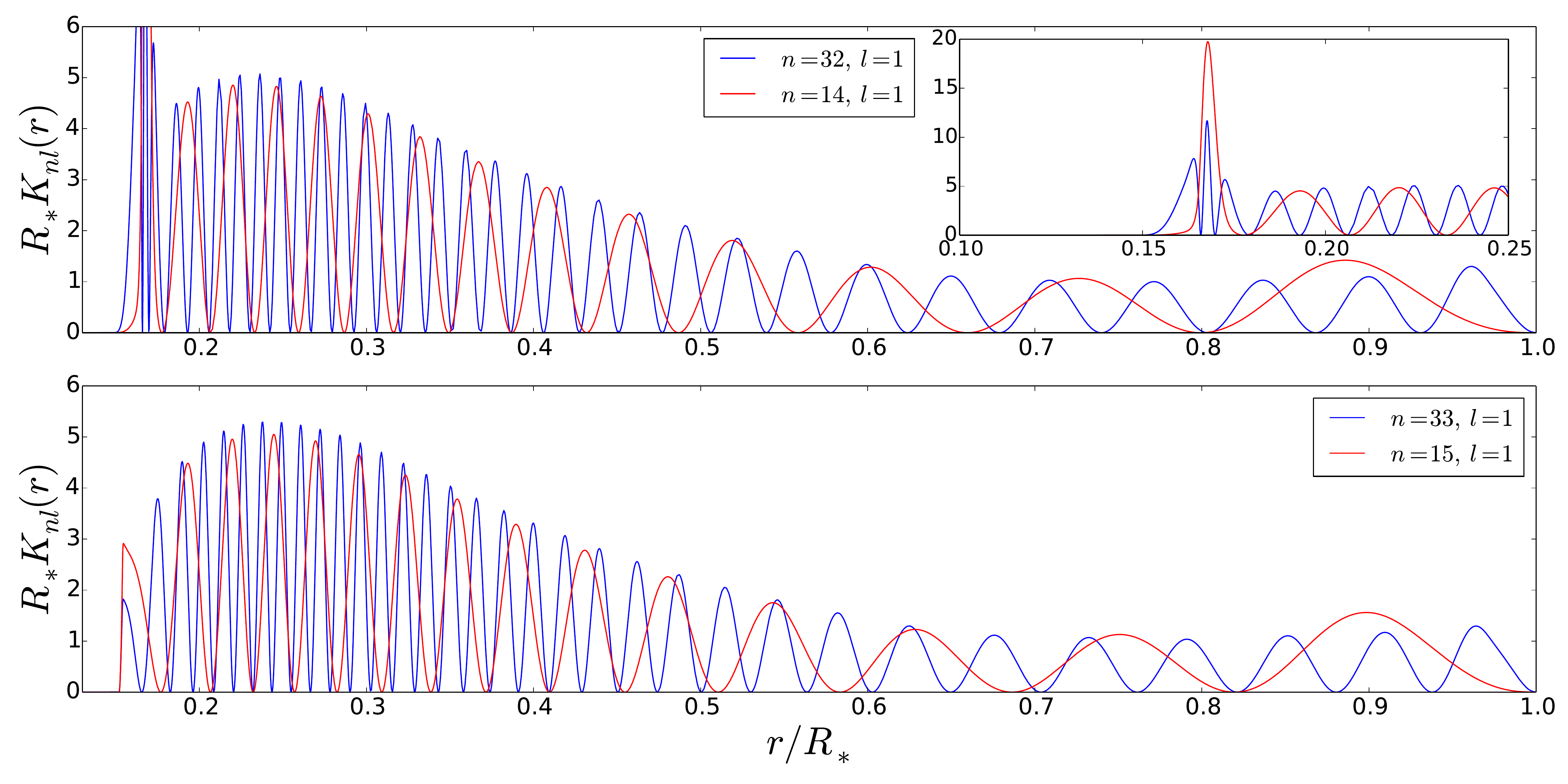}
\caption{The rotational kernels of dipole zonal modes of highest and lowest
  radial order, plotted against the stellar fractional radius $r/R_*$, for 
  the best matching models found by \citet[][lower
  panel]{papics2014} and \citet[][upper panel]{moravveji2015} --- cf.\ lower
  panel of Fig.\,\ref{fig:data}.  Some of the modes in the upper panel are
  trapped near the convective core as evidenced by the large peak just outside
  the convective core, as visible in the inset.  The kernels from the best
  seismic model found by \citet{papics2014} shown in the bottom panel do not
  exhibit trapping, but the overall shapes are similar otherwise. Inversion
  results from both of these models are qualitatively similar.}
\label{kernels}
\end{figure}

In a non-rotating star, oscillation mode families are characterized by their
radial order $n$ and degree $l$, with individual family members corresponding to
different values of $m$, the azimuthal wave number, sharing the same
eigenfrequency. Rotation lifts this degeneracy.  The identification of the
radial order $n$ of the 19 detected modes of KIC\,10526294 was achieved by
comparing the periods of the observed zonal ($m=0$) dipole ($l=1$) g modes with
those predicted by equilibrium models computed with the MESA stellar structure
and evolution code \citep{Paxton2011,Paxton2013} and coupled to the GYRE stellar
oscillation code \citep{Townsend2013}.  This comparison is possible thanks to
detected triplets on the one hand, and the almost equally-spaced sequence of
dipole g modes with consecutive order as theoretically expected for such
modes on the other hand.  The radial orders matching the observations range from
$n=32$ to $n=14$.

\begin{figure}[hb!]
\centering
\includegraphics[width=0.7\linewidth]{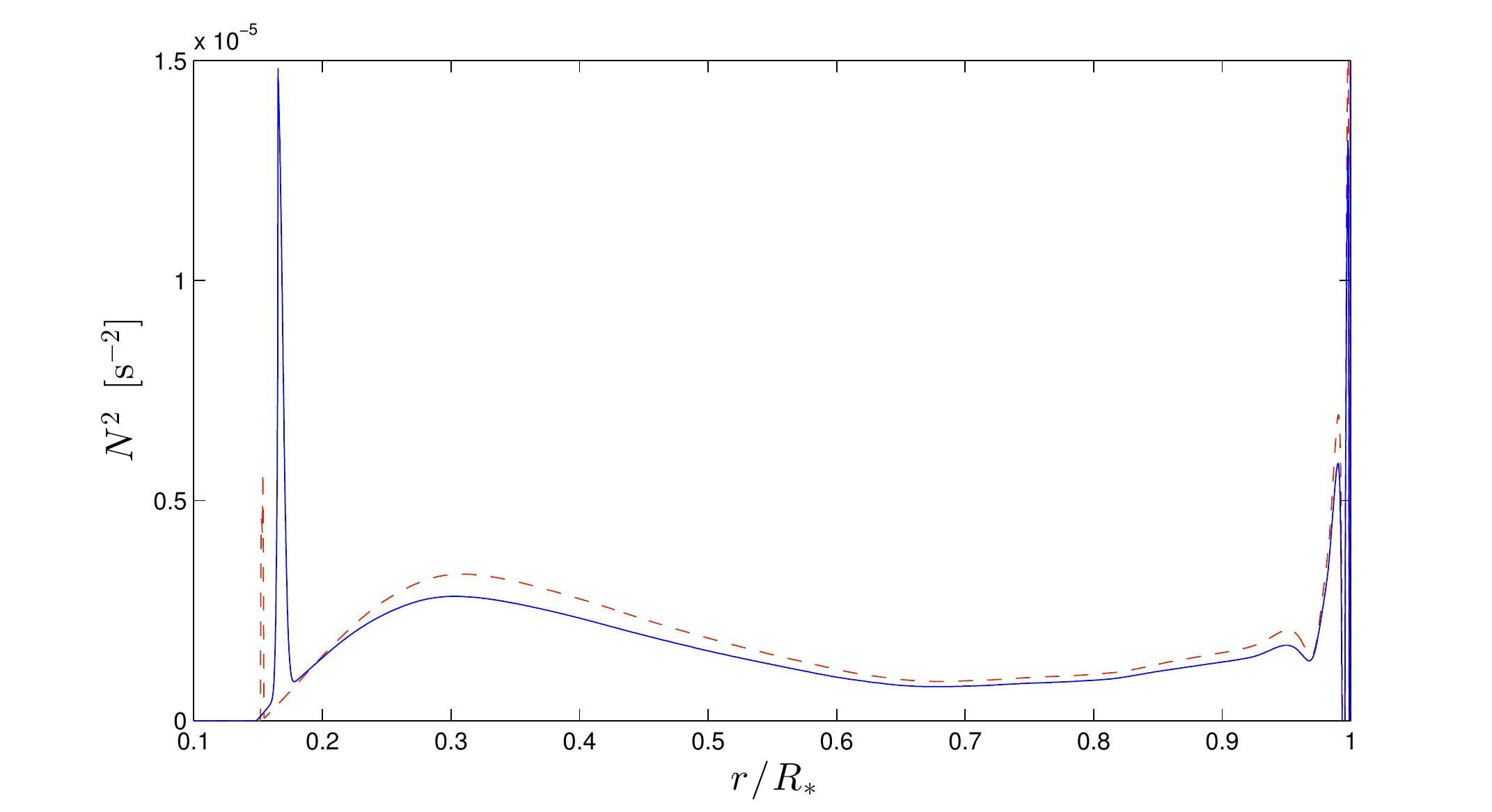}
\caption{The squared Brunt-V\"{a}is\"{a}l\"{a} frequency $N^2$ as a function of 
 stellar fractional radius
 for the best model from \cite{moravveji2015} (continuous blue curve) and for the best model from \cite{papics2014} (red-dashed curve). The larger peak of the former model near the core boundary (at $r\sim0.15\,R_*$) is comparatively further away from that boundary than the peak of the latter model, thus allowing some modes to
 be ``trapped'' between the core boundary and the peak. See also Fig.\,\ref{kernels}.}
\label{fig:bv}
\end{figure}

Most of the 19 dipole modes of KIC\,10526294 reveal a very narrow
rotationally-split triplet structure.  If we assume that the {\em cyclic\/}
rotation frequency, which we denote as $\Omega$, depends on the radial
coordinate $r$ only, the frequency splitting of a mode with degree $l$ and
radial order $n$, denoted as $\delta_{nlm}$, can be written as
\begin{equation}
\label{eq:rotsplit}
\delta_{nlm}=m \beta_{nl} \int_0^{R_*} K_{nl}(r)\, \Omega(r)\, \mathrm{d}r,
\label{eq:forward}
\end{equation}
where the unimodular mode kernel 
$K_{nl}(r)$ is a function of the mode's
displacement amplitudes $\xi_r(r)$ (vertical) and $\xi_h(r)$ (horizontal), while
$\beta_{nl}$ is given by
$$
\beta_{nl}=\frac{\int_0^{R_*}\left[ \xi_r^2+l(l+1)\xi_h^2 -2\xi_r\xi_h-\xi_h^2\right]r^2\rho\, \mathrm{d}r}{\int_0^{R_*}\left[ \xi_r^2+l(l+1)\xi_h^2\right]r^2\rho\, \mathrm{d}r}
$$
and is connected with the Ledoux splitting as $\beta_{nl}=1-C_{nl}$
\citep[e.g.,][Chapter 3]{Aerts2010asteroseismology}.  
The rotationally-induced splittings for the
19 detected modes are shown in Fig.\,\ref{fig:data} (top panel, adapted from
Fig.\,8 in \citet{papics2014}), together with the mode periods derived from
 Model\,1 and Model\,2 (bottom panel). The mode kernels for both models are
 shown in Fig.\,\ref{kernels} and their squared
Brunt-V\"{a}is\"{a}l\"{a} frequency $N^2$ is shown in
Fig.\,\ref{fig:bv}. The latter illustrates the slightly more advanced stage of Model\,1
compared to Model\,2.

\begin{deluxetable}{cllllllcl}
  \tablecolumns{9} \tablewidth{0pc} \tablecaption{\label{modelparameters}
Fundamental stellar parameters
    of KIC\,10526294 from the best matching theoretical model (Model\,1) found by
    \citet{moravveji2015} and by \citet{papics2014} (Model\,2).}
  \tablehead{ \colhead{Model} & \colhead{$T_{\rm eff}$ [K]} &
    \colhead{$M_*/M_\odot$} & \colhead{$R_*/R_\odot$} & \colhead{$f_{ov}$} &
    \colhead{$Z$} & \colhead{$X_c$} & \colhead{Age [Myr]} &
    \colhead{$\chi^2$}} \startdata
  1 & 13000  & 3.25  & 2.215  & 0.017  & 0.014  & 0.627  & 63  &  1.42 \\
  2 & 12470 & 3.20 & 2.100 & $<0.015$ & 0.020 & 0.693 & 12 & 10.9
\enddata
\end{deluxetable}

Some of the observed splittings are not symmetric with respect to the central
$m=0$ peak. These asymmetries are usually related to mechanisms capable of lifting partially the $2l+1$ 
degeneracy of a given multiplet, such as deviations from sphericity or
large-scale magnetic fields. We assume to be dealing with a physical phenomenon that
causes asymmetries without lifting the degeneracy between the retrograde ($m<0$)
and prograde ($m>0$) modes. Rotationally-induced splittings
 as expressed by Eq.\,(\ref{eq:rotsplit}) are only capable of lifting the $\pm m$ 
degeneracy. Hence, forward modeling including only differential rotation up to
first order cannot fully capture these observed asymmetries. On the other hand,
Eq.\,(\ref{eq:rotsplit}) is still perfectly valid even in the presence of mechanisms mentioned above, with the caveat
that it then accounts for one-half of the splitting between the $+m$ and the $-m$ modes and not
for the splitting between a mode with a given $m$ and the corresponding central $m=0$ peak.
 It is known that the presence of
 a magnetic field near the convective core, as discussed by \citet{hasan2005},
  can give rise to such effects. In fact,
the mode kernels considered here have substantial amplitudes precisely in that
zone, making them particularly susceptible to this effect. Given that we have no
information on the presence or absence of a magnetic field in KIC\,10526294, we
will address only the symmetric components of the splittings in this study,
assuming rotation to be the dominant mechanism responsible for the observed
splittings and leaving the modeling of the asymmetries for a future more
specialised study. As explained earlier, the presence of the asymmetries leads to increased errors in the
splittings.

\section{Cumulative kernel integrals}
\label{sec:apx}

\begin{figure*}[ht!]
\centering
\includegraphics[width=\linewidth]{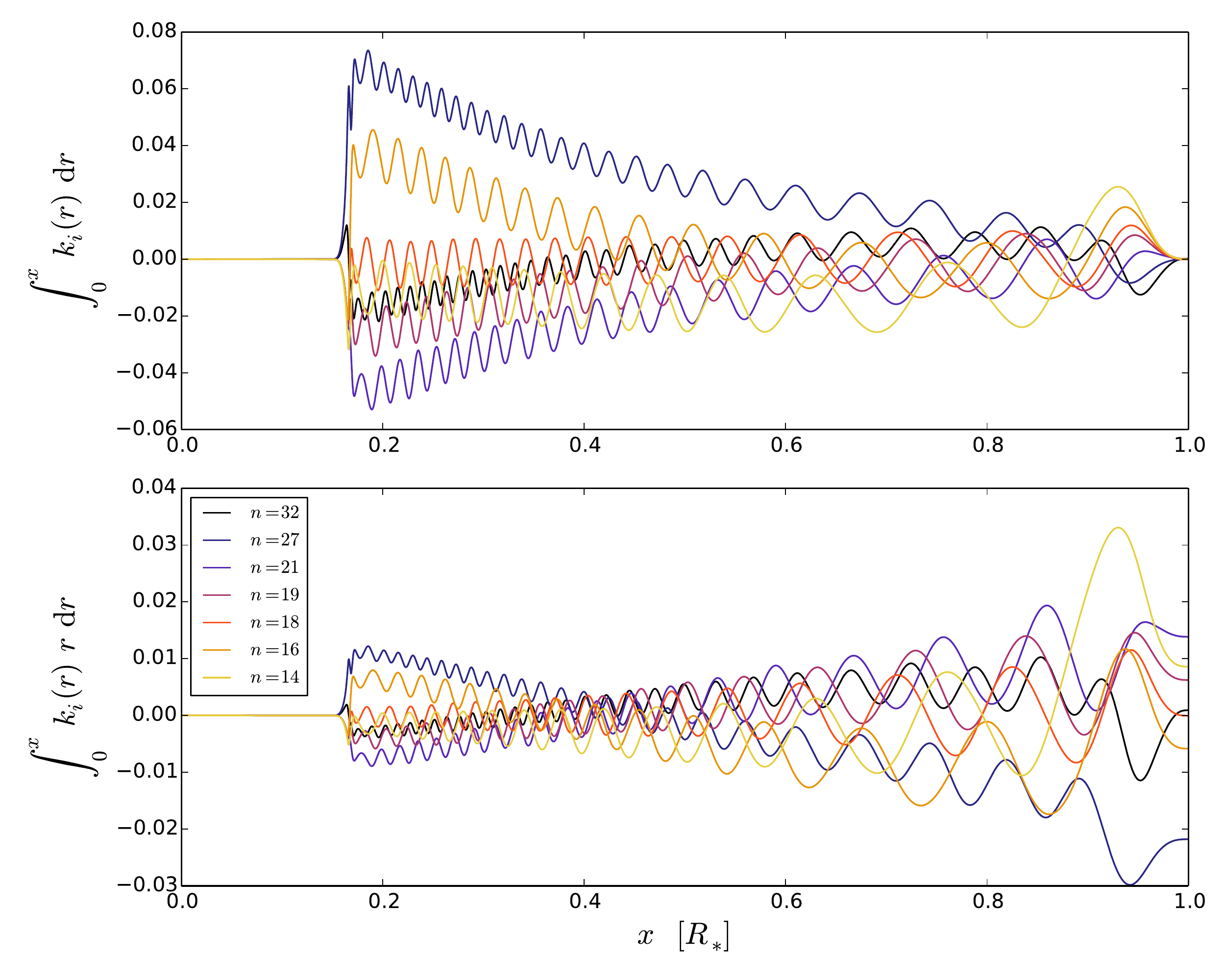}
\caption{Cumulative integrals of $k_i(r)$ (top) and $r\,k_i(r)$ (bottom) for a sample of dipole ($l=1$) modes from Model\,1 with various degrees of trapping. Mode with $n=27$ is the ``most'' trapped near the core as compared with the $n=21$ which is the ``least'' trapped.}
\label{fig:cuint}
\end{figure*}

Each mode samples differently the internal rotation of the star. This is usually
illustrated by plotting the cumulative integral of the kernels $K_i$. For
increased contrast we plot instead in Fig.\,\ref{fig:cuint} the cumulative
integral of $k_i(r)$, which we define by
\begin{equation}
\label{eq:k}
k_i(r)=K_i(r)-\left<K(r)\right>,
\end{equation}
where $\left<K(r)\right>$ is the average of the kernels across modes
at each radius $r$. This $\left<K(r)\right>$ is the `common' kernel and its integral will
only contribute to the average of the splittings. Similarly we can express a
given profile $\Omega(r)$ as a sum of its mean value $\bar \Omega$ (across the radial
coordinate), plus a fluctuating part $\omega(r)$ (with zero mean):
\begin{equation}
\Omega(r)=\bar \Omega+\omega(r).
\end{equation}
Therefore we can write the \emph{scaled} splittings as:
\begin{equation}
\label{eq:kw}
\Delta_i=\frac{\delta_{nlm}}{m\beta_{nl}}= \bar\Omega + \int_0^{R_*} \left< K(r)\right>\, \omega(r) \, \mathrm{d}r + \int_0^{R_*} k_i(r)\omega(r)\, \mathrm{d}r.
\end{equation}
The first two terms on the right hand side comprise the average splitting, while
differences across modes come into play in the last term.
Figure\,\ref{fig:cuint} shows the cumulative integrals of $k_i(r)$ and of
$k_i(r)\,r$ for a sample of modes from Model\,1 with various degrees of trapping near the
core. From the figure it is evident that the ``least'' trapped mode (the one with $n=21$) would have the largest splitting in response to a rotation profile that increases linearly with radius. Conversely, the ``most'' trapped mode near the core ($n=27$) would have the smallest splitting under the same condition.

\section{Trapped modes and linear rotation profiles}
\label{trapped}
\begin{figure}[h!]
\centering
\includegraphics[width=\linewidth]{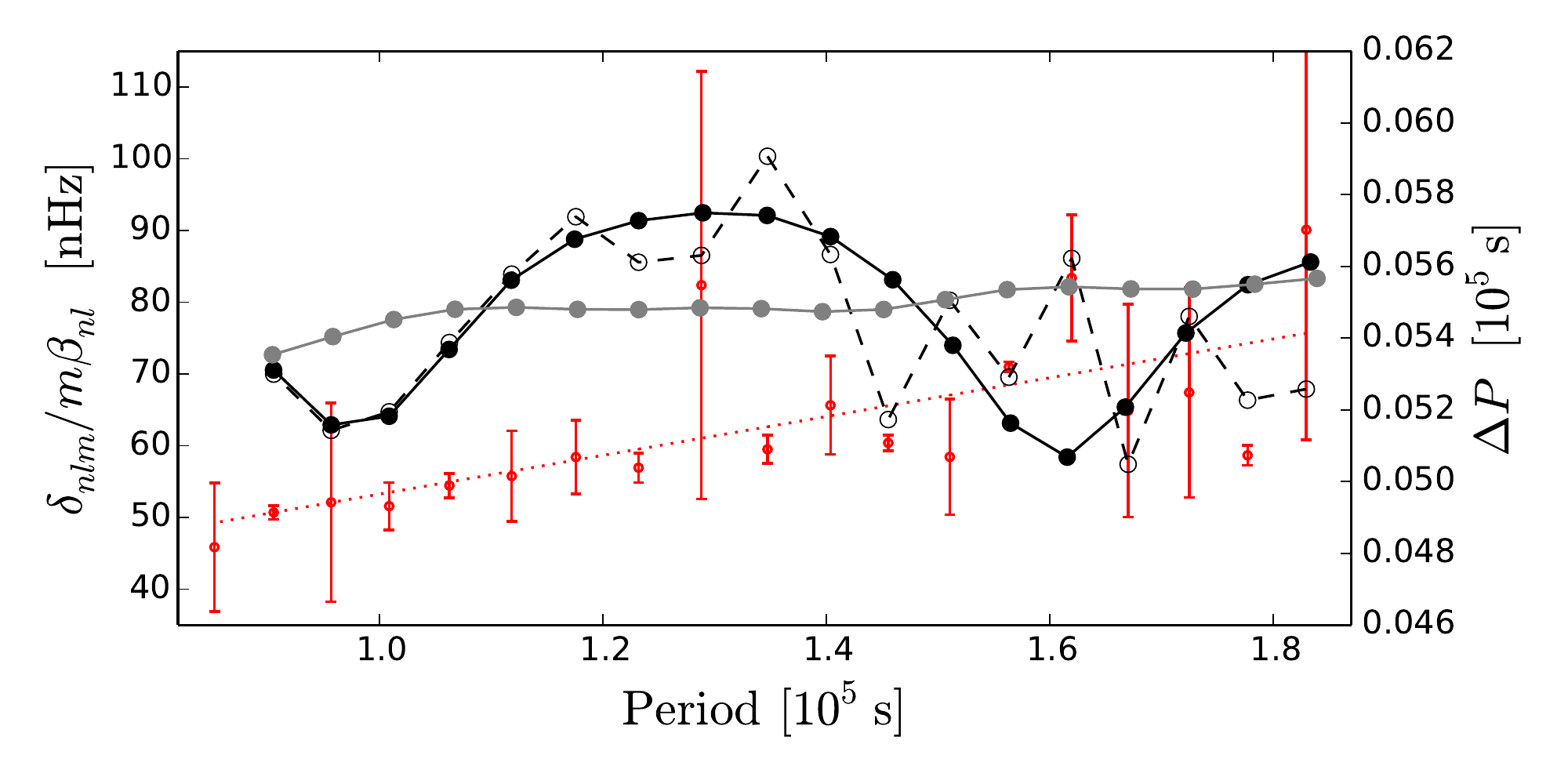}
\caption{The scaled rotational splittings $\delta_{nlm}/m\beta_{nl}$ deduced from
 the observations (red error bars) as a function of the mode period together with a
  linear fit (dotted red line, slope:  
$27.04\,\mathrm{nHz}/(10^5\mathrm{s})$, 
intercept: $26.20\,\mathrm{nHz}$, scale is on the left). 
The dashed line connects the
  observed period spacings defined as $\Delta P_n = \Pi_n-\Pi_{n-1}$. The 
black continuous line connects the
  spacings derived from the seismic Model 1 in Table\,1 while the grey continuous
  line connects the spacings derived from Model 2.
  (scale is on the right). 
The trapped modes of Model 1 are those with periods
  close to $0.95\times10^5\,\mathrm{s}$ and $1.6\times10^5\,\mathrm{s}$.}
\label{fig:spacings}
\end{figure}

Here we follow closely the analysis done by \citet{kawaler1999} for g modes in
white dwarf pulsators. We take advantage of the fact that some modes are trapped
as revealed by the kernel amplitudes. They are trapped very close to the
overshooting zone while other modes have more spread-out amplitudes,
comparatively. This trapping manifests itself as reduced period spacings if we
plot them as functions of period \citep{kawaler1994}. The observed spacings of
KIC\,10526294 are shown in Fig.\,\ref{fig:spacings} together with the {\em scaled}
splittings $\delta_{nlm}/m\beta_{nl}$ (see Eq.\,\ref{eq:forward}). A linear fit to the scaled splittings results in a slope of $27.04\,\mathrm{nHz}/(10^5\mathrm{s})$ and an intercept of  $26.20\,\mathrm{nHz}$.

\begin{figure}[ht!]
\centering
\includegraphics[width=0.8\linewidth]{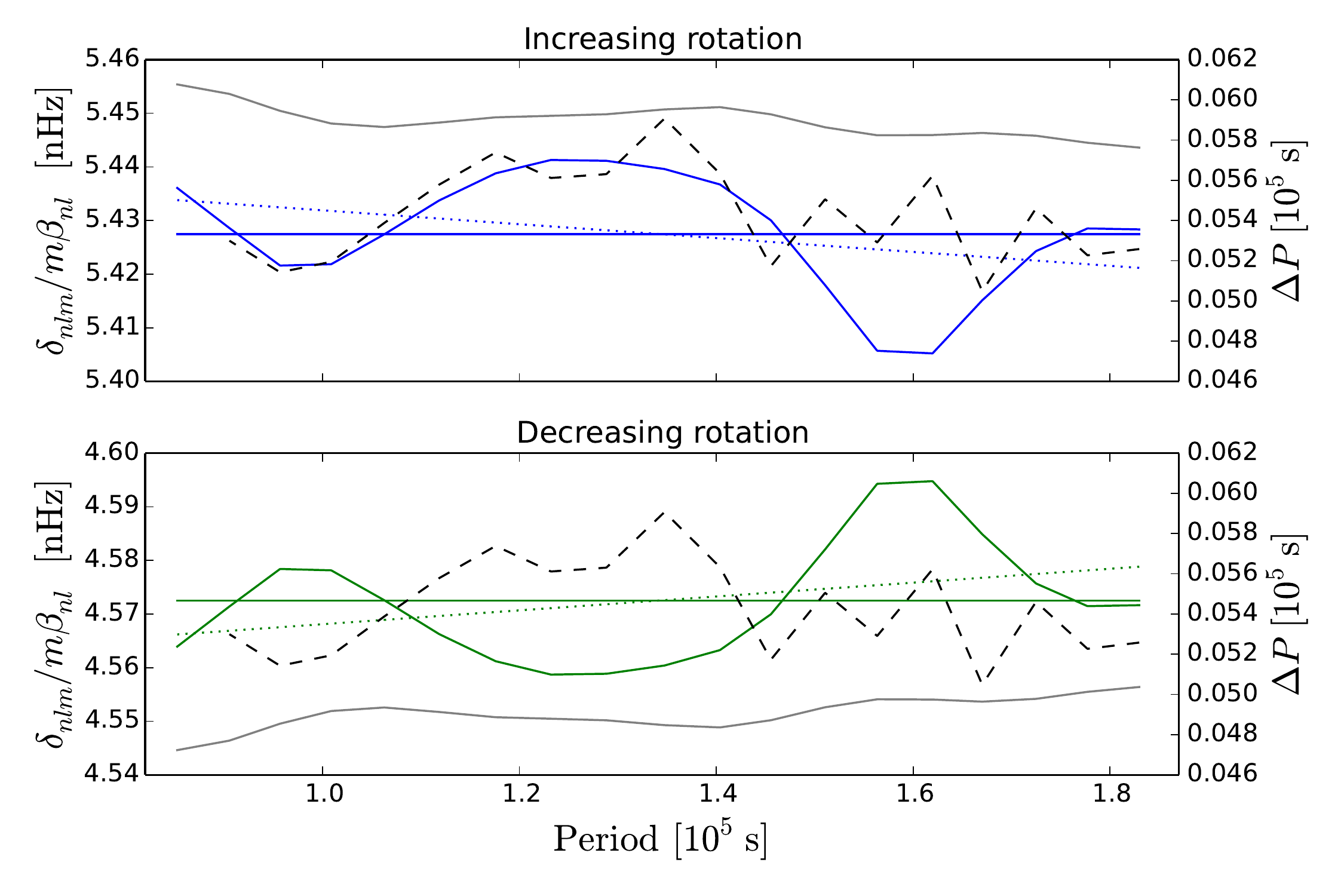}
\caption{Top: the scaled rotational splittings for a linearly increasing
  rotation profile derived from Model 1 and 2 (continuous blue and grey curves,
  respectively), the blue dotted line is a linear fit to the splittings from Model 1,
   scale is on the left. Bottom: the splittings for a linearly decreasing
  rotation profile derived from both Model 1 and 2 (continuous green and grey
  curves, respectively), the green dotted line is a linear fit to the splittings
  from Model 1,
  scale is on the left. For comparison, the horizontal lines show
  the splittings from constant rotation profiles having the same mean splittings
  as the splittings from the increasing or decreasing profiles (top, blue horizontal
  line and bottom, green horizontal line, respectively). The black dashed lines on 
  both plots show the period
  spacings from observation, scales are on the right.}
  
\label{fig:inc_dec}
\end{figure}

Let us now do forward modeling to find the predicted splittings using 
synthetic, linear rotation test profiles. Two test profiles have 
$\Omega=5\,\mathrm{nHz}$ at $r=0$ and have the same slope in absolute value,
$1\,\mathrm{nHz}/R_*$, but opposite in sign. 
The results are shown in Fig.\,\ref{fig:inc_dec},
where the top and bottom plots correspond to increasing and decreasing rotation
profiles, respectively.  We see the clear signature of mode trapping. Trapped
modes are closer to the core, so if the rotation profile increases with radius, then
the corresponding splittings are comparatively smaller. Analogously, if the
rotation decreases with radius then the trapped modes will show comparatively
larger splittings than the other modes. The latter situation 
is precisely what we can see in
Fig.\,\ref{fig:inc_dec}.
A similar situation occurs for the white dwarf PG\,1159-035, as reported by
\citet{kawaler1999}, the only difference being that some modes in the white
dwarf are trapped close to the surface such that the results are reversed
compared to those for KIC\,10526294.

This simplified analysis is helpful because it gives a first idea of the
sign of the slope of an unknown rotation profile just by plotting the splitings
and the period spacings and see if they vary in phase. In our case it is not
very clear if they vary in phase or not, so instead we make use of the linear
trends of the splittings as indicated with the linear fits (dotted lines) in
Fig.\,\ref{fig:inc_dec}. When period spacings and splittings vary in phase, the
linear trend is downward (negative slope). Conversely, when period spacings and
splittings are in anti-phase, the linear trend is upward (positive slope). The
observed splittings of KIC\,10526294 have an {\em increasing\/} trend, as
evidenced by the linear fit (dotted red line) in Fig.\,\ref{fig:spacings}. 
We associate this with a {\em decreasing\/} rotation rate.

We can now do a simple calculation to estimate the optimal slope $a$ of a linear
rotation profile $\Omega(r)=a\,r+b$ that best matches the observed 
{\em slope}. We assume that the slope of the fit to the measured 
splittings is linear with
respect to the slope of the rotation profile, which seems to be the case. The
{\em ratio\/} of the linear slopes associated with the observed
splittings and the splittings of the linearly decreasing test profile discussed
above is 
$\sim 2.0918\times10^3$. 
Therefore, if we use a linearly decreasing rotation profile with a
slope of $2.0918\times10^3$ times the original slope of our (linearly decreasing) test
profile and adjust its mean level to match the {\em mean\/} of
the observed splittings, we might get an idea of the underlying rotation
profile. The slope we obtain in this way is $a\sim-2.0918\,\mu\mathrm{Hz}/R_*$. 
Adjusting the slope of the test profile so as to match the trend
of the observed splittings also requires adjusting $b$ according to Eq.\,(\ref{kw0}) if we
are to match the average of the observed splittings.
We calculate 
the intercept $b$ as 
(see also Eq.\,\ref{eq:kw})
\begin{equation}
\label{kw0}
b=\left<\delta\right>-a \int_0^{R_*}\left<K(r)\right> r\,\mathrm{d}r,
\end{equation}
where $\left< \right>$ denotes the average over all modes, and get 
$b\sim0.957\,\mu\mathrm{Hz}$. These values of $a$ and $b$ imply that 
part of the rotation profile becomes negative. Hence, in the
  case that 
the splittings are actually caused by a linear rotation profile or any other
profile closely resembling a linear one, we deduce that the mean observed
splittings should have been considerably higher in magnitude 
in order to obtain a rotation
profile that would not include counter-rotation inside the star.
Of course, with this exercise, we
were only trying to match the {\em linear trend\/} and the {\em mean\/} 
of the observed splittings through
 a linear test profile.  In this case the {\em rms\/} deviation of the predicted
 splittings from the observed ones is around $22.42\,\mathrm{nHz}$, comparatively
 large compared to the mean error from the observations which is $8.41\,\mathrm{nHz}$ (scaled, from Error Set\,1). 
 We can make use as well of the {\em reduced\/} $\tilde \chi^2$ values which we compute
  throughout
this work as
\begin{equation}
\label{eq:chi}
\tilde \chi^2=\frac{\chi^2}{\nu}=\frac{1}{\nu}\sum_{i=1}^M \left(\frac{\bar \delta_i- \delta_i}{\epsilon_i}\right)^2,
\end{equation}
where $\delta_i$ and $\bar \delta_i$ are the measured and predicted splittings,
 respectively, $\epsilon_i$ are the errors, $\nu$ is the \mbox{{\em effective\/}} number
  of degrees of freedom 
and $M$ is the number of observed splittings. In the case just discussed above we fitted
 two parameters, $a$ and $b$, so the number 
of degrees of freedom is $\nu=M-2=17$ which leads to $\tilde\chi^2\sim 194.8$.
 
\begin{figure}[h]
\centering
\includegraphics[width=0.7\linewidth]{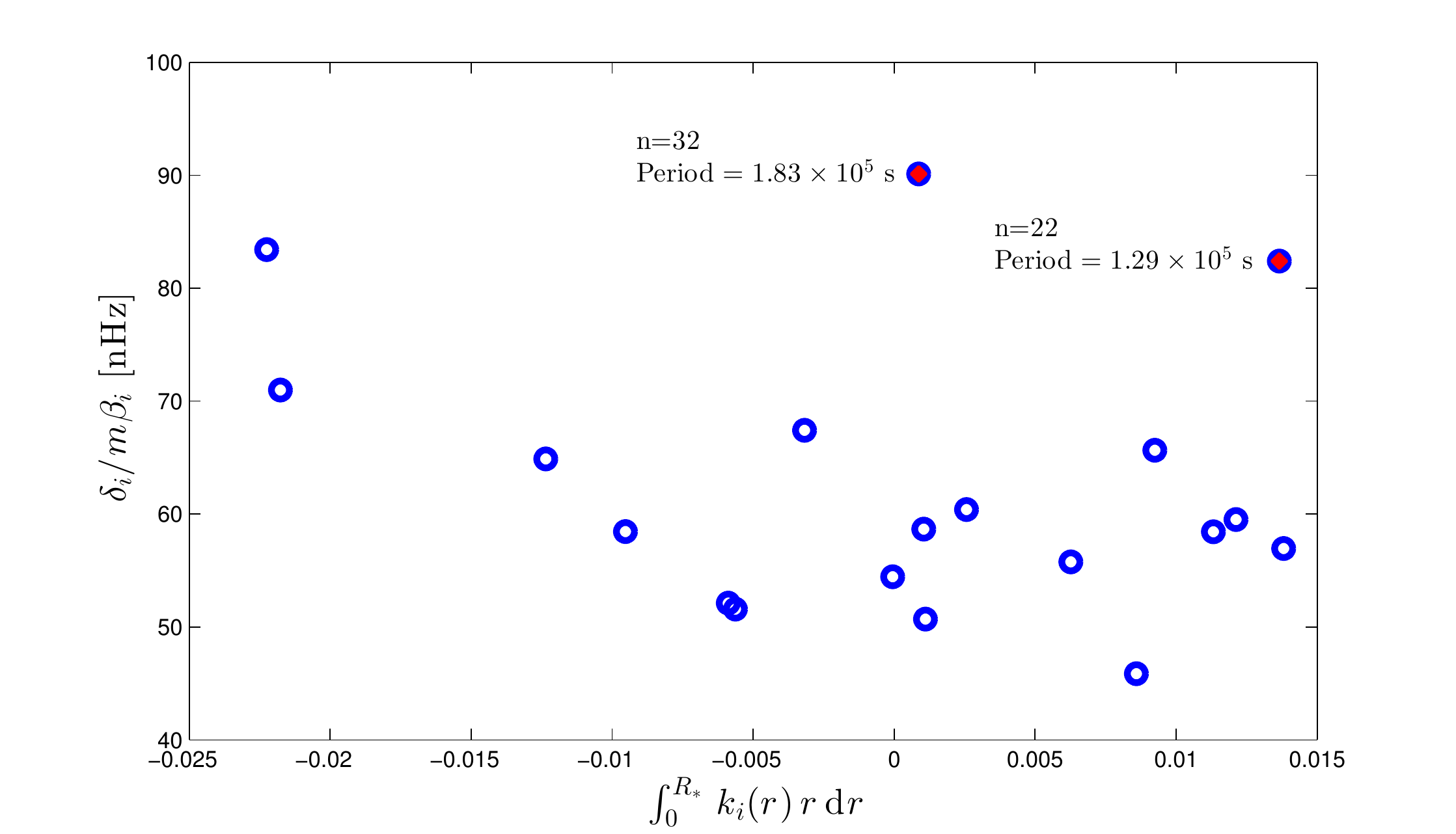}
\caption{The observed splittings (scaled) as a function of
  $\int_0^{R_*}k_i(r)\,r\,\mathrm{d}r$. Even considering the two splittings
  marked in red (which are coincidentally the ones with largest errors), 
there is a downward trend which hints at decreasing
  rotation as the radius increases.}
\label{fig:corr}
\end{figure}

{Figure\,{\ref{fig:corr}} presents another hint towards the decreasing rotation as the radius
increases. We plot the observed splittings as a function of $\int_0^{R_*}k_i(r)\,r\,\mathrm{d}r$, where
$k_i(r)$ is defined in Eq.\,({\ref{eq:k}}).
This should have been a straight line with negative slope if the splittings were actually caused by a linear
rotation profile. As the figure shows, the splittings have
an overall downward trend which is a rough indication that the rotation profile decreases as we move
towards the star's surface.}

 In a next exercise, we searched for the linear profile that minimizes
 $\tilde\chi^2$. This profile has a slope of
$a=-530.7\,\mathrm{nHz}/R_*$ and an intercept $b=285.1\,\mathrm{nHz}$,
leading to $\tilde\chi^2=6.74$. This profile again leads to negative values for the
rotation frequency in the outer envelope. To have an idea of the comparative
statistical significance of this result, we computed the optimal \mbox{{\em constant}}
rotational profile, as well as the optimal linear profile {\em restricted to
  positive values} (using a Lagrange multiplier as an additional fitting parameter).
 We obtained $\tilde\chi_{\mathrm{const}}^2=19.86$ and
$\tilde\chi_+^2=16.86$, respectively. None of the 
positive linear rotational profiles have a $\tilde\chi^2$ similar to the
linear profile with counter-rotation.

\section{Linear, piece-wise rotation models}
\label{piece-wise}

\begin{figure*}[h!]
\centering
\includegraphics[width=\linewidth]{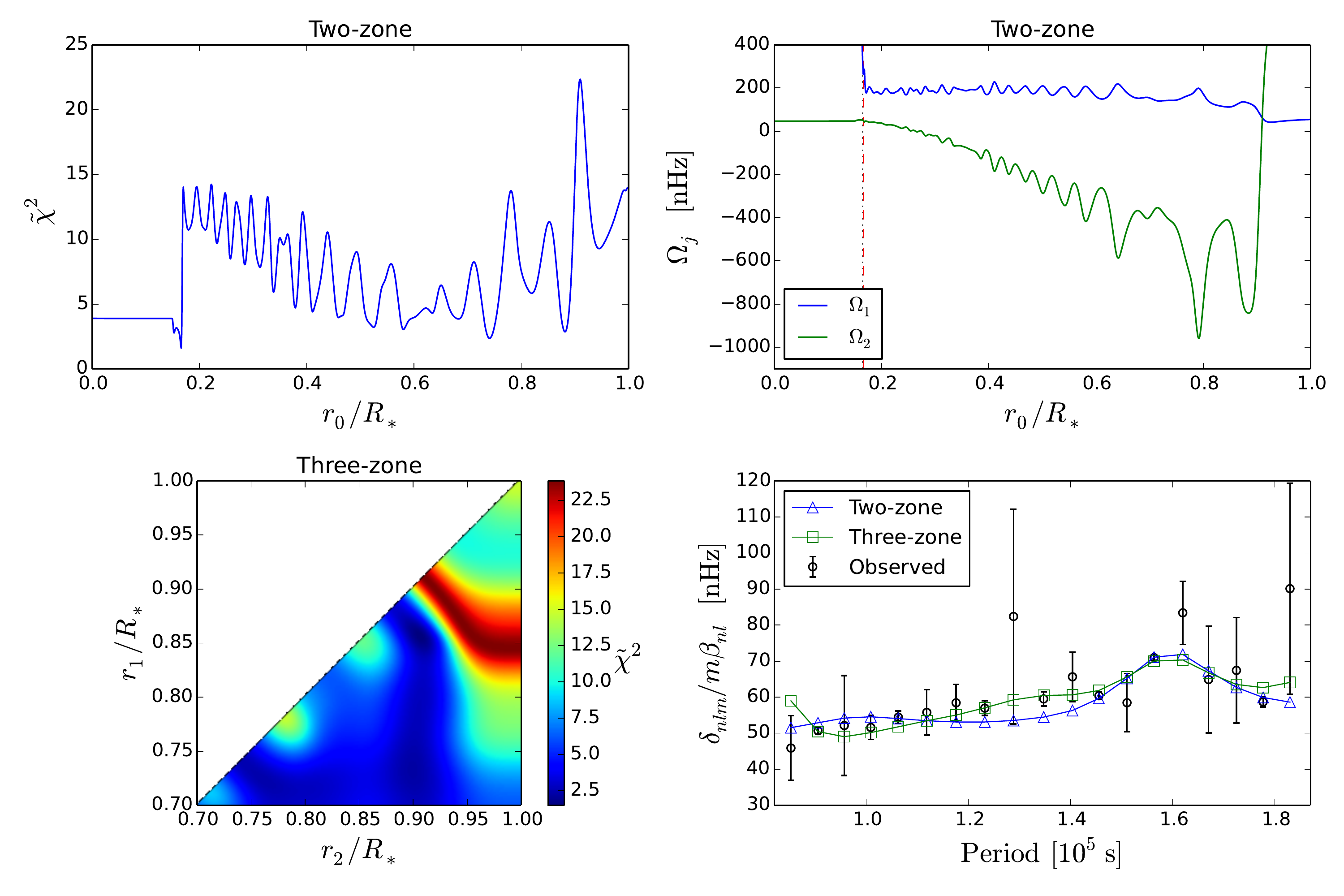}
\caption{Top left: $\tilde\chi^2$ from the difference between the predicted and observed
  splittings as a function of the two-zone parameter $r_0$.  Top right: the
  values of  $\Omega_1$ (blue) corresponding to $r<r_0$ and $\Omega_2$
  (green) for $r>r_0$, both as functions of $r_0$. The vertical dotted line in black
  indicates the extent of the core and the overshoot region, it coincides
  closely with the optimum $r_0$. Bottom left: $\tilde\chi^2$
  computed from a three-zone model with $r_1$ and $r_2$ as parameters. The lowest
  $\tilde\chi^2$ is achieved when $r_1=0.85\,R_*$ and $r_2=0.91\,R_*$. Bottom right: 
  Observed and predicted splittings from the two-zone model together with
  the predicted splittings corresponding to the lowest $\tilde\chi^2$ from the three-zone 
  model (green squares).}
\label{fig:2z}
\end{figure*}
We now assume a two-zone, piece-wise rotational profile such that its value is
$\Omega_1$ if $0<r<r_0$ and $\Omega_2$ if $r_0<r<R_*$.  The parameter $r_0$ is
variable and we optimize $\tilde\chi^2 (r_0)$ so as to best match the
observations, whose errors are derived from Error Set\,1.  The resulting
$\tilde\chi^2$ versus $r_0$ is shown in the top left panel of
Fig.\,\ref{fig:2z}.  {The minimum occurs at $r_0\sim 0.166\,R_*$.}  The values
of $\Omega_1,\Omega_2$ are shown as functions of $r_0$ in the top right panel of
Fig.\,\ref{fig:2z}.
The bottom right panel of Fig.\,{\ref{fig:2z}} shows the observed splittings (scaled) and the
splittings for the two-zone model with the minimum 
at $r_0=0.166\,R_*$, corresponding to $\Omega_1=262.71$ nHz and 
$\Omega_2=49.33$ nHz. We note that the boundary of the convective core of Model\,1,
extended with the core overshoot zone, is situated at $0.1652\,R_*$ \citep{moravveji2015}.
The second 
deepest minimum in the top left panel of
Fig.\,\ref{fig:2z} occurs 
at $r_0=0.741\,R_*$ and corresponds to $\Omega_1=142.0$ nHz and
$\Omega_2=-418.2$ nHz; this solution leads to an $r_0$ that does not play a
special role in Model\,1 in terms of physical quantities.
 
\begin{figure*}[h!]
\centering
\includegraphics[width=0.9\linewidth]{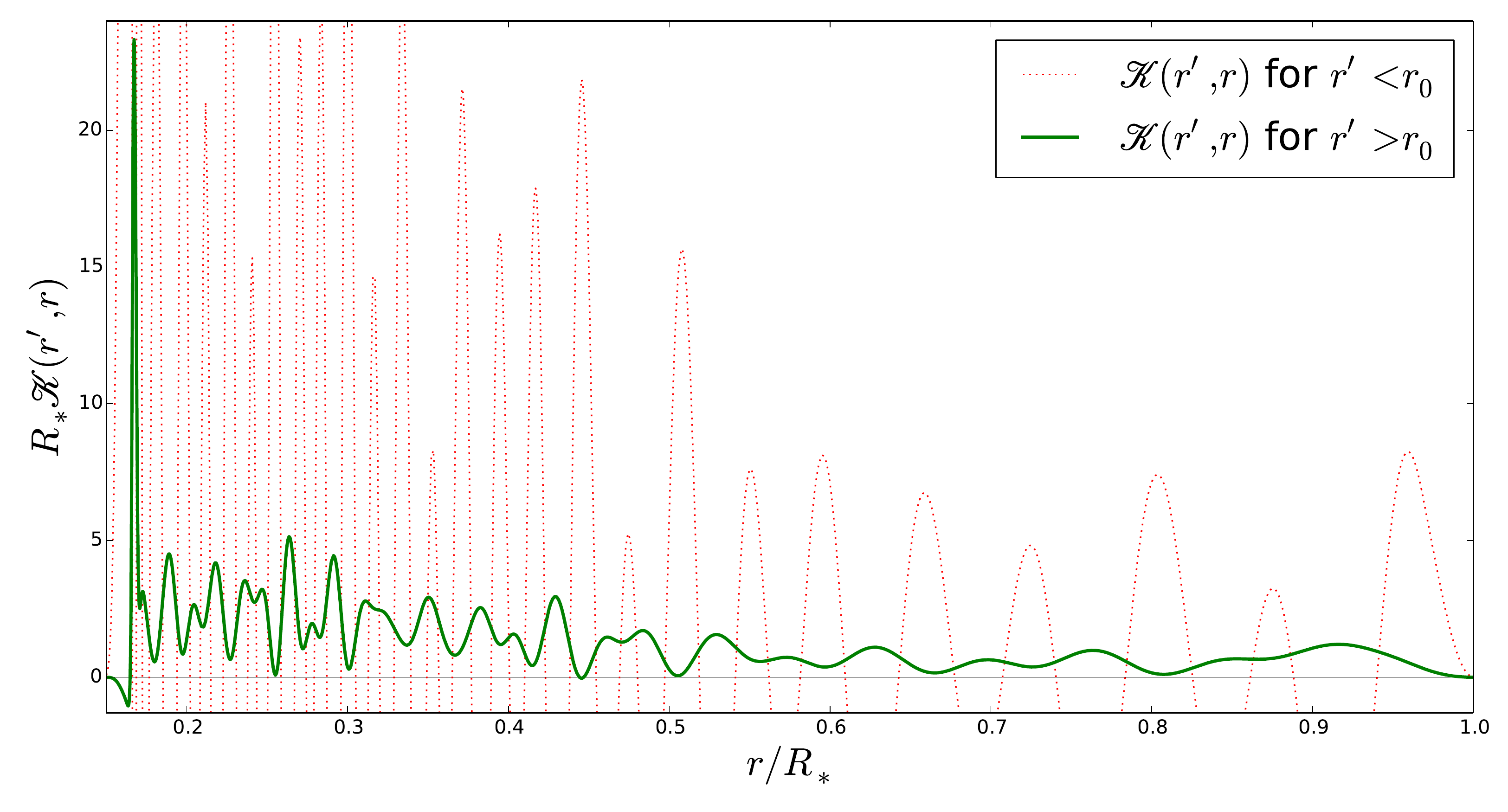}
\includegraphics[width=0.9\linewidth]{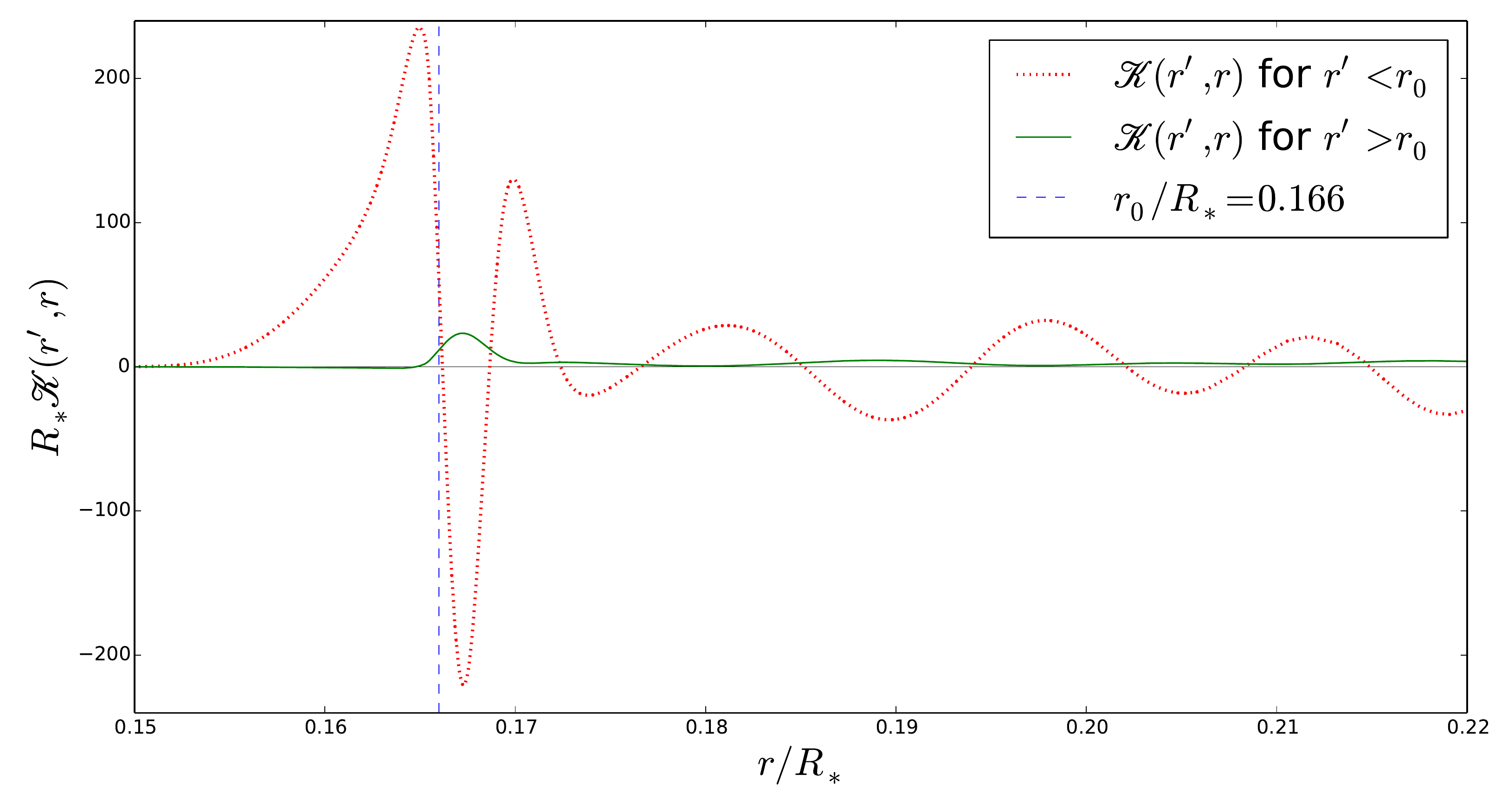}
\caption{Top: averaging kernels $\mathscr{K}(r',r)$ of the optimum two-zone
    model with a  discontinuity at $r_0/R_*=0.166$. Bottom: zoom on the
    core-envelope boundary region. }
\label{fig:2z_avker}
\end{figure*}

 The two-zone model thus favors a region rotating with a period of 42\,d
  near the core overshoot zone and a co-rotating envelope with a period of
  254\,d ($\tilde\chi^2=1.59$).  The second minimum has $\tilde\chi^2=2.36$ and
  corresponds to a counter-rotating profile.  The averaging kernels (for a
  definition, see Section\,6) associated with the best two-zone model are presented
  in Fig.\,\ref{fig:2z_avker} and reveal that the outer zone averaging kernel
  probes mostly the radiative envelope, while the inner zone averaging kernel
  exhibits a large maximum just before reaching the core-envelope boundary and
  rapid oscillations around zero away from it.

\FloatBarrier

A slightly more complex version of this two-zone model is implemented by
introducing a third middle zone where the rotation profile changes linearly from
$\Omega_1$ at the end of the inner zone to $\Omega_2$ at the start of the outer
zone. In this case, we have two linear parameters $\Omega_1,\Omega_2$ and
two non-linear parameters $r_1,r_2$ defining the
zone boundaries. The lowest $\tilde\chi^2=1.51$ for this three-zone model is achieved
when $r_1=0.85\,R_*$ and $r_2=0.91\,R_*$ (see bottom left panel of Fig.\,{\ref{fig:2z}}) and indicates
counter-rotation $(\Omega_1=151.8\,\mathrm{nHz},\,\Omega_2=-1069.5\,\mathrm{nHz})$. The 
corresponding splittings are also shown
in the bottom right panel of Fig.\,\mbox{\ref{fig:2z}}.

Very similar results are obtained when the uncertainties are derived from Error
Set\,2 or when the most asymmetric modes are excluded as evidenced by
Figs.\,\ref{fig:2z_2} through \ref{fig:2z_4}. All the three-zone $\tilde\chi^2$
minima in the four cases, i.e., using Error Set\,1 or 2 and with or without the
most asymmetric splittings, correspond to cases with counter-rotation.
  According to their $\tilde\chi^2$ values, these three-zone models have a
  statistical significance comparable with the best two-zone model. However,
the position of their discontinuities $r_1$ and $r_2$ have no obvious physical
meaning.

To estimate the performance of different models, with the aim to choose the best
  one in a statistical sense, one can only rely on likelihood ratios in the case
  of {\it nested\/} models, i.e., for models where all terms of a simpler
  model
version also occur
  in a more complex version of the model \citep[e.g.,][]{hastie2009elements}.
  We are not in such a situation here because we wish to compare linear,
  discontinuous linear multi-zone, and continous non-linear inversion profiles
  (the latter will be discussed in the next Section). In such a case of
  non-nested models, an adequate statistical measure for model selection is the
Akaike Information Criterion (AIC), which assigns a score to a given model
rewarding goodness-of-fit (e.g., as measured by $\chi^2$) but penalizing
overfitting, thus discouraging the use of complex models with too many
adjustable parameters \citep{burnham2002model, hastie2009elements}. For our
purposes, having only a relatively small number of measurements ($M=19$), it is
appropriate to use the {\em corrected\/} AIC, denoted AICc, which we define
according to its common use in the literature \citep[e.g.,][]{hurvich1989}:
 \begin{equation}
 \label{eq:aic}
 \mathrm{AICc}=2k+\chi^2+\frac{2k(k+1)}{M-k-1},
 \end{equation}
where $k$ is the number of parameters to fit and $M$ is the number of observations.
As advocated by \cite{burnham2002model}, $k$ should include the variance of the residuals as a
parameter to be fitted, e.g.\ $k=3$ for a linear regression. 
With this definition, $k=M-\nu+1$. 
The preferred model among a set of models is the one with the lowest AICc value, where we
limit proper model comparison to the case $M-k > 1$. These AICc
values are only intended for model inter-comparison and have no absolute meaning
by themselves.
Table\,{\ref{aic}} lists all the rotation models considered in this work together
with some of their associated statistical measures, including their AICc's.
Similar tables based on Error Set\,2 or by avoiding the most asymmetric
splittings, are given in Appendix\,\ref{apx_A}.
We can see from the AICc values in Table\,{\ref{aic}} 
that the two-zone piece-wise model  outperforms the three-zone model.

\begin{deluxetable}{lrrrr}
  \tablecolumns{5} \tablewidth{0pc} \tablecaption{\label{aic}
Comparison of rotation profiles for Model\,1 (except for one entry, where we used
Model\,2). The profiles marked with + are enforced to be positive-definite.
The effective number of degrees of freedom $\nu$ for the inversions 
were computed as $M-\kappa$, where $\kappa$ is trace of the `hat' matrix $\mathbf{G}\mathbf{C}^\top$
 (see Section \ref{sec:theory}), following the same
approach as \cite{Deheuvels2014} and explained in
\cite{hastie2009elements}. Error Set\,1 is used here.}
  \tablehead{ \colhead{Rotation Profile} & 
\colhead{{\it rms} error [nHz]} & \colhead{$\nu$}
   & \colhead{$\tilde\chi^2$} & \colhead{AICc} }
    \startdata
    Constant     & 11.63 & 18.00 &  19.86 &  362.3 \\ 
    Linear       & 12.32 & 17.00 &   6.74 &  122.1 \\ 
    Linear+      & 11.46 & 16.00 &  16.86 &  280.7 \\ 
    Two-zone     & 10.88 & 16.00 &   1.59 &  36.36 \\ 
    Three-zone   &  9.48 & 15.00 &   1.51 &  37.35 \\
     
    RLS, $N=8$   &  9.31 & 14.43 &   0.52 &  24.52 \\ 
    RLS, $N=14$  & 10.46 & 14.32 &   1.44 &  38.18 \\
     
    RLS, $N=8$ (Model 2) & 11.35 & 14.74 & 6.42 & 110.3 \\
    
    RLS+, $N=8$  & 12.05 &  10.43 & 17.16 & 222.2 \\
    RLS+, $N=14$ & 11.09 &  5.32 &  3.07 & 184.3 \\

    SOLA, $N=8$  &  9.77 & 12.64 &   0.58 &  33.59 \\ 
    SOLA, $N=14$ &  8.19 &  9.28 &   0.45 &  60.11 \\ 
       \enddata
\end{deluxetable}

We end this Section by noting that the
overall spectral line broadening of 18\,km\,s$^{-1}$ 
measured for KIC\,10526294, which is the combination of rotational and
pulsational broadening \citep{papics2014}, is compatible with all the
$\Omega_2$-values found from the minima listed in this Section 
and does not allow any
discrimination among those solutions as it was the case for the subgiant
studied by \citet{Deheuvels2012}. In the following, we investigate
rotation profiles obtained from inversion methods.
%\FloatBarrier

\section{Inversions}
\label{sec:theory}
We first introduce the basic concepts and terminology behind the inversion
approaches we applied.  We are interested in the approximate determination of
$\Omega(r)$ based on a set of observed rotational splittings
$\delta_{nlm}$, see, e.g., \cite{gough1985} for one of the earliest applications
of this method in the solar case and \citet{kawaler1999} and
\citet{Deheuvels2012} for applications to white dwarfs and subgiants,
respectively. 
This constitutes a linear problem and the 
approximate solution
$\bar \Omega(r)$ can be written as
\begin{equation}
\label{eq:c_i}
\bar \Omega(r)=\sum_{i=1}^M c_i(r) \Delta_i,
\end{equation}
where $\{i\}$ represents the collective index $\{nlm\}$, $\Delta_i \equiv
\delta_{nlm}/m \beta_{nl}$ are the scaled splittings, $c_i(r)$ are the
yet-unknown {\em inversion coefficients}, and $M$ is the number of observed
modes.  It is convenient to express the approximate rotational profile $\bar
\Omega(r)$ in terms of the true profile $\Omega(r)$ by means of the {\em
  averaging kernels} $\mathscr{K}(r',r)$. They are related to the kernels
$K_i(r)$ through $ \mathscr{K}(r',r)=\sum_{i=1}^M c_i(r') K_i(r)$ and fulfill
\begin{equation}
\label{eq:avker}
\bar \Omega(r')=\int_0^{R_*} \mathscr{K}(r',r)\, \Omega(r)\, \mathrm{d}r.
\end{equation}
From the preceding relation it is clear that the averaging kernels
$\mathscr{K}(r',r)$ should be as much as possible localized around $r'$, ideally
resembling a delta function $\delta(r',r)$.

We consider now in what follows a radial grid (scaled with the stellar radius
$R_*$) of $N+1$ uniformly spaced points $\{r_0,r_1,\ldots,r_N\}$ with $r_0=0$
and $r_N=1$, covering the full range of fractional radius. The goal of the
inverse problem is to determine the $N$ unknowns $\bar \Omega_j$, which represent
the predicted angular velocity $\bar \Omega$ at radius $r$ such that
$r_{j-1}\leq r \leq r_j$, where $j=1,\ldots,N$ is the grid index. This
discretization of $\bar\Omega$ on a radial grid allows us to write an expression
for the corresponding predicted splittings $\bar\Delta_i$, based on
Eq.\,(\ref{eq:rotsplit}), as:
\begin{equation}
\label{eq:gmatrix}
\bar\Delta_i=\sum_{j=1}^N G_{ij}\bar\Omega_j,\,\mathrm{where}\,\, G_{ij}=
\int_{r_{j-1}}^{r_j}K_i(r)\, \mathrm{d}r,
\end{equation}
or simply $\mathbf{\bar\Delta}=\mathbf{G} \mathbf{\bar\Omega}$ in matrix
form. Analogously, we express Eq.\,(\ref{eq:c_i}) as
\begin{equation}
\bar\Omega_j=\sum_{i=1}^M c_{ij} \Delta_i,
\end{equation}
with the inversion coefficients $c_{ij}$ constituting the matrix
$\mathbf{C}$. It is instructive to put the relation above in terms of the
discrete version of the {\it true\/} rotation profile $\Omega(r)$. It is
straightforward to show that the matrix $\mathbf{A}$, defined through
$\mathbf{A}=\mathbf{C}^\top \mathbf{G}$, accomplishes such task by fulfilling
\begin{equation}
\label{eq:Ajk}
\bar\Omega_j=\sum_{k=1}^N A_{jk} \Omega_k.
\end{equation}
Ideally $A_{jk}$ should resemble a Kronecker-delta $\delta_{jk}$ indicating that
the recovered profile at a given radius (specified by the grid index $j$) does
not suffer from `leakage' coming from other radial regions. The matrix
$\mathbf{A}$ is thus the discrete equivalent of the averaging kernels
$\mathscr{K}(r',r)$.

Note that the observed splittings are linearly related to the predicted
 splittings through
 the matrix $\mathbf{G}\mathbf{C}^\top$, which is known as the `hat' matrix.
 The trace of this matrix is an estimate of the effective number of 
adjustable parameters \citep[][p. 232]{hastie2009elements}.

If the observational errors $\epsilon_i$ are uncorrelated, as we assume here,
the variance of the recovered profiles can be estimated as
\begin{equation}
\label{eq:variance}
\sigma^2\left(\bar\Omega_j\right)=\sum_{i=1}^M c_{ij}^2 \epsilon_i^2.
\end{equation}
The relation above accounts for the errors on the measurements only and its impact 
on the inversion results; it does not account for the errors inherent to the inversion 
process itself.

Below we describe two different inversion techniques we used to obtain an
approximation to the internal rotation of KIC\,10526294, as well as quantitative
estimates of the uncertainties originating from the measurement errors.

\subsection{Regularized least squares method}

A technique commonly used in helio- and asteroseismology is the regularized
least squares (RLS or Tikhonov) method \citep[e.g.,][]{craig1986}, 
which seeks to minimize 
the quantity
$\mathscr{T}$ defined as
\begin{equation}
\mathscr{T}=\sum_{i=1}^{M} \frac{\left(\Delta_i - \bar\Delta_i \right)^2}{\epsilon_i^2}+
\mu_{\rm RLS} \int_0^{R_*}\left(\frac{\partial^2 \bar \Omega}{\partial r^2}\right)^2 \mathrm{d}r,
\label{eq:Tik}
\end{equation}
where $\Delta_i$ are the observed splittings, $\epsilon_i$ the corresponding
measurement error, $\bar\Delta_i$ are the predicted splittings, and $\mu_{\rm RLS}$ is a
free parameter (known as the regularization or smoothing parameter) used to
limit the norm of the second derivative on the predicted $\bar\Omega(r)$.  Using
our discrete radial grid described earlier, minimization means $\partial
\mathscr{T}/\partial \bar \Omega_j=0$ for all $j$. This condition can be written
more explicitly using Eq.\,(\ref{eq:gmatrix}) and Eq.\,(\ref{eq:Tik}) as
\begin{equation}
\label{eq:minmatrix}
\mathbf{H}^\top\left(\mathbf{G} \mathbf{\bar \Omega}-\mathbf{\Delta}\right)+
\frac{\mu_{\rm RLS}}{\epsilon_0^2\delta^3}\mathbf{L}^\top \mathbf{L}\, \mathbf{\bar\Omega}=0,
\end{equation}
where $H_{ij}= G_{ij}/\epsilon_i^2$, $\bf L$ being the discrete second derivative 
operator, $\delta$ the radial grid spacing, and $\epsilon_0^2$ is a scale factor
 (introduced for convenience) which we set equal to the squared mean of the errors $\epsilon_i$. A formal solution is
\begin{equation}
\label{eq:formal}
\mathbf{\bar\Omega}=\left(\mathbf{H}^\top\mathbf{G}+\frac{\mu_{\rm RLS}}{\epsilon_0^2\delta^3}
\mathbf{L}^\top \mathbf{L}\right)^{-1} \mathbf{H}^\top \mathbf{\Delta}.
\end{equation}
Obviously, the inverse matrix on the right hand side of Eq.\,(\ref{eq:formal}) 
might not exist and in practice we seek a solution in the {\it least squares sense} 
instead.

\subsection{The SOLA method}

The Subtractive Optimally Localized Averaging (SOLA) method \citep{pijpers1994} 
consists in
determining a linear combination of the inversion coefficients $c_i(r')$ such
that the averaging kernels resemble as much as possible a {\em target function}
$T(r',r)$ while keeping the variance of the predicted profiles,
$\sigma^2(\bar \Omega)$, low. To implement this method we minimize
\begin{equation}
\int_0^{R_*} \left[ \mathscr{K}(r',r)-T(r',r) \right]^2 \mathrm{d}r + \frac{\mu_{\rm SOLA}}
{\epsilon_0^2} \sum_{i=1}^M c_i^2(r')\, \epsilon_i^2
\end{equation} 
at each $r'$, with the constraint $\int_0^{R_*} \mathscr{K}(r',r)\,
\mathrm{d}r=1$. In addition to the free parameter $\mu_{\rm SOLA}$ we can also adjust the
shape of the target function $T$. The problem reduces to solving the linear set
of $M$ equations ($i=1,\ldots,M$ and for each radial location $r'$)
\begin{equation}
\label{eq:sola}
\sum_{k=1}^M W_{ik}\, c_k(r')=\int_0^{R_*} K_i(r)\, T(r',r)\, \mathrm{d}r,
\end{equation}
where $W_{ik}=\int_0^{R_*} K_i(r)\, K_k(r)\, \mathrm{d}r + (\mu_{\rm SOLA}/\epsilon_0^2)\, \delta_{ik}\,
\epsilon_i^2$, together with the constraint $\sum_k c_k(r')=1$. Using the
discrete radial grid with $N$ segments and choosing the target functions as
\begin{displaymath}
\label{eq:target}
T(r',r) = \left\{
\begin{array}{ll}
 N & \mathrm{if}\,\, r' \in (r_{j-1},r_j]\\
 0 & \mathrm{otherwise,}
\end{array}
\right.
\end{displaymath}
the problem to solve becomes
\begin{equation}
\sum_{k=1}^M \left[N\left(\mathbf{G}\,\mathbf{G}^\top\right)_{ik}+\frac{\mu_{\rm SOLA}}
{\epsilon_0^2}\,\delta_{ik}\,\epsilon_i^2\right]c_{kj} = NG_{ij},
\end{equation}
with the constraint $\sum_{k=1}^M c_{kj}=1$. In this case the target function
will approach a Dirac-$\delta$ as $N$ increases. 
On the other hand, $N$ can be low or moderate as long as $\Omega(r)$ can
be assumed not to vary appreciably over a radial segment of the grid.

\subsection{Profiles from inversion}
\label{ProfilesInversion}

We present the internal rotation profiles obtained by the two methods described
above. For both the RLS and SOLA methods, we scanned different resolutions
ranging from $N=2$ to 14, each covering a wide range of smoothing parameters
$\mu_{RLS}$ and $\mu_{SOLA}$, to examine the resulting inversion profiles. To
choose the appropriate parameters $N$ and $\mu$ is not an easy task and in practice it depends
on the {\em a-priori\/} information we might have. For example, if we can assume
that the rotation profile does not change appreciably over a radial distance
$\lambda$, then we can safely use a resolution such that
$N=1/\lambda$. Alternatively, we can interpret the inversion result as providing
information only on those components of the rotation profile that do not change
appreciably over a distance $1/N$. Generally the higher the change of the
profile over a given length scale is, the lower the accuracy of the
inversion. If we consider that the rotation profile can be represented as a
superposition of profiles with increasing detail (e.g.\ like a Fourier
expansion), then the inversion methods can provide useful information at least
on those components that change slowly with radius. Below we discuss the results
for the $N$ values with the best statistical score in terms of the AICc.

\begin{figure*}[ht!]
\centering
\includegraphics[width=0.9\linewidth]{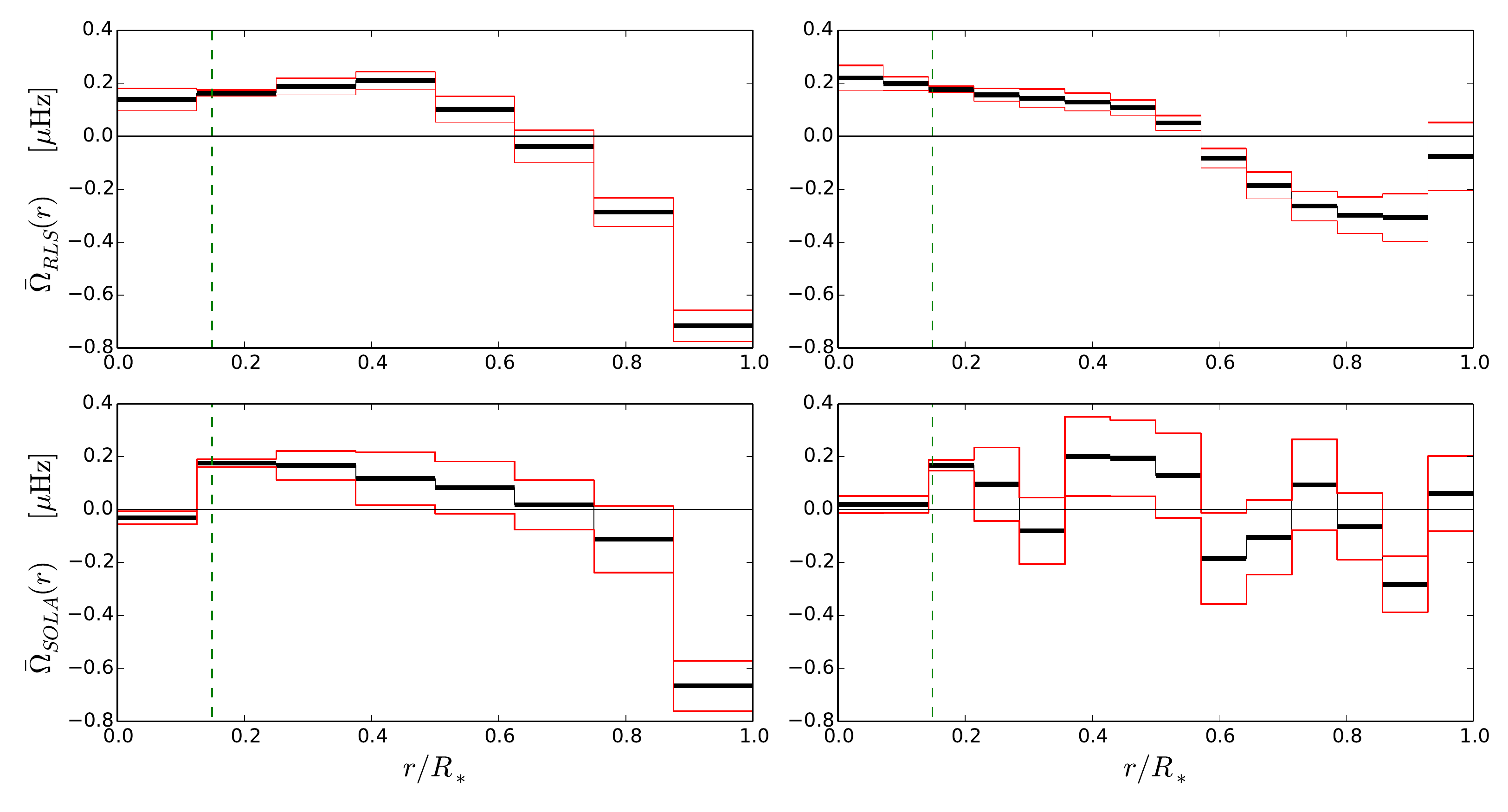}
\caption{Inversion profiles resulting from the RLS method (top row) with 
$\mu_{RLS}=10^{-5}$ and the SOLA method (bottom row) with $\mu_{SOLA}=10^{-2}$ 
for two different resolutions $N={8,14}$ (left to right). Vertical dashed 
lines indicate the approximate location of the convective core's radius. 
Error Set\,1 has been used.}
\label{fig:pom1}
\end{figure*}

\begin{figure}[t!]
\centering
\includegraphics[width=0.7\columnwidth]{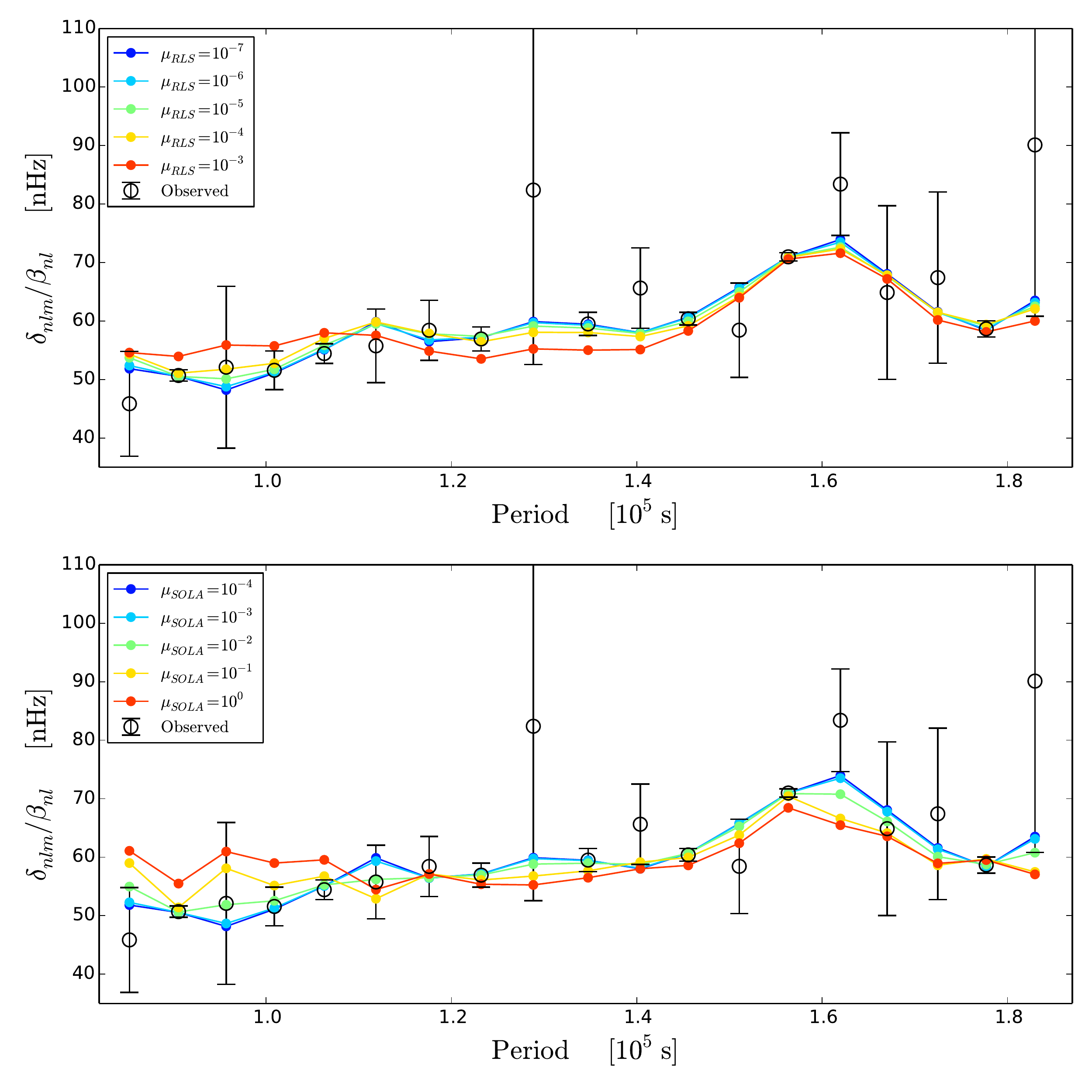}
\caption{The predicted splittings for both the RLS (top) and SOLA methods (bottom)
 using a resolution of $N=14$ and a range of $\mu$ parameters for each
  method. The predicted splittings were computed based on the uncertainties from 
Error Set\,1.}
\label{fig:ps_full}
\end{figure}

We show in Fig.\,\ref{fig:pom1} the inversion profiles with $\mu_{RLS}=10^{-5}$
and $\mu_{SOLA}=10^{-2}$, which represent a good balance between error and
resolution for Error Set\,1. Similar figures for Error Set\,2 and/or ignoring
the asymmetric splittings are shown in Figs.\,\ref{fig:pom2}-\ref{fig:pom2_mask}
in Appendix\,\ref{apx_A}.  We see that both methods give qualitatively similar
results for $N=8$ and $N=14$. The uncertainty of the SOLA method grows quickly
as $N$ is increased, as opposed to the RLS method. The averaging kernels, as can
be judged from Figs.\,\ref{fig:ctg_rls} and \ref{fig:ctg_sola} in
Appendix\,\ref{apx_B}, are better localized using the SOLA method (as expected, by design),
although the uncertainties are somewhat larger compared to the RLS method.  The
kernels feature larger amplitudes near $r\sim0.2\,R_*$, corresponding to small
uncertainties at that location.  Further out in radius, near $r \sim 0.9\,R_*$,
the uncertainty is larger but the rotation rate is still constrained to be
opposite in sign. Note that the kernels provide no information for $r <
0.15\,R_*$ so the inversion results within this radial range have no meaning
(cf.\ Appendix\,\ref{apx_B}).  An appropriate assessment of the inversion's
accuracy is provided by the averaging kernels, or their discrete counterpart
represented by the matrix $\bf\mathrm{A}$ (see Figs.\,\ref{fig:ctg_rls} and
\ref{fig:ctg_sola}).
Generally speaking, as $\mu_{RLS}$ or $\mu_{SOLA}$ increases, the predicted
splittings $\bar \Delta_i$ will deviate more from the measured ones.  The
predicted splittings for $N=14$ using both RLS and SOLA methods and Error Set\,1
is displayed in Fig.\,\ref{fig:ps_full}. The result in the case of the omission
of the asymmetric splittings is provided in Fig.\,\ref{fig:ps_mask} of
Appendix\,\ref{apx_A}.

The best two-zone model from Section\,5 might seem in disagreement with the
  profiles obtained from inversion but upon further inspection they are actually
  in good agreement, at least for the outer zone. Indeed, if we take the profile
  from RLS inversion with $N=8,\,\mu_{RLS}=10^{-5}$ and interpolate it
  appropriately, the resulting average rotation using the outer zone averaging
  kernel displayed in Fig.\,\ref{fig:2z_avker} amounts to $50.34$ nHz, which is
  in very good agreement with the $49.33$ nHz value obtained in Section\,5 for
  the outer zone. The rotation rate computed from the inversion profile and the
  inner zone averaging kernel amounts to $206.01$ nHz, which is to be compared
  with the $262.71$ nHz value for the inner zone in the two-zone model.
  Although both the best RLS model and the best two-zone are mutually
  consistent, the RLS model resolves the outer zone better. The AICc values also
  give preference to the RLS inversions as we deduce from Table\,\ref{aic}.

  To add yet another model comparison, we have performed the so-called
  leave-one-out cross-validation technique \citep[see
  e.g.][]{hastie2009elements}, which does not rely on $\tilde\chi^2$
  values. This technique consists in omitting one of the measurements when
  fitting a model and comparing the predicted splitting based on the fitted
  model $\bar \Delta_i$ with the measurement that has been omitted $\Delta_i$:
  $e_i=(\bar \Delta_i -\Delta_i)/\epsilon_i$. We performed this procedure for
  each of the measurements at a time and obtained a final score by computing the {\it
    rms\/} value of all the $e_i$. The preferred model is then the one with the
  lowest score. In the case of the two-zone model this score is $1.51$ while for
  the best RLS inversion the score is $0.79$. 

 Using a synthetic profile without counter-rotation as a test,
 we found that the RLS inversion method works properly and that it is not prone
to give spurious counter-rotation solutions. See Appendix \ref{apx_D} for further details.

\FloatBarrier

\begin{figure}[h!]
\centering
\includegraphics[width=\columnwidth]{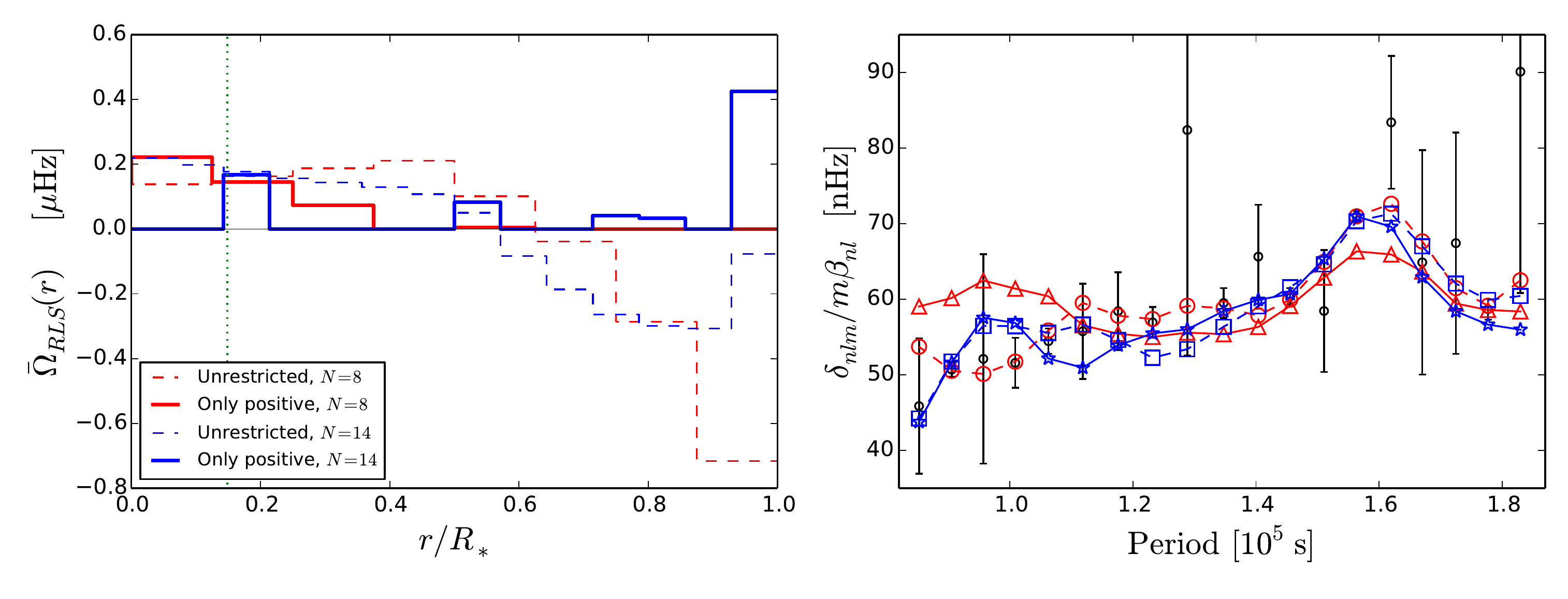}
\caption{Left: comparison of inversion profiles obtained by restricting the rotation
to be positive for all $r$ and those using unrestricted RLS regularization for two
 different resolutions $N=8,14$ and for $\mu_{RLS}=10^{-5}$. Right: the 
 corresponding predicted splittings
compared to observations. See main text for details.}
\label{fig:pprior}
\end{figure}

It is possible to perform RLS regularized inversions enforcing {\em a-priori\/} a rotation
profile that does not exhibit counter-rotation. We can achieve this by including
 additional terms to the quantity $\mathscr{T}$ defined in Eq.\,(\ref{eq:Tik}) in such 
 a way that they represent our prior knowledge of the rotation profiles being definite
 positive. Loosely speaking, we can express the probability of obtaining a value $\Omega_j$
 at a given radial location as being proportional to $e^{\lambda_j \Omega_j}$, with
 $\lambda_j\ge0$. A concrete technique to solve such minimization problem is provided
 by the Karush-Kuhn-Tucker conditions \citep{kuhn1951}, which consists in minimizing
 $\mathscr{T}-\sum_j \lambda_j\Omega_j$ together with the constraints $\lambda_j\Omega_j=0$
 for all radial locations $j$ \citep[see e.g.][]{boyd2004}. The $\lambda_j$ are treated
 as Lagrange multipliers. The result of this exercise is shown
 in Fig. \ref{fig:pprior}.

A qualitative idea of how well these restricted-positive profiles represent
 the data is provided by the plot on the right of Fig.\,\ref{fig:pprior}. Quantitatively,
  the $\tilde\chi^2$ for the restricted-positive profiles are $\tilde\chi_{8+}^2=112.9$ 
and $\tilde\chi_{14+}^2=46.1$ corresponding to resolutions of $N=8$ and $N=14$, 
respectively. These values are
to be compared with $\tilde\chi_8^2=0.52$ and $\tilde\chi_{14}^2=1.44$ from the unrestricted case.
Note here that the effective number of degrees of freedom $\nu$ is reduced considerably when
restricting the profiles to be positive definite given the additional parameters
included as Lagrange multipliers. In Table\,{\ref{aic}} we assembled all the AICc scores
of these and other inversions along with results from previous sections.
 
Although it is possible to increase the resolution $N$ beyond the number of
observations $M$ without overfitting for the inversion result itself, 
following the principle of regularization
(see Appendix \ref{apx_C}), the value of 
$\nu$ computed as the trace of
the `hat' matrix $\mathbf{G}\mathbf{C}^\top$  might become smaller than unity
if the regularization parameters are small, pushing 
the AICc  used for model comparison
beyond meaningful values. For this reason we limit
ourselves to moderate resolutions for the sake of 
meaningful model comparisons. The
best models are the RLS or SOLA inversions with $N=8$ with large
statistical margin over all the other rotational profiles we obtained.

 The profiles obtained using the Error Set\,2 (Figs\,\ref{fig:pom2} and \ref{fig:pom2_mask})
 are very similar to those
obtained with Error Set\,1. Obviously, the inversion uncertainties are larger but still
very similar counter-rotating profiles result. Note that the corresponding $\tilde\chi^2$ 
and AICc values (Table\,\ref{aic2}) give clear indication of appreciable error
over-estimation in the case of Error Set\,2 as already stressed before. Indeed,
$\tilde\chi^2<1$ in virtually all cases. This is already evident from Fig.\,\ref{fig:data} where
the point-to-point variation of the rotational splittings is visibly much smaller than the
average uncertainty from Error Set\,2. The inversion profiles are equally unaffected
if the inversion procedure is carried out excluding the most asymmetric splittings as 
evidenced by Figs.\,\ref{fig:pom1_mask} and \ref{fig:pom2_mask}. This is not surprising since
the large errors associated with the most asymmetric splittings already give them less
weight compared to the others when minimizing $\tilde\chi^2$.

\begin{figure}[h!]
\centering
\includegraphics[width=0.9\columnwidth]{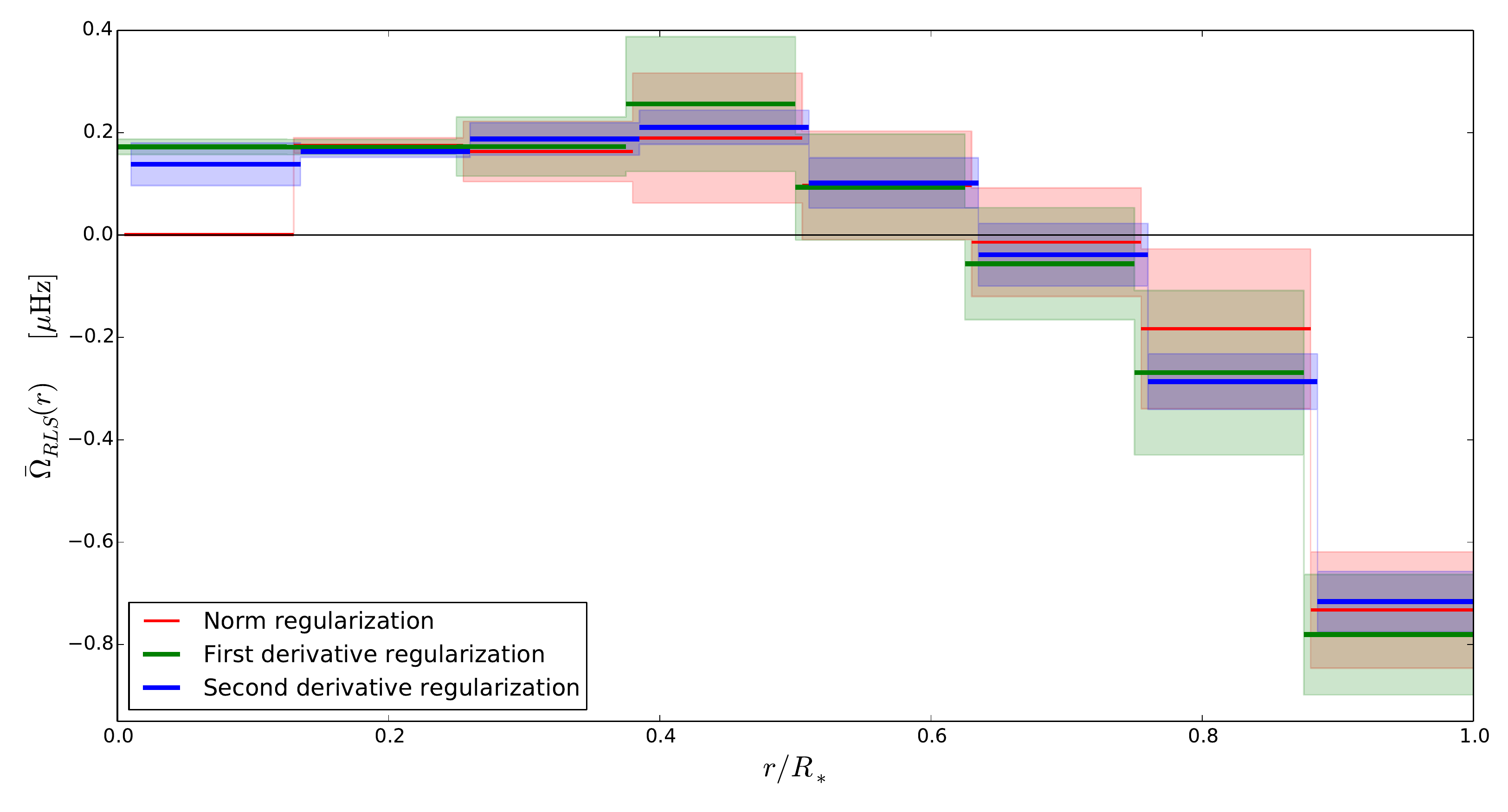}
\caption{ The inversion profiles obtained by regularizing the norm of the
  profile (red, $\mu_{\rm RLS=10^{-6}}\, ,\tilde\chi^2=0.53$), regularizing the
  norm of the first derivative (green, $\mu_{\rm RLS=10^{-6}}\,
  ,\tilde\chi^2=0.51$) and regularizing the norm of the second derivative (blue,
  $\mu_{\rm RLS=10^{-5}}\, ,\tilde\chi^2=0.52$). For visual aid, the profiles
  have been slightly shifted horizontally with respect to each other.  }
\label{fig:012}
\end{figure}

The RLS inversions can be carried out also by regularizing the norm of the first derivative
of the profile or even the norm of the profile itself. it turns out they are all consistent and
have similar properties as the RLS inversions using the norm of the second derivative that we presented
earlier. Figure\,\ref{fig:012} shows the corresponding inversion profiles and their associated uncertainties
derived from the measurement errors.  

\begin{figure}[h]
\centering
\includegraphics[width=0.7\columnwidth]{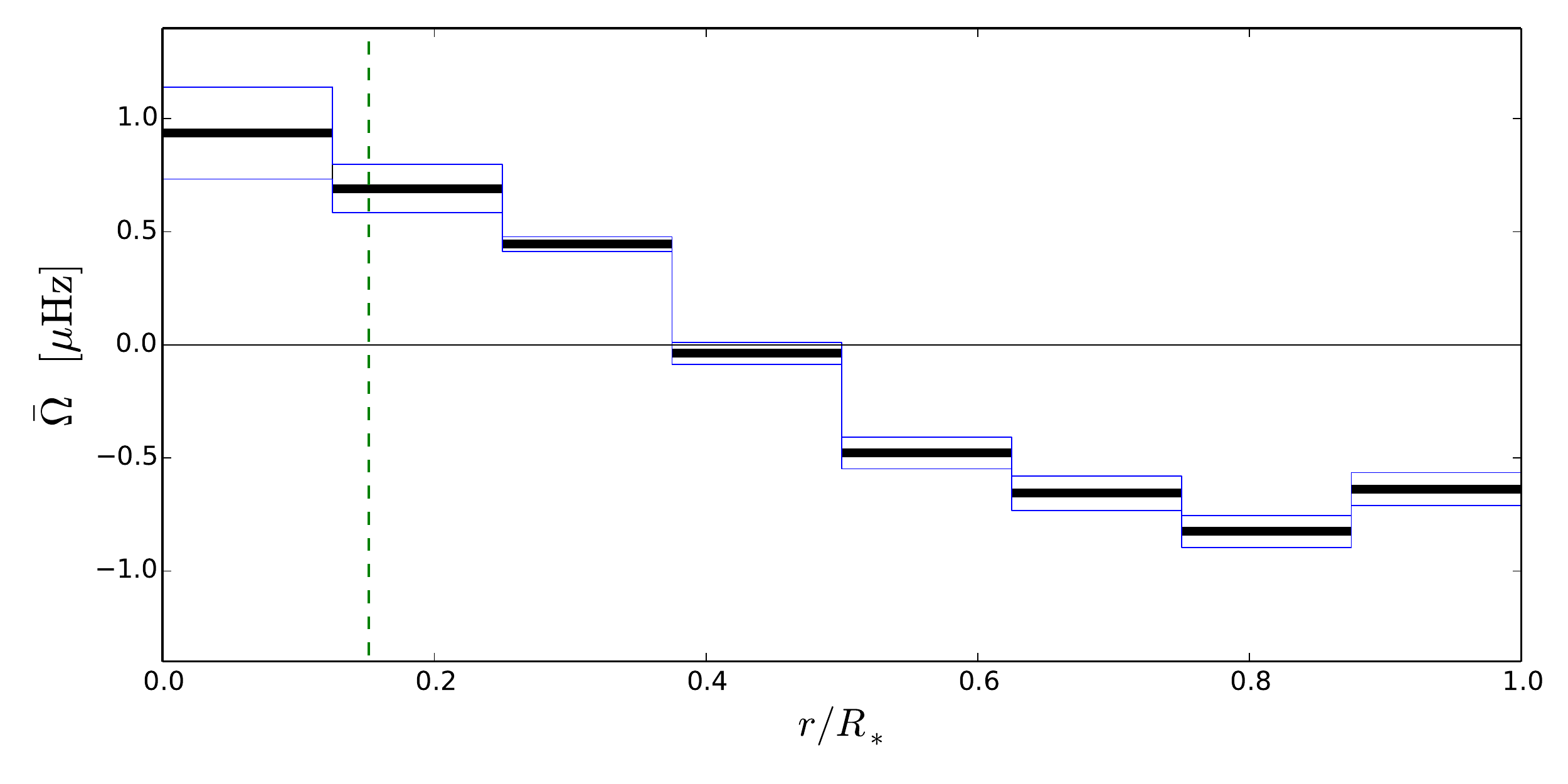}
\caption{The inversion profile obtained using kernels from Model\,2 as found by 
  \citet{papics2014}. Here the parameters are identical to those used for the
  top, left panel in Fig.\,\ref{fig:pom1}
  ($\mu_{RLS}=10^{-5},\,N=8$). The vertical dashed line marks the convective core
  radius. The counter-rotation within the radiative zone is a characteristic
  feature of all the models we examined.}
\label{fig:pom_P1}
\end{figure}

So far we have based our inversions on the Model\,1 found by
\citet{moravveji2015}. However, the main qualitative characteristics of the
inversion profiles are robust under different model choices.  We have
  explicitly tested that all the models among the best ones from the forward
  modeling in \citet{papics2014} and \citet{moravveji2015} produce qualitatively
  similar results for the inverted rotation profiles, i.e., counter-rotation in
  the radiative envelope. To provide a specific example, we show in
Fig.\,\ref{fig:pom_P1} the resulting profile using Model\,2 found by
\citet{papics2014}. Although the kernels based on this model do not exhibit a
large peak right outside the convective core (i.e., no trapped modes), they have
a similar shape otherwise (see also Fig.\,\ref{fig:bv}). Again, the inversion
profile hints at counter-rotation within the star's radiative zone. The
uncertainty on the recovered profile is somewhat larger overall compared to the
uncertainty associated with Model\,1 (Fig. \ref{fig:pom1}, top left panel). We
attribute this to the lack of variation of the kernels in this model, i.e.,
there are no trapped modes to make the kernels more `different' from mode to
mode. The more similar kernels for the modes of Model\,2 result in higher
  $\tilde\chi^2$ and AICc values, which we have added for the case of RLS and
$N=8$ in Table\,\ref{aic}.

\section{Monte Carlo simulations}
\label{sec:mc}

Yet another method to obtain an approximation to the real profile $\Omega(r)$
under the assumption of a smooth profile, 
consists in generating a large collection of random, synthetic test profiles and
assigning a score to each measuring how close the predicted splittings are to the
observed ones. 
This method is straightforward but very inefficient computationally
since the number of synthetic profiles that need to be calculated is necessarily
large.

\begin{figure*}[h!]
\centering
\includegraphics[width=0.7\linewidth]{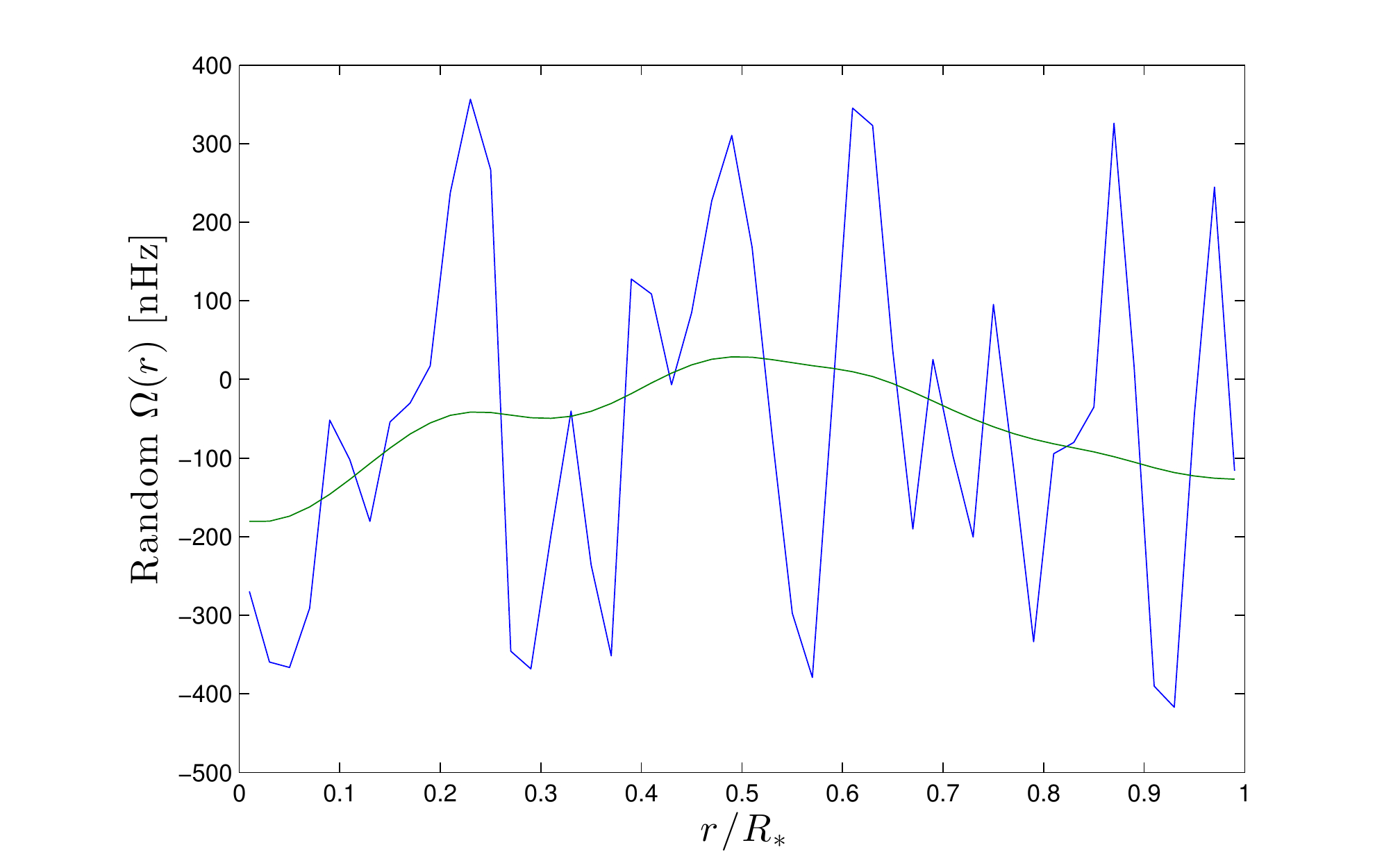}
\caption{A random profile (blue) before any smoothing with $Q=100$ and
  $h=500\,\mathrm{nHz}$.  The `low pass' filtering produces the green curve. The
   correlation length is $\lambda=0.3$.}
\label{fig:smooth}
\end{figure*}

A random, synthetic profile on a radial grid with $Q$ points can be generated by
choosing a random rotation value $\Omega_j$ at each radial location $r_j$. The
rotation values are to be picked from a random (uniform) distribution extending
from $-h$ to $h$ (in $\mathrm{nHz}$), where the range $h$ is to be
chosen appropriately as described below. In general, such a profile will
exhibit strong fluctuations along the radius, i.e. it will be a
'noisy' spiky 
profile,
particularly if $Q$ is large (we used $Q\geq100$ in our simulations). There are
a number of ways  to smooth out the profile, a simple one being  to
use a `low pass' filter to remove the `high frequency' components of the profile (if
we think of it as a time series). The filter cutoff point defines a characteristic length
scale $\lambda$ below which the profile can be considered to have only smooth variations.
 Some padding at each end of the profile is
necessary to avoid end effects when filtering. See Fig.\,\ref{fig:smooth} for an illustrative
example.
After the smoothing the profile will have spatial fluctuations only on length
scales larger than $\lambda$. Since generally $Q>N$, i.e., the random profiles
are defined on a finer grid than the one used to define $\mathbf{G}$, we should
resample the profile to match the radial grid with $N$ points. We do this by
calculating the integral average of the random profile along each segment in the
radial grid associated with $\mathbf{G}$ after an appropriate
interpolation. Other methods can be used to achieve the same result. What is
essential here is that the smoothing should be chosen in accordance to the final
resolution $N$ so that $\lambda N\geq1$.

Once the smoothing is performed, the rotation values are not any longer distributed
uniformly on the $[-h,h]$ interval, resembling instead a Gaussian
distribution. This is simply because we have introduced short range correlations
with our smoothing. Therefore we should adjust the range $h$ to ensure that the
rotation values at a given radial location are more or less equally probable,
thus covering uniformly the expected range of $\Omega(r)$. This expected
range can be roughly estimated from the mean of the observed rotational
splittings $\left< \delta_i/m \beta_i \right>$. As a concrete example we found that
for $\lambda=0.3$ and $h=4\,\,\mu\mathrm{Hz}$, the random profiles visited
more or less uniformly the range $[-0.2, 0.2]\,\,\mu\mathrm{Hz}$ within a
14\% margin.

If we interpret the rotation value $\Omega_j$ at each radial location as a
random variable, and given a large collection of random rotation profiles, it is
possible to calculate the associated covariance matrix. A given row $j$ of this
matrix will resemble a Gaussian distribution centered at $r_j$. The mean of the
FWHM of the Gaussians in all rows is then an (after-the-fact, of course)
estimate of $\lambda$. In practice, the random profiles can be considered
approximately constant over radial scales not larger than $\sim
\lambda/3$.

\begin{figure}[h!]
\centering
\includegraphics[width=0.7\columnwidth]{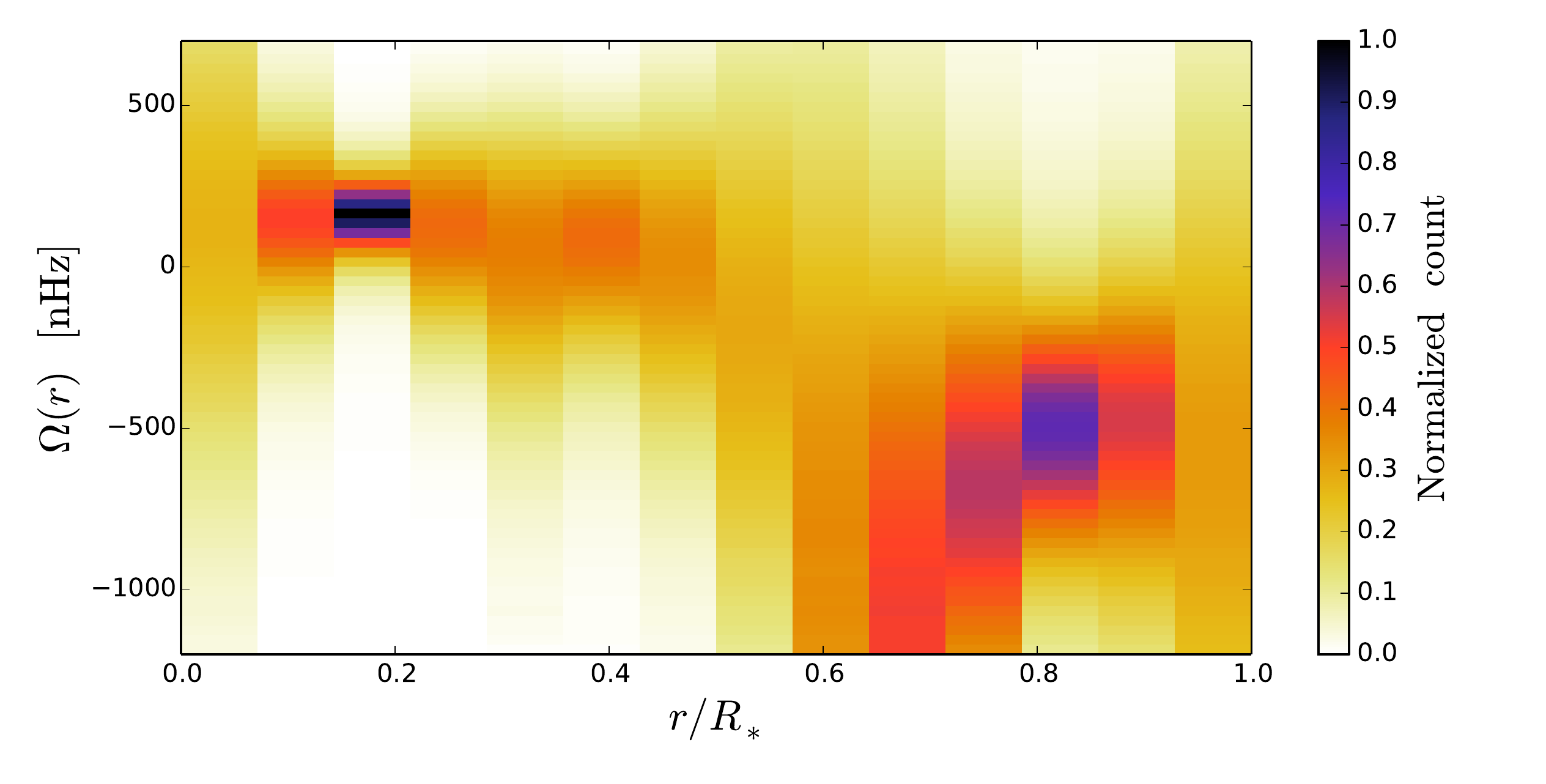}
\caption{ {Color coded \mbox{{\em  weighted\/}} histogram count for each radial location of 
a large sample of random rotation profiles using $1/\chi^4$ as weights (see text
for details).
The original random profiles have been smoothed out so that the 
typical length scale is $\lambda=0.3$.}}
\label{fig:mc}
\end{figure}

Once a random profile has been generated and smoothed out, its
associated splittings are calculated via Eq.\,(\ref{eq:rotsplit}).  We compute then
 a score which is proportional to $\chi^{-4}$ (using Error Set\,1). The lower $\chi^2$ is the higher
 the score becomes. After scoring a large number of profiles ($10^9$),
 we compute the histogram of the rotation rates at each radial location {\em  weighted 
 with their corresponding scores} and then normalized by an ordinary histogram count.
 In this way, for each radial location and each rotation rate interval we obtain 
 a number indicative of its likelihood to explain the observed splittings.

For the results shown in Fig.\,\ref{fig:mc}, the random profiles
have a resolution of $N=14$ and have been smoothed so that $\lambda=0.3$. At each
radial segment $r_j<r<r_j+1$ we computed the  weighted histogram of the ocurrences of $\Omega_j$
(as explained above) over an interval starting from -1500 up 1200 nHz and subdivided 
in 71 bins. By comparing Fig.\,{\ref{fig:mc}} and Fig.\,{\ref{fig:pom1}} we see
 that the Monte Carlo method reproduces very well the rotation rates at $r\sim0.2\,R_*$ while
 giving only a broad distribution of rotation rates centered around negative values at 
 $r\sim0.8\,R_*$. We note that virtually none of the high scoring random profiles are strictly positive 
(or negative), they all involve at least one sign change
along the radial coordinate.
 
This use of random profiles is also suitable to establish the `quality' of a set of kernels. To do this
we take first a random profile (the reference profile) and calculate its associated splittings
via Eq.\,(\ref{eq:rotsplit}). These splittings are then `inverted' and we compare the resulting
profile with the reference profile. Some random noise could in principle be added to the splittings before attempting
the inversion in order to simulate measurement errors, but this is unnecessary
here since Eq.\,(\ref{eq:variance})
already describes properly the effect of the measurement variance on the inversion profiles. The differences 
between inverted and reference profiles can therefore be attributed solely to the inadequacy of the
kernel set to fully recover the solution.

\begin{figure}[h!]
\centering
\includegraphics[width=0.7\columnwidth]{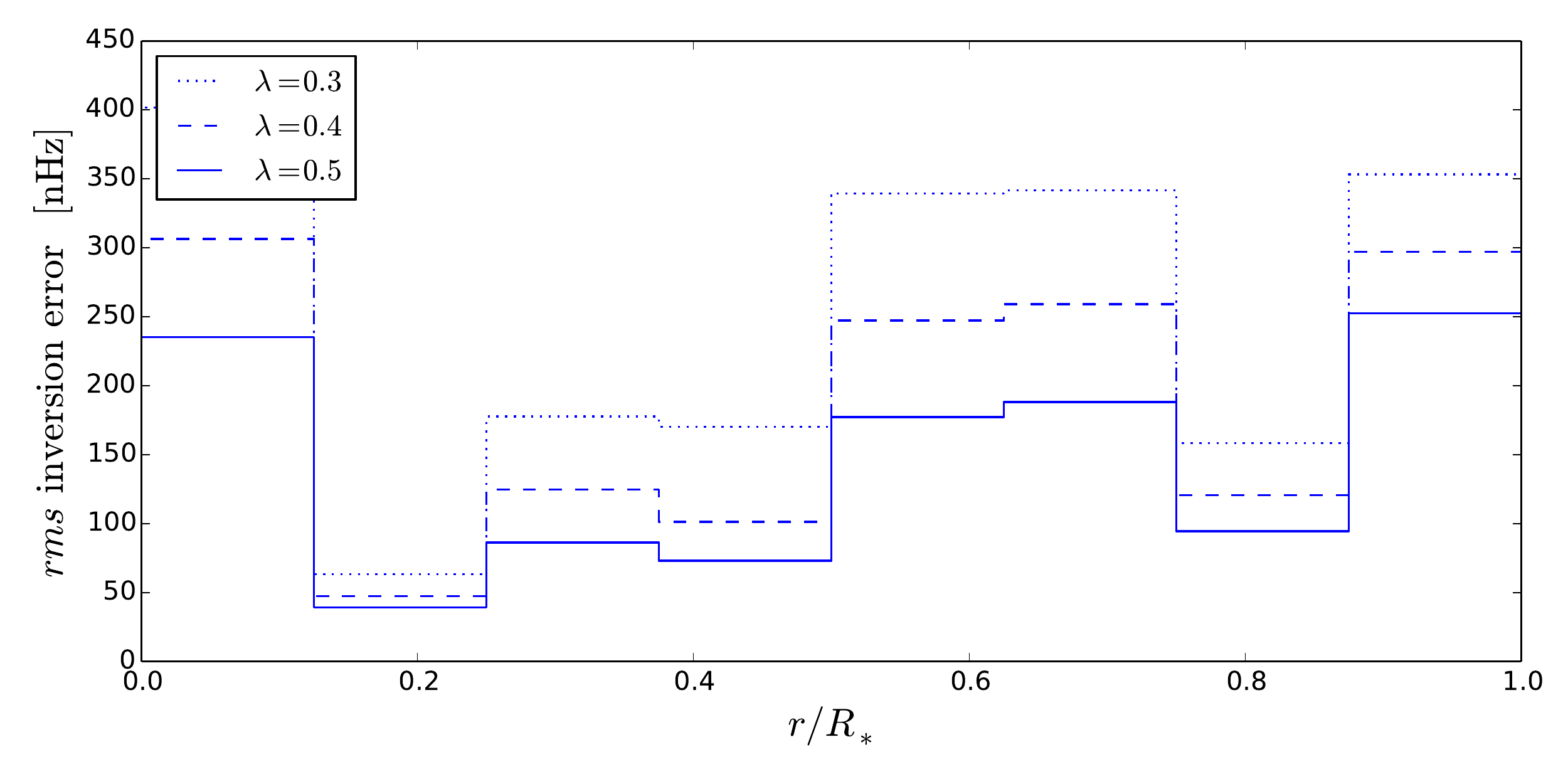}
\caption{ {\it rms} error incurred by RLS inversions with $\mu_{RLS}=10^{-5}$ and $N=8$ computed from random profiles with
various degrees of smoothing. See text for details.}
\label{fig:mc_new}
\end{figure}

To implement the above, we computed three sets of smooth random profiles ($\lambda=0.3,0.4,0.5$,
respectively and with
$10^6$ profiles each). We then rescaled the amplitude of each individual profile so as to make the corresponding
splittings have a mean that equals the mean of the observed splittings. After discarding those profiles whose
splittings had mixed signs, we proceeded to perform the inversions ($\mu_{RLS}=10^{-5}$). We computed inverted profiles with $N=8$ intervals
of radial resolution and compared them with the reference profiles (integral-averaged over the same radial intervals).
The standard errors at each radial interval calculated 
from all the profiles in the set are shown in Fig.\,\ref{fig:mc_new}.
We see clearly that the error becomes larger as the profiles have more variability. The radial locations where
the errors are comparatively smaller coincide roughly with the locations where the $\bf A$ matrices have better
localization (see Fig.\,\ref{fig:ctg_rls}).
\begin{figure}[h!]
\centering
\includegraphics[width=0.7\columnwidth]{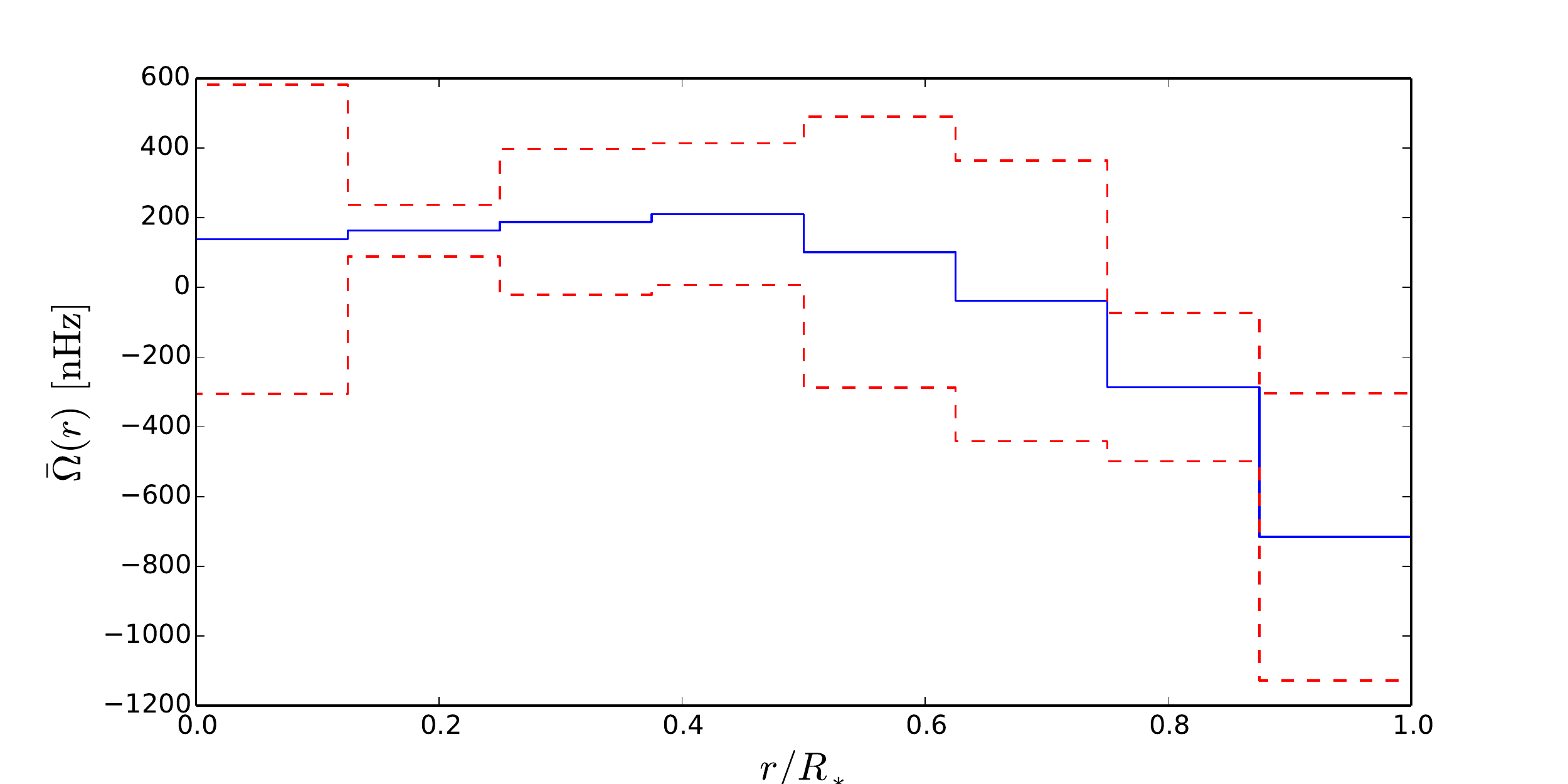}
\caption{ RLS inversion profile (solid line) of KIC\,10526294 ($N=8,\,\mu_{RLS}=10^{-5}$). The error margins
incorporate both the measurements errors (from Error Set\,1) and the errors related to the inadequacy of the kernels to recover the `true' rotation profile. This latter error was estimated by using smooth random profiles with $\lambda=0.3$.}
\label{fig:final}
\end{figure}

To conclude this Section we present the inversion profile (from the
real KIC\,10526294 data) together with an overall ($1\sigma$) uncertainty (derived from both the measurement errors
\emph{and\/} the kernel error as explained above) in Fig.\,\ref{fig:final}. This profile represents a balance
between good overall statistical measures and good localization properties, at least near the bottom of the radiative
zone and close to the stellar surface. Note that a fully positive rotation
  profile is possible at $2\sigma$ level.

\FloatBarrier

\section{Summary and conclusion}

Numerical models and their pulsation properties (based on the MESA
  evolution code and the GYRE pulsation code) have allowed us to obtain kernels
  of oscillation modes whose frequencies closely match the identified zonal
  dipole mode frequencies of the B8V star KIC\,10526294
  \citep{papics2014,moravveji2015}.  Based on these kernels, we computed
  rotational profiles explaining the detected rotationally split dipole mode
  frequencies by assuming different functional forms (constant, linear, two-zone
  and three-zone). We also performed RLS and SOLA inversions and implemented a
Monte Carlo approach to obtain an approximate rotational profile and to estimate
the errors incurred by the inversion process.
We relied on the optimal equilibrium model 
found so far for this pulsator \citep{moravveji2015} (Model\,1) but other
seismically derived equilibrium models were also 
examined and lead to qualitatively similar results. 

While the most likely rotation profiles depend on the assumptions made about the
functional form of the profile, we were able to constrain the average rotation
rate near the overshoot region to be about $163\pm89$ nHz, a value supported by
almost all the rotational models we considered.  Towards the surface of the star our
results are less constrained since they are sensitive to the a-priori
assumptions on the shape of the rotational profile.  If a smooth and continuous
profile is assumed, our results point to a  mild counter-rotating region in the
envelope towards the surface of the star rotating at frequency
$-717\pm412\,\mathrm{nHz}$ with the sign change occurring around
$r\sim0.7\,R_*$.  On the other hand, if we assume a discontinuous two-zone
profile, we find an outer envelope rotating about six times slower than the
overshoot region, at $49$nHz. 
The averaging kernel associated with the outer zone of this two-zone model
  leads to a weighted average over most of the radiative envelope. The best
  counter-rotating profiles from inversion, when averaged over the radiative
  zone using this outer zone averaging kernel, lead to rotation rates entirely
  consistent between the two models.

We performed model comparisons based on the Akaike Information Criterion as
  well as the leave-one-out cross-validation technique, which are both better
  suited than the reduced $\tilde\chi^2$ when comparing the performance of
  models that are not-nested, as is the case for the models we considered in
  this study.  Both methods give preference to the inversion models with the
  presence of a mild counter-rotation in the radiative envelope at $1\sigma$
  level. The Monte Carlo simulations, fully independent of the above, are
  consistent with such result. Current stellar structure models have so far not
considered this type of physical ingredient.

Following the first rough estimates of $\Omega_{\rm core}/\Omega_{\rm envelope}$
for three core-hydrogen burning B stars prior to the asteroseismology space era
\citep{Aerts2003,Pamyatnykh2004,Briquet2007}, the recent studies by
\citet{Kurtz2014} and \citet{Saio2015} made the first high-precision
asteroseismic measurement of surface-to-core rotation in two $\sim
1.5\,$M$_\odot$ main-sequence hybrid heat-driven pulsators from four years of
{\it Kepler\/} photometry.  They found the star KIC\,11145123 to have slightly
faster envelope than core rotation and an average rotation period $\sim$100\,d,
while KIC\,9244992 has slightly faster core than envelope rotation and an
average rotation period of $\sim$65\,d.  The authors deduced these results from
the measured rotationally split g-mode triplets and p-mode triplets and
quintuplets without relying on forward seismic modeling of the zonal modes as we
have done in the present work. Our study of the 3.2\,M$_\odot$ main-sequence
B-type star KIC\,10526294 hints to an envelope whose inner rotation rate is
opposite to its outer rate with a small factor ranging {from $\sim -0.06$ to
  $\sim -0.2$} taking into account the uncertainties, while the star has a
depth-averaged rotation period of about 186\,d. In these three cases, even after
taking into account that rotation rates at stellar birth have been largely
overestimated \citep[e.g.,][]{Zwintz2014b}, a strong and efficient mechanism
must have been at work to slow down these stars' rotation after their
birth. Moreover, an efficient mechanism must be active to transport angular
momentum within the star. Internal gravity waves (IGWs) could be viable as such
a mechanism. Indeed, numerical simulations based on IGWs for a 3\,M$_\odot$ star
by \citet{Rogers2013} have shown that such waves can transfer angular momentum
on short timescales and over the appropriate distances in stars with a
convective core and a radiative envelope. Additionally, the study by
\citet{Rogers2013} led to the conclusion that IGWs can lead to either a slightly
faster envelope than core rotation, or an outer envelope rotating opposite to
the inner regions. This mechanism thus could be the natural cause of the
observational results on the rotational properties of KIC\,10526294,
KIC\,11145123, and KIC\,9244992.

The type of rotation profile found for KIC\,10526294 and KIC\,11145123 is not
achieved in any standard stellar evolutionary scenario. A similar but much
stronger discrepancy between models and observations occurs for the core
rotation of red giants \citep[e.g.,][]{Cantiello2014}. In a next step, we
  do not only plan to perform similar studies as this one for OB-type stars with
  various stellar parameters, but we will also investigate how the stellar
  structure, and in particular the density profile, behaves during the evolution
  of the star in the presence of the most likely rotation profiles we found in
  this study, testing new physical ingredients such as internal gravity waves
  that were not yet included to describe the physics in the radiative envelope
  of massive stars. Only an extension of the sample of stars with seismic
  inversion treated with appropriate statistical model selection and coupled to
  an iterative procedure to upgrade the input physics can deliver a meaningful
  improvement in the stellar models. Our study is a first step in this
  direction for massive stars.

  \acknowledgments The authors thank Profs. Geert Molenberghs and Marc Aerts
    of the Center for Statistics, University of Hasselt, Belgium, for
    enlightening discussions and valuable advice on statistical model selection.
    They also acknowledge detailed comments from a referee, which helped
    them to present the results in a more consistent way.  Further, they
  are grateful to the {\it Kepler\/} team and everybody who has
  contributed to making this mission possible. Funding for the {\it Kepler\/}
  Mission was provided by NASA's Science Mission Directorate.  The research
  leading to these results has received funding from the Fund for Scientific
  Research of Flanders (FWO, grant number G.0728.11).  E.M.\ is beneficiary of a
  postdoctoral grant from the Belgian Federal Science Policy Office (Belspo)
  co-funded by the Marie Curie Actions FP7-PEOPLE-COFUND2008 no.\,246540
  MOBILITY GRANT from the European Commission.  Funding for the Stellar
  Astrophysics Centre is provided by The Danish National Research Foundation
  (Grant DNRF106). The research is supported by the ASTERISK project
  (ASTERoseismic Investigations with SONG and Kepler) funded by the European
  Research Council (Grant agreement no.\, 267864).

%% in parentheses, after the keyword, but they will not be verified.

%{\it Facilities:} \facility{Kepler}.

\pagebreak

\appendix
%In the following appendices we further illustrate some properties of the kernels
%as related to the inversions methods. We base our discussion on the best model
%obtained by \cite{moravveji2015}.

\section{Appendix A: Results for Error Set 2 and/or for limited triplet sets}
\label{apx_A}

\begin{figure*}[h!]
\centering
\includegraphics[width=\linewidth]{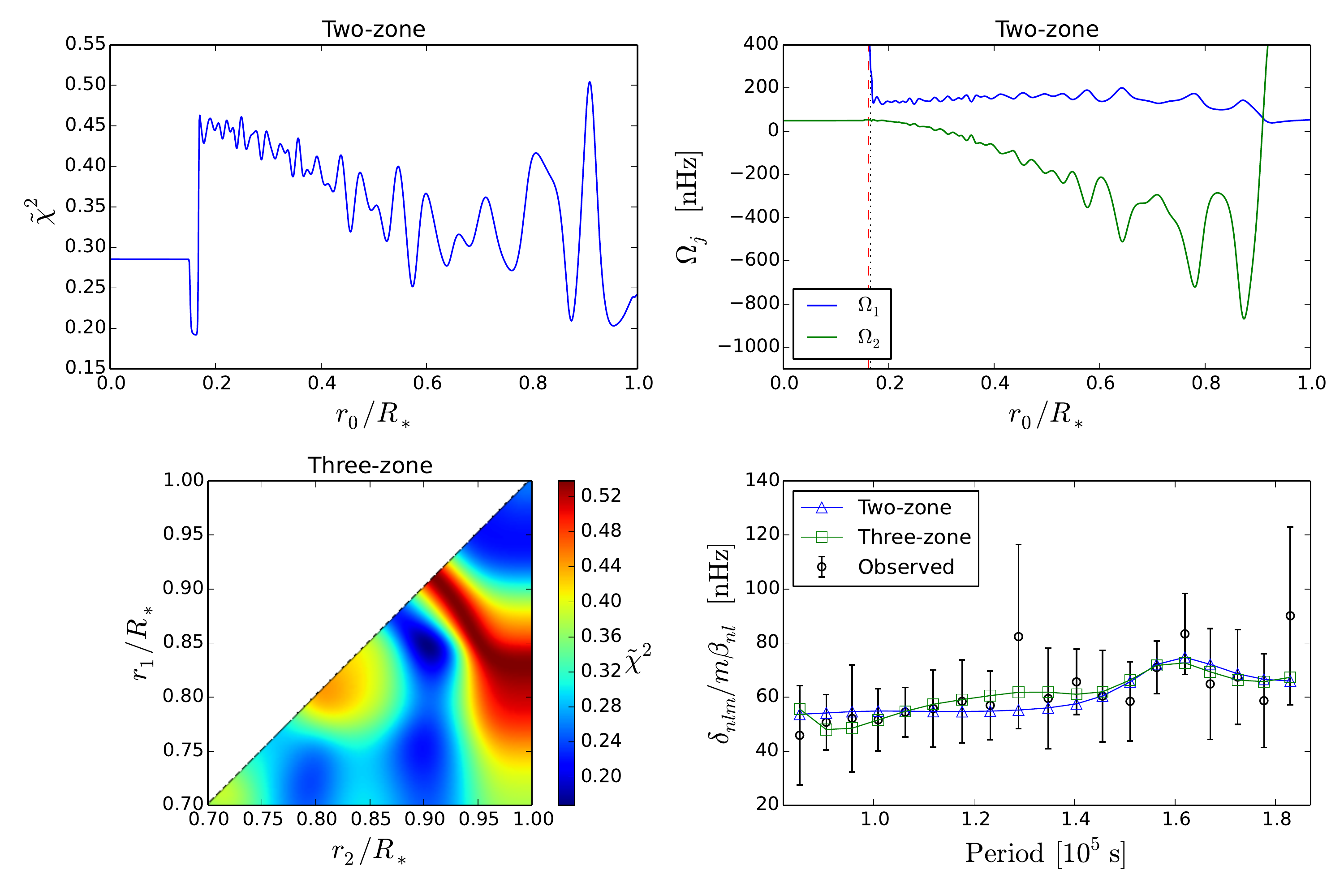}
\caption{Same as Fig.\,\ref{fig:2z} with the exception that Error Set\,2 has been used.
The vertical dashed line in the top, right panel marks the location of the minimum of $\tilde\chi^2$
for the two-zone model.}
\label{fig:2z_2}
\end{figure*}

\begin{figure*}[h!]
\centering
\includegraphics[width=\linewidth]{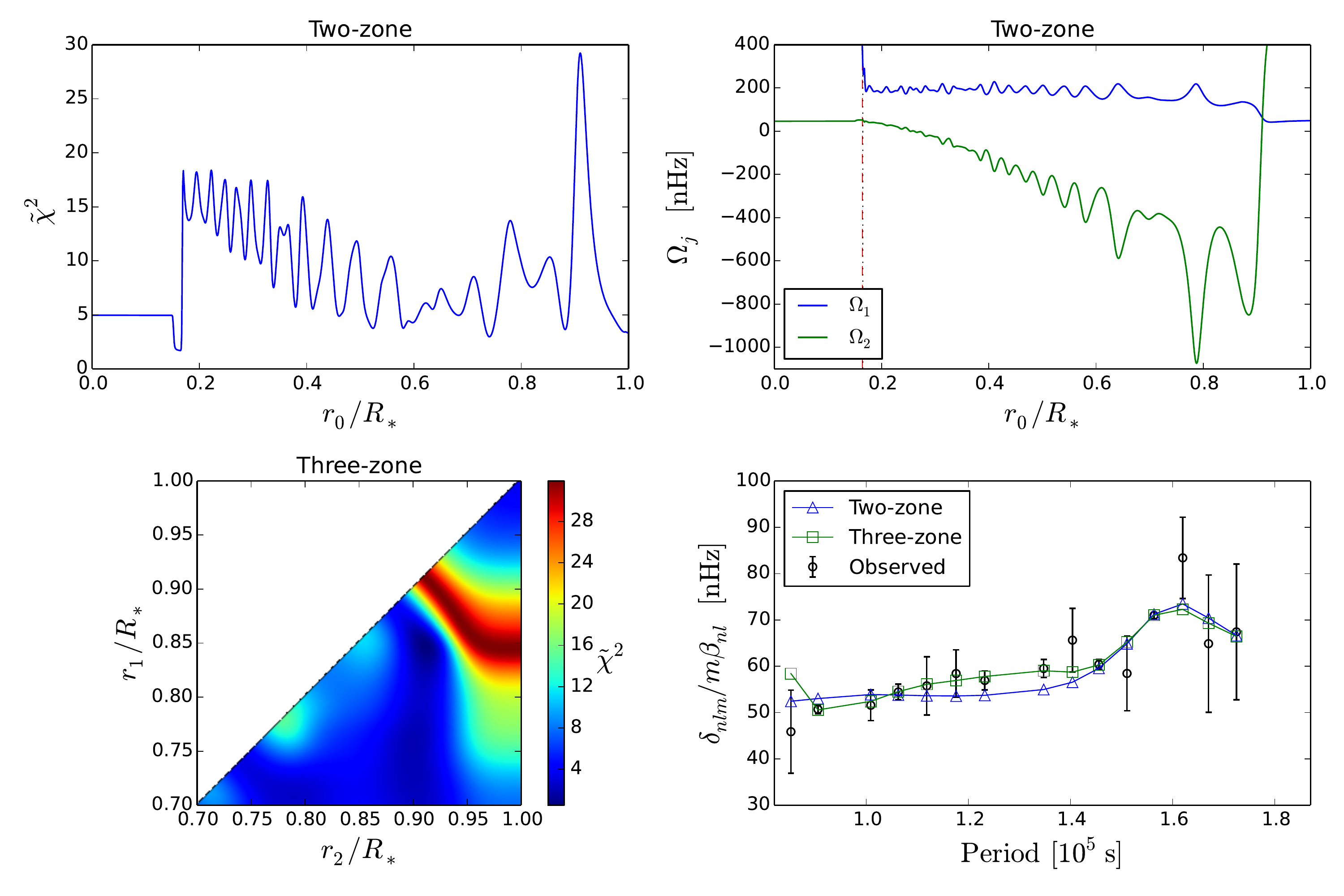}
\caption{Same as Fig.\,\ref{fig:2z_2} with the exception that the three most asymmetric
 splittings have been excluded from the $\tilde\chi^2$ minimization, i.e. those with
 periods near $0.96\times 10^{5}$s, $1.29 \times 10^{5}$s and $1.83\times 10^{5}$s.
 The single $m=+1$ splitting with period near $1.78\times 10^{5}$s was also excluded.
 Error Set\,1 has been used.}
\label{fig:2z_3}
\end{figure*}

\begin{figure*}[h!]
\centering
\includegraphics[width=\linewidth]{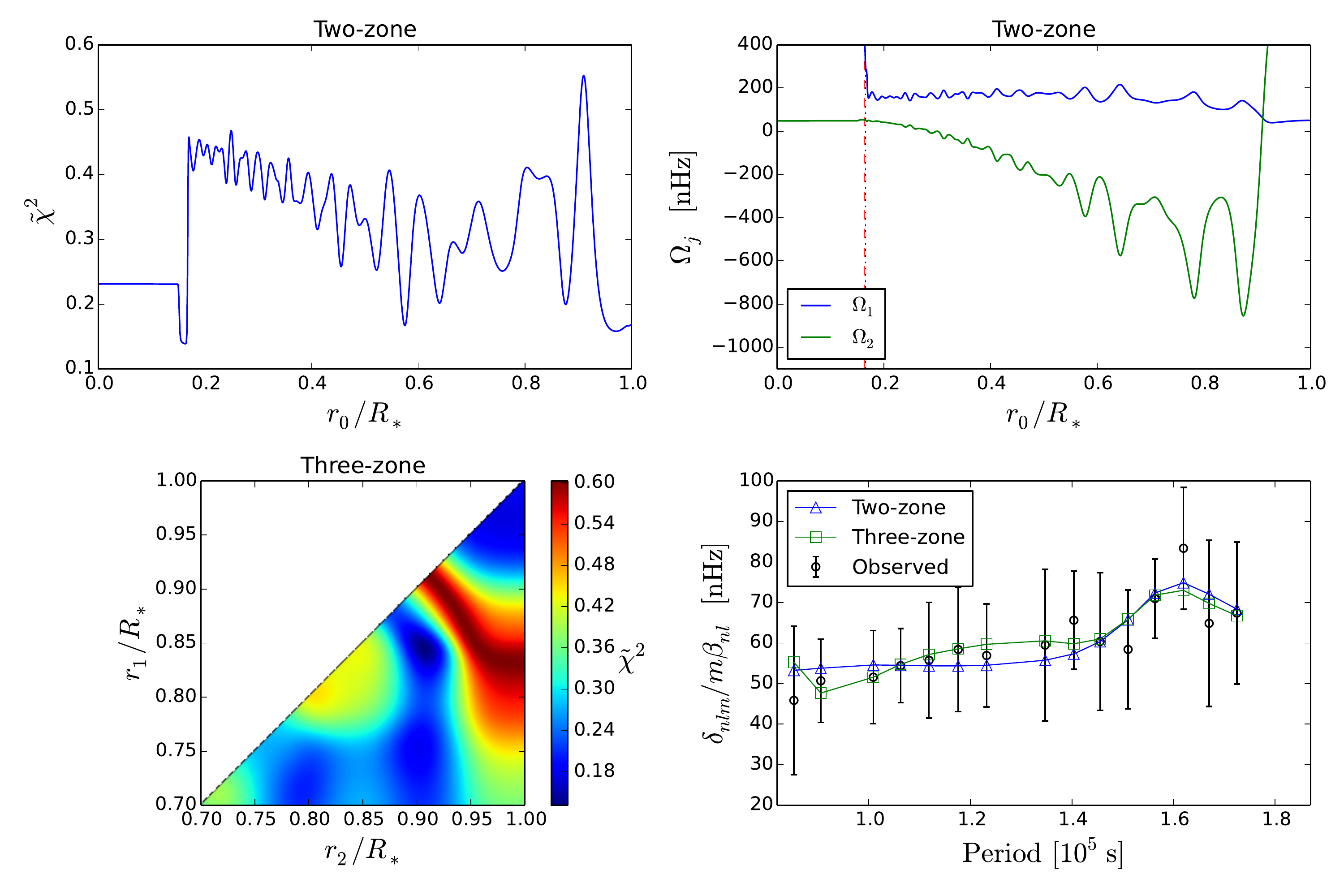}
\caption{Same as Fig.\,\ref{fig:2z_3}, i.e., omitting the three most asymmetric modes
and the single $m=+1$ splitting, but this time using Error Set\,2.}
\label{fig:2z_4}
\end{figure*}

\begin{deluxetable}{lrrrr}
  \tablecolumns{5} \tablewidth{0pc} \tablecaption{\label{aic2}
Same as Table\,\ref{aic} but using the observation uncertainties 
from Error Set\,2.}
  \tablehead{ \colhead{Rotation Profile} & 
\colhead{{\it rms} error [nHz]} & \colhead{$\nu$}
   & \colhead{$\tilde\chi^2$} & \colhead{AICc} }
    \startdata
    Constant     & 11.85 & 18.00 &   0.45 &  12.82 \\ 
    Linear       & 11.78 & 17.00 &   0.36 &  13.67 \\ 
    Linear+      & 11.66 & 16.00 &   0.45 &  18.05 \\ 
    Two-zone     &  9.61 & 16.00 &   0.19 &  13.93 \\ 
    Three-zone   &  8.44 & 15.00 &   0.17 &  17.13 \\ 

    RLS, $N=8$   &  9.17 & 15.93 &   0.17 &  13.83 \\ 
    RLS, $N=14$  & 10.29 & 16.09 &   0.26 &  14.66 \\ 
%    RLS+, $N=8$  & 11.23 &  8.44 &   1.85 &  69.57 \\
%    RLS+, $N=14$ &  9.61 &  6.39 &   0.80 &  93.97 \\
    SOLA, $N=8$  &  9.46 & 14.30 &   0.22 &  20.80 \\ 
    SOLA, $N=14$ &  7.59 & 11.48 &   0.15 &  35.90 \\ 
   \enddata
\end{deluxetable}

\begin{deluxetable}{lrrrr}
  \tablecolumns{5} \tablewidth{0pc} \tablecaption{\label{aic3}
Same as Table\,\ref{aic} but excluding
 the three most asymmetric splittings, i.e., those with periods near $0.96\times 10^{5}$s,
 $1.29 \times 10^{5}$s, and $1.83\times 10^{5}$s. The single $m=+1$ mode with period
near $1.78\times 10^{5}$s was also excluded. Error Set\,1 is used.}
  \tablehead{ \colhead{Rotation Profile} & 
\colhead{{\it rms} error [nHz]} & \colhead{$\nu$}
   & \colhead{$\tilde\chi^2$} & \colhead{AICc} }
    \startdata
    Constant     &  9.07 & 14.00 &  25.07 &  355.9 \\ 
    Linear       &  7.22 & 13.00 &   8.55 &  119.4 \\ 
    Linear+      &  8.29 & 12.00 &  22.08 &  277.0 \\
    Two-zone     &  4.95 & 12.00 &   1.69 &  32.27 \\ 
    Three-zone   &  5.16 & 11.00 &   0.53 &  22.46 \\ 

    RLS, $N=8$   &  4.84 & 11.29 &   0.63 &  22.33 \\ 
    RLS, $N=14$  &  4.77 & 11.20 &   2.35 &  42.00 \\ 
%    RLS, $N = 14$ (Model 2) & 6.71 & 9.83 & 8.47 & 106.87 \\
%    RLS+, $N=8$  &  7.85 &  3.63 &  73.04 &  493.1 \\
%    RLS+, $N=14$ &  5.08 &  0.47 &  31.94 &  - \\
    SOLA, $N=8$  &  5.86 &  9.45 &   0.86 &  34.48 \\ 
    SOLA, $N=14$ &  3.58 &  7.21 &   0.40 &  53.45 \\ 
   \enddata
\end{deluxetable}

\begin{deluxetable}{lrrrr}
  \tablecolumns{5} \tablewidth{0pc} \tablecaption{\label{aic4}
Same as Table\,\ref{aic3} but using the observation uncertainties 
from Error Set\,2.}
  \tablehead{ \colhead{Rotation Profile} & 
\colhead{{\it rms} error [nHz]} & \colhead{$\nu$}
   & \colhead{$\tilde\chi^2$} & \colhead{AICc} }
    \startdata
    Constant     &  8.95 & 14.00 &   0.47 &  11.63 \\ 
    Linear       &  7.05 & 13.00 &   0.32 &  12.36 \\ 
    Linear+      &  8.27 & 12.00 &   0.47 &  17.66 \\
    Two-zone     &  4.91 & 12.00 &   0.14 &  13.67 \\ 
    Three-zone   &  4.70 & 11.00 &   0.13 &  18.11 \\
     
    RLS, $N=8$   &  4.96 & 12.15 &   0.14 &  13.11 \\ 
    RLS, $N=14$  &  5.85 & 12.39 &   0.24 &  13.42 \\ 
%    RLS, $N = 14$ (Model 2) & 5.93 & 11.24 & 0.42 & 20.13 \\
%    RLS+, $N=8$  &  7.86 &  5.98 &   1.26 &  83.10 \\
%    RLS+, $N=14$ &  5.12 &  0.95 &   2.99 &  - \\
    SOLA, $N=8$  &  6.04 & 10.99 &   0.23 &  19.27 \\ 
    SOLA, $N=14$ &  3.49 &  8.59 &   0.10 &  34.55 \\ 
       \enddata
\end{deluxetable}

\begin{figure*}[h!]
\centering
\includegraphics[width=\linewidth]{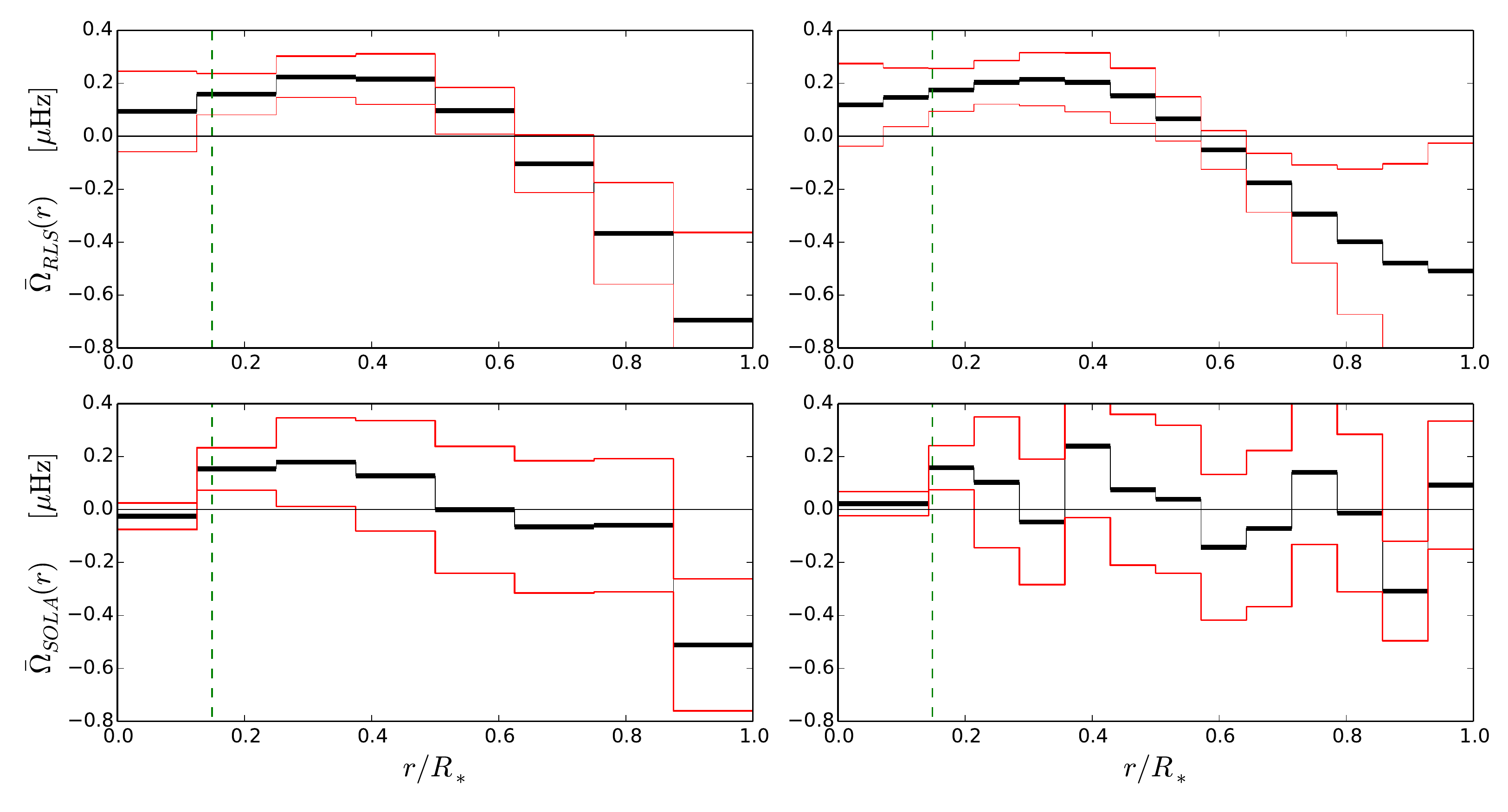}
\caption{Same as Fig.\,\ref{fig:pom1} with the exception that Error Set\,2 has been used.}
\label{fig:pom2}
\end{figure*}

\begin{figure*}[h!]
\centering
\includegraphics[width=\linewidth]{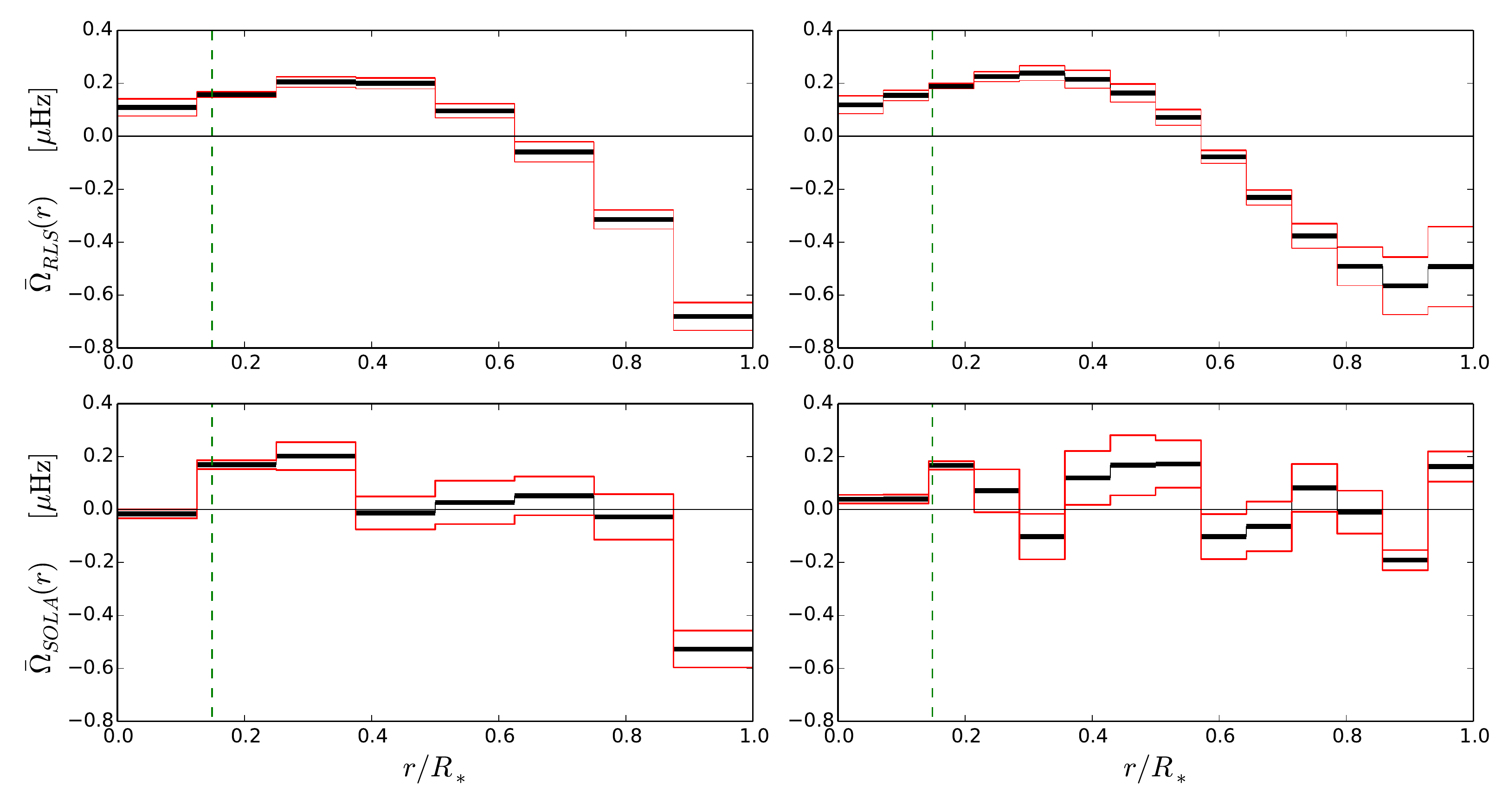}
\caption{Same as Fig.\,\ref{fig:pom1} but this time excluding
 the three most asymmetric splittings, i.e., those with periods near $0.96\times 10^{5}$s,
 $1.29 \times 10^{5}$s and $1.83\times 10^{5}$s. The single $m=+1$ mode with period
near $1.78\times 10^{5}$s was also excluded from the computation.}
\label{fig:pom1_mask}
\end{figure*}

\begin{figure*}[h!]
\centering
\includegraphics[width=\linewidth]{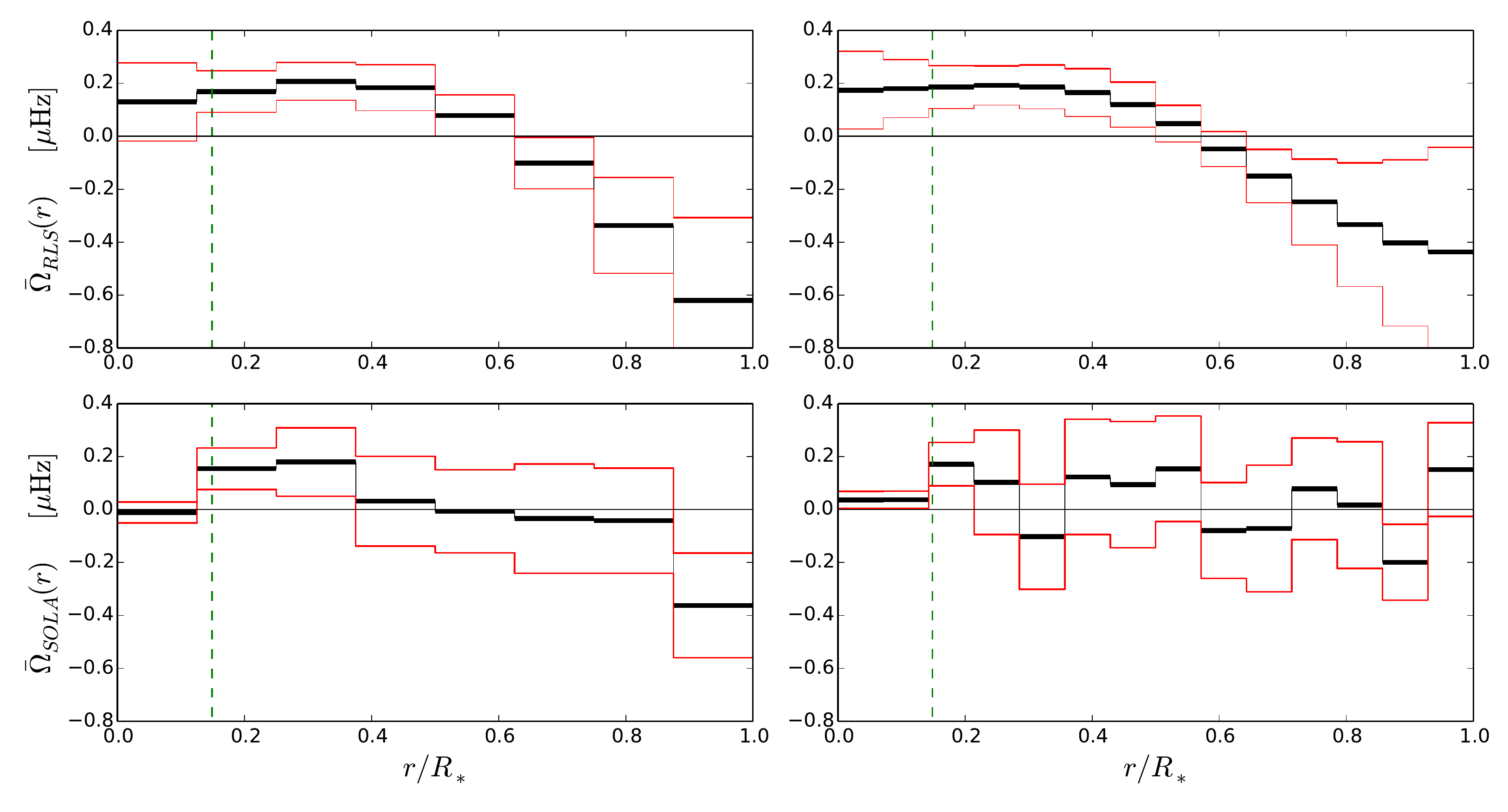}
\caption{Same as Fig.\,\ref{fig:pom1_mask} with the exception that Error Set\,2 has been used.}
\label{fig:pom2_mask}
\end{figure*}

\begin{figure}[h!]
\centering
\includegraphics[width=0.7\columnwidth]{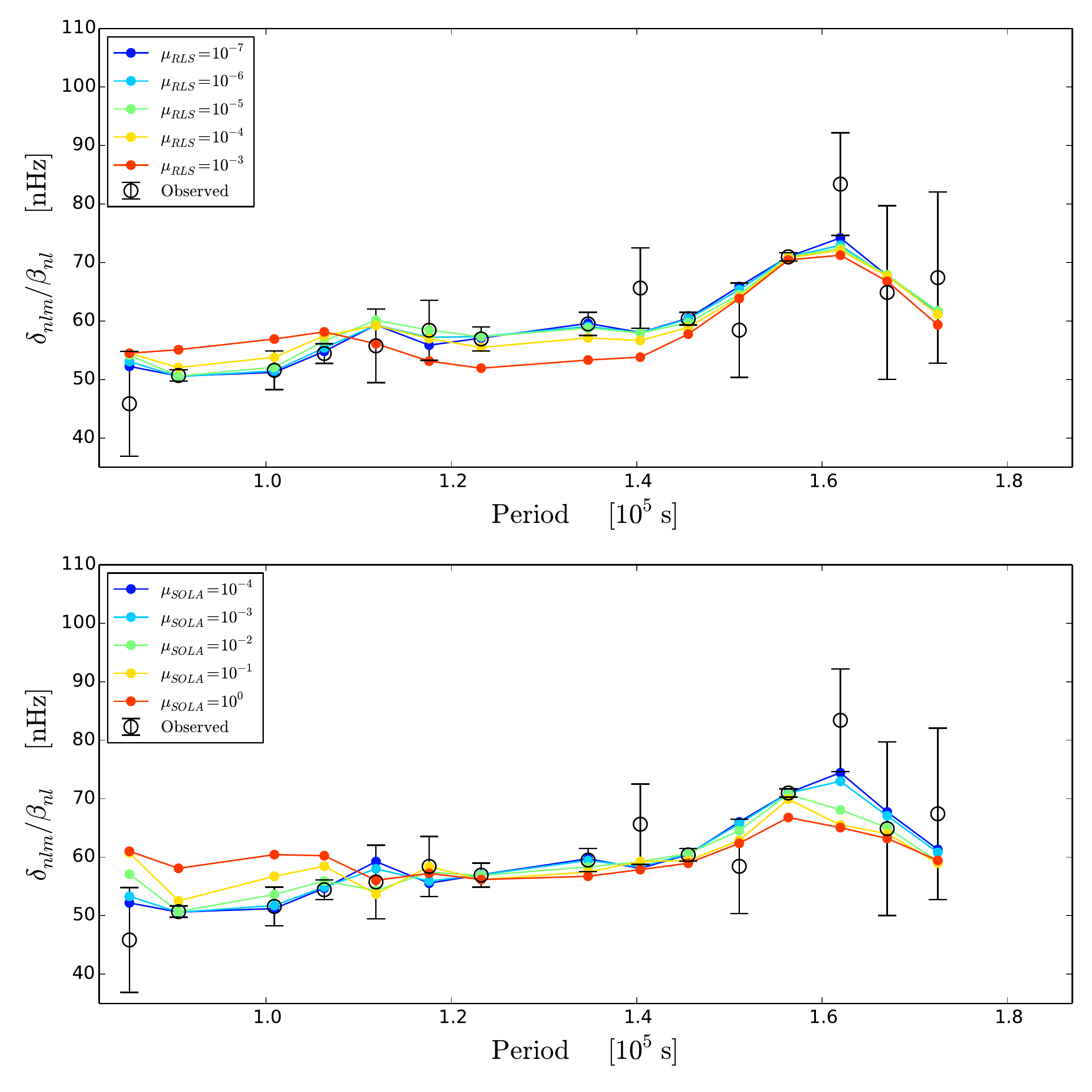}
\caption{Same as Fig.\,\ref{fig:ps_full} but excluding
 the three most asymmetric splittings, i.e., those with periods near $0.96\times 10^{5}$s,
 $1.29 \times 10^{5}$s and $1.83\times 10^{5}$s. The single $m=+1$ mode with period
near $1.78\times 10^{5}$s was also excluded from the computation.}
\label{fig:ps_mask}
\end{figure}

\FloatBarrier

\pagebreak

\section{Appendix B: The $\bf{A}$ matrices}
\label{apx_B}

\begin{figure*}[h!]
\centering
\includegraphics[width=0.95\linewidth]{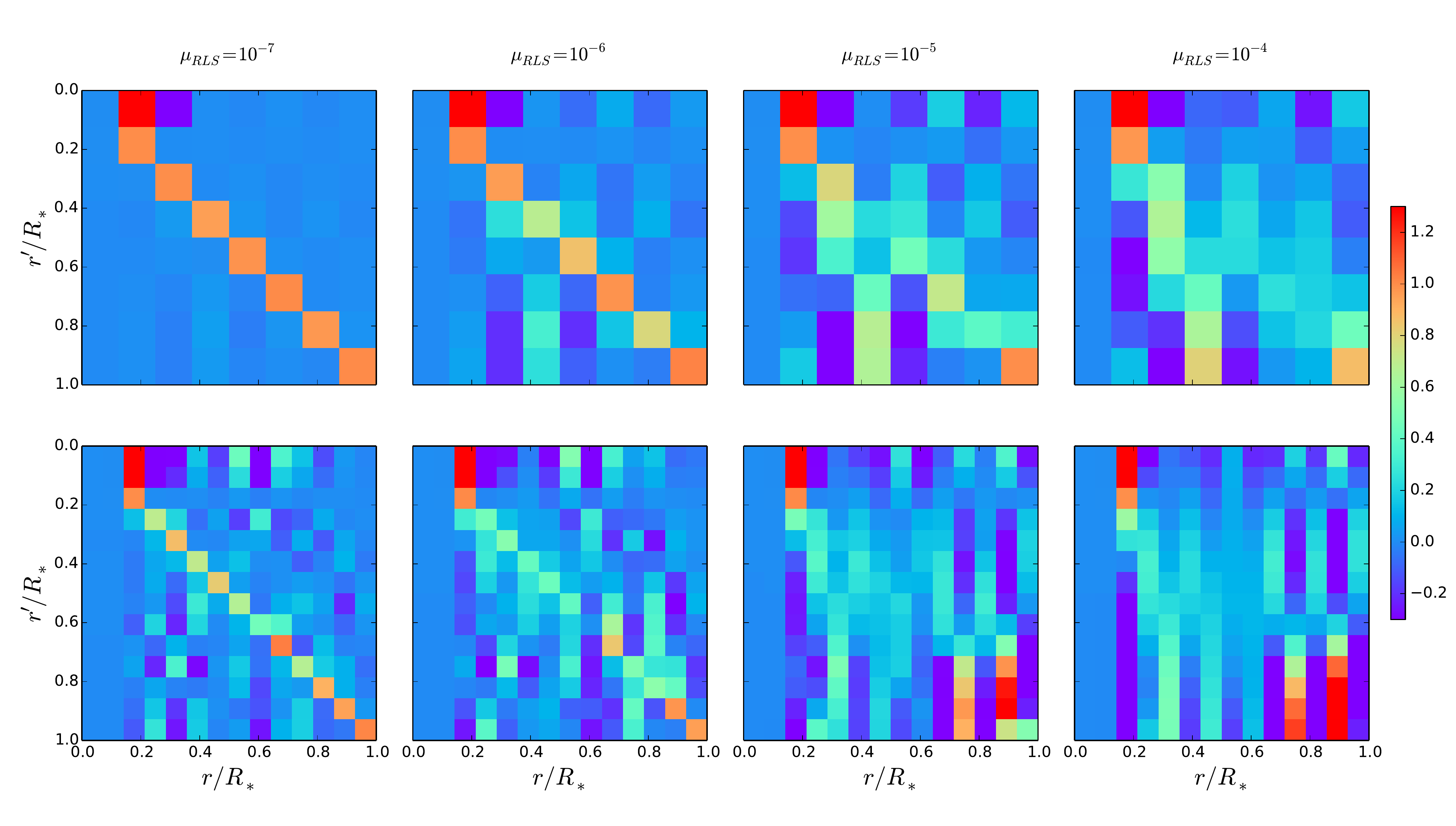}
\caption{The matrix $\bf{A}$ for the RLS method using the kernels from the best
 model in \cite{moravveji2015} with two different
  resolutions ($N=\{8,14\} $; top and bottom row respectively) and a
  range of smoothing parameters $\mu_{RLS}$. Each row on these matrices is the
  discrete version of the averaging kernel $\mathscr{K}(r',r)$. The $\bf{A}$
  matrix should resemble as much as possible the identity matrix, as
  Eq.\,(\ref{eq:Ajk}) indicates, if we want a faithful reconstruction of the
  rotation rate (at the radial location defined by the row index, i.e., first
  row corresponds to the first bin in the radial grid). Note how near $r/R_*\sim
  0.17$ the rotation rate is well recovered almost independently of the choice of
  $N$ or $\mu_{RLS}$.}
\label{fig:ctg_rls}
\end{figure*}

\begin{figure*}[h!]
\centering
\includegraphics[width=0.95\linewidth]{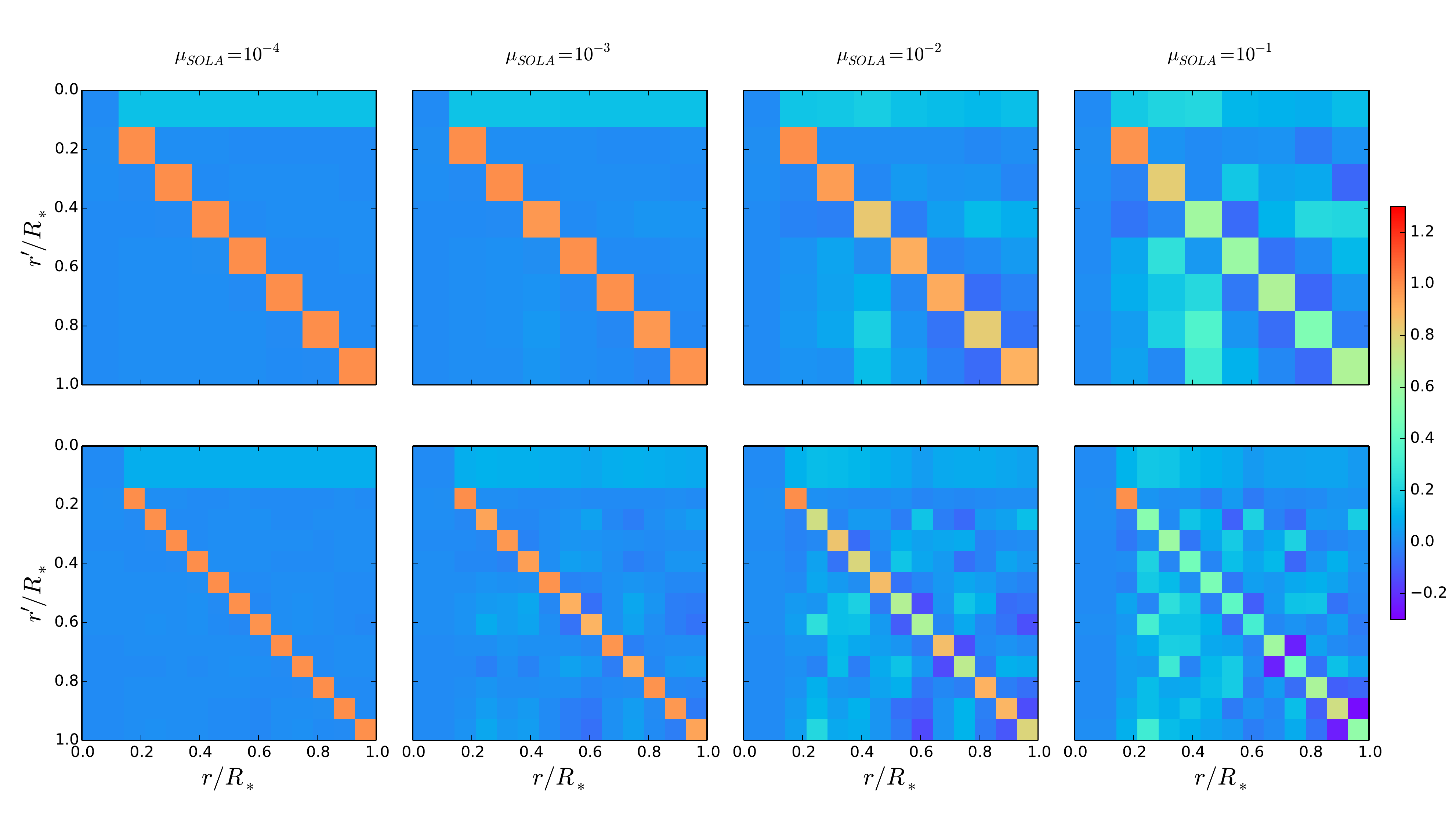}
\caption{Same as in Fig.\,\ref{fig:ctg_rls} but using the SOLA method. The
  matrices are very well localized except at higher resolutions and higher
  $\mu_{SOLA}$.  At $\mu_{SOLA}=10^{-2}$ and $N=8$ there is a small amount of leakage from
  regions close to $r=0.4\,R_*$ into the outer radial bins ($r'\sim 0.9\,R_*$). We estimate from this that the rotation
  rate there is overestimated by the inversion by about 3\%.}
\label{fig:ctg_sola}
\end{figure*}

As mentioned in Section \ref{sec:theory}, the matrix $\bf{A}$ gives an
indication of how well the inversion profile recovers the true profile (in the
ideal case of no measurement error in the splittings). Using the best model from \cite{moravveji2015}, we can see this fact at
work very clearly in both Figs.\,\ref{fig:ctg_rls} and \ref{fig:ctg_sola} as the
respective $\mu$ parameter is varied. The more $\bf{A}$ resembles the identity
matrix the better the reconstruction is, thus providing a qualitative assessment
of the inversion.

\pagebreak

\section{Appendix C: Averaging kernels in the continuous limit}
\label{apx_C}

The RLS and SOLA inversions behave very differently when the resolution is
increased. In this section's experiments we kept $\mu_{RLS}=10^{-5}$ and
$\mu_{SOLA}=10^{-2}$ fixed but used three different resolutions,
$N=14,28,5000$. From the viewpoint of the inversion, 
it is possible to increase the resolution beyond the 
number of observations $M$ since regularization
 keeps the {\em effective\/} number
of fitted parameters  below $M$. Here we do
not include the variances on the inversions in the discussion.
We have chosen the Error Set\,1 as the uncertainties on the splittings.

\begin{figure*}[h!]
\centering
\includegraphics[width=\linewidth]{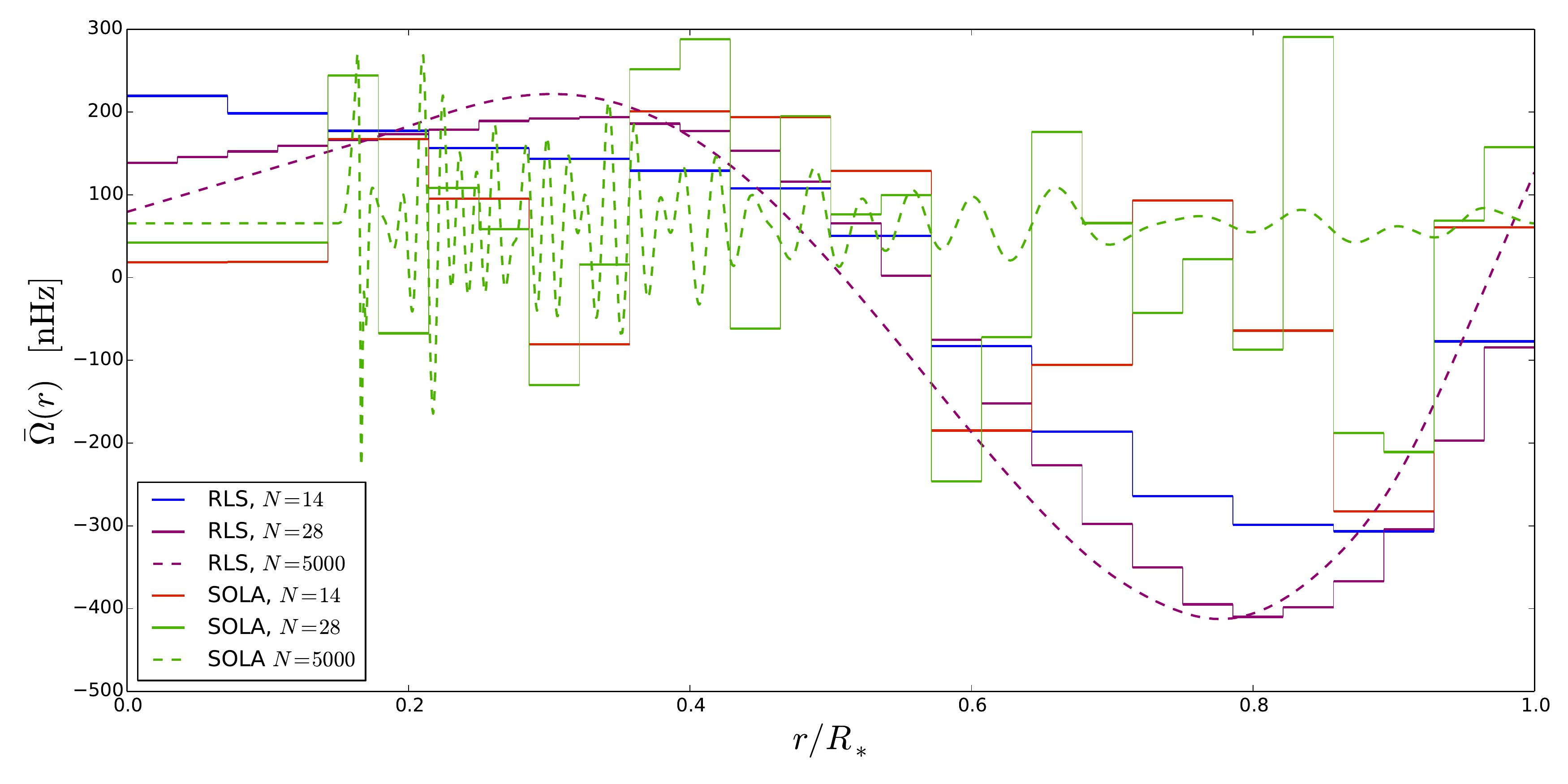}
\caption{Rotation profiles from RLS ($\mu_{RLS}=10^{-5}$) and SOLA 
($\mu_{SOLA}=10^{-2}$) methods and three different resolutions $N=14,28,5000$.
 The inversions are based on the kernels from the best model of \cite{moravveji2015}.
 The measurement uncertainties are taken from the error set 2.}
\label{fig:apx6}
\end{figure*}

In Fig.\,\ref{fig:apx6} we show the resulting inversion profiles. 
There are two radial locations
 where most
of the rotation rates roughly coincide. One is at $r\sim 0.17$ and the other one
 is at $r\sim 0.92$ where the rotation
values are not too far from each other except for the SOLA inversions with 
$N=28,5000$.

\begin{figure*}[h!]
\centering
\includegraphics[width=0.9\linewidth]{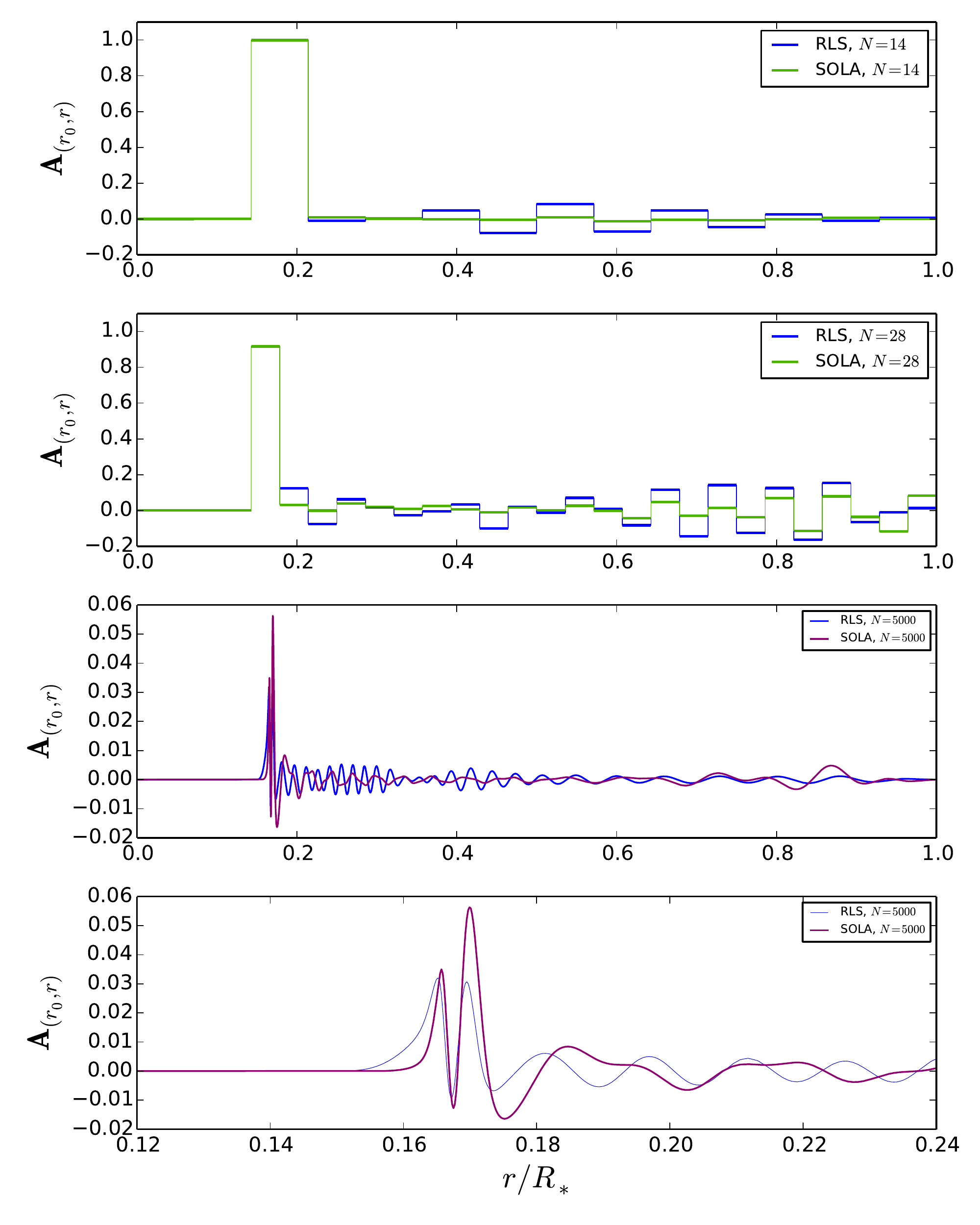}
\caption{The row of $\mathbf{A}$ corresponding to $r_0=0.17$ at three resolutions (top three plots). Bottom plot is a zoom of the one immediately above.}
\label{fig:apx9}
\end{figure*}

\begin{figure*}[h!]
\centering
\includegraphics[width=0.9\linewidth]{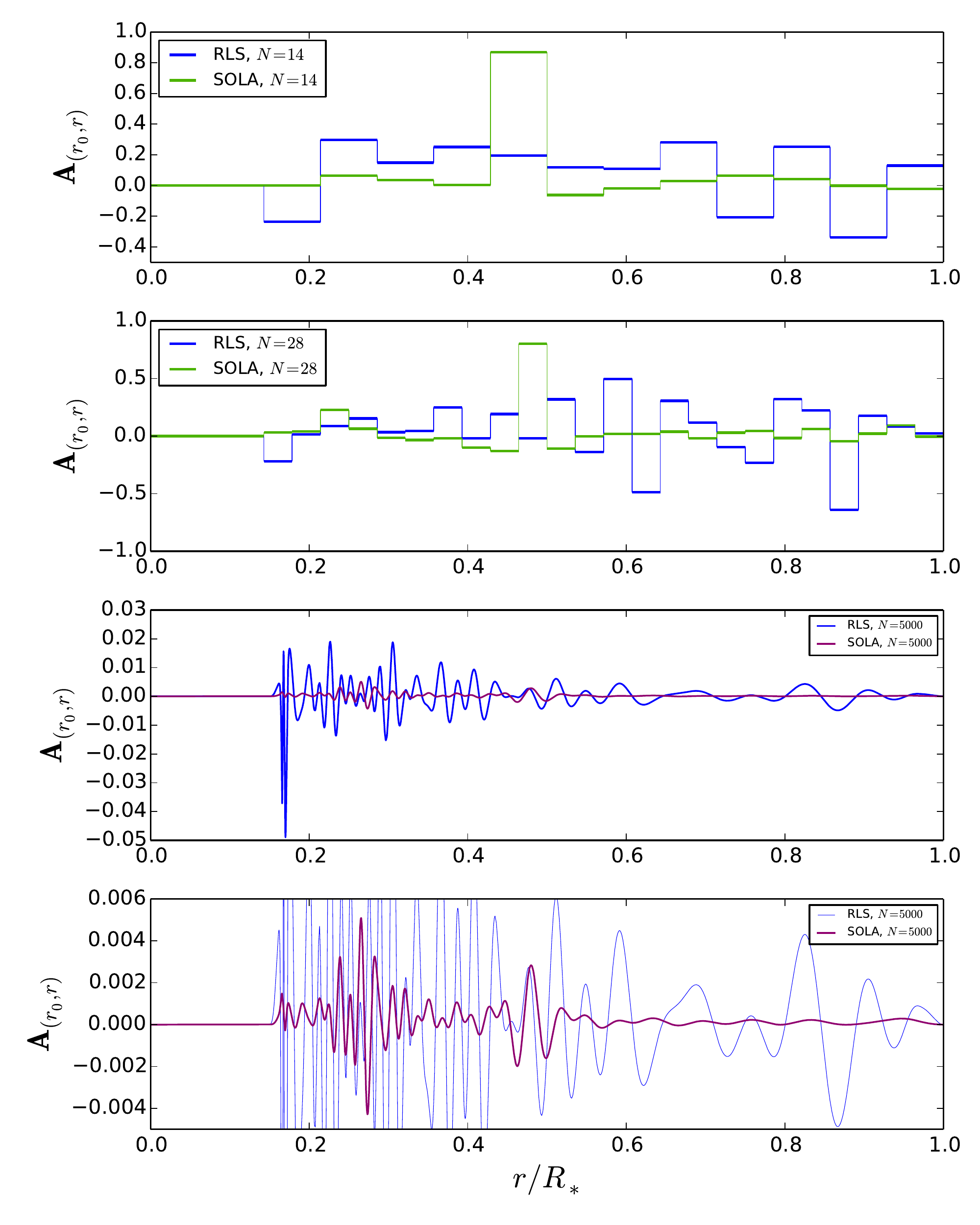}
\caption{Same as Figure \ref{fig:apx9} but for $r_0=0.48$.}
\label{fig:apx10}
\end{figure*}

\begin{figure*}[h!]
\centering
\includegraphics[width=0.9\linewidth]{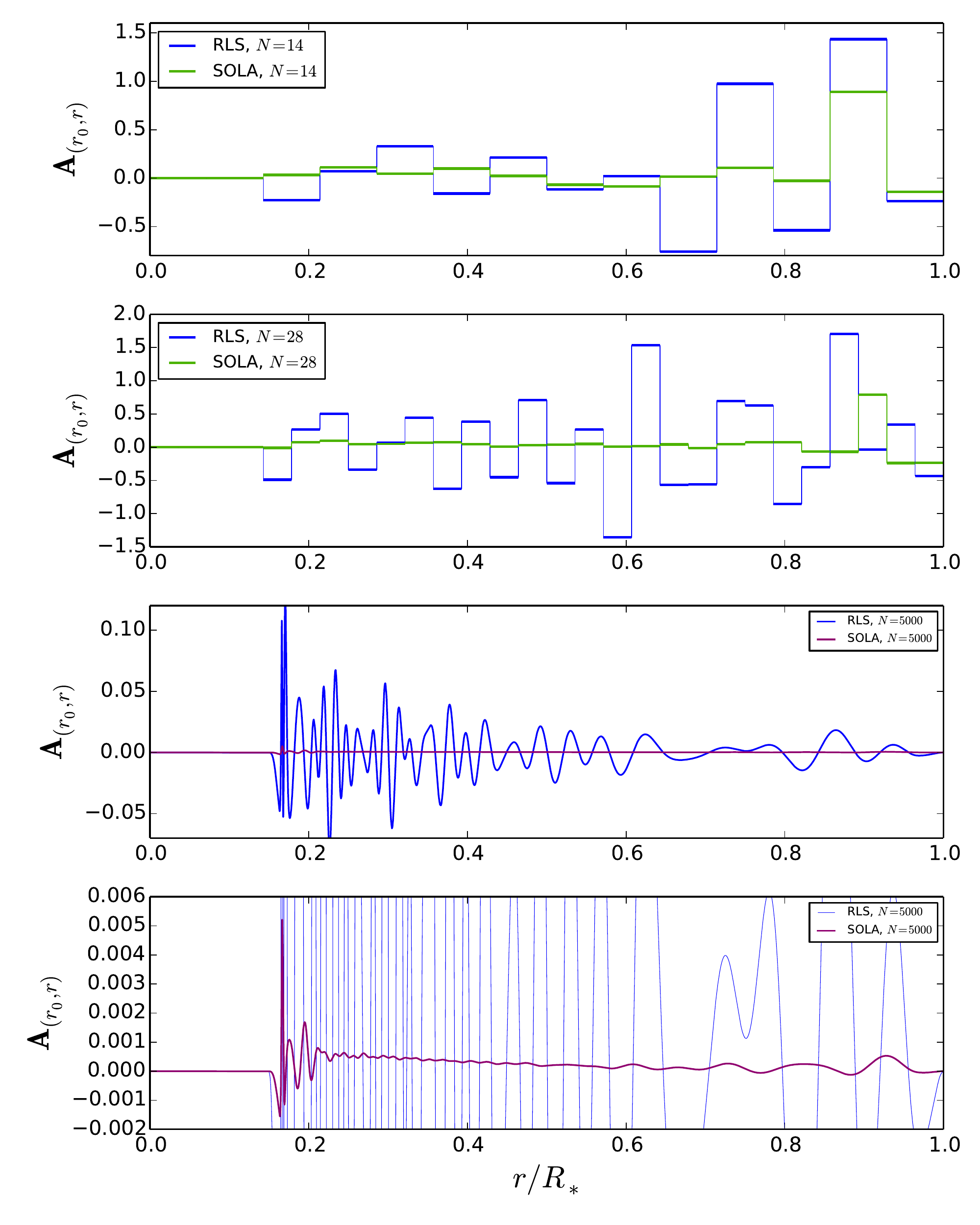}
\caption{Same as Figure \ref{fig:apx10} but for $r_0=0.92$.}
\label{fig:apx11}
\end{figure*}

Let us examine the averaging kernels from three selected radial locations
$r_0=0.17\,R_*$, $r_0=0.48\,R_*$ and $r=0.92\,R_*$ as shown in Figures \ref{fig:apx9},
\ref{fig:apx10} and \ref{fig:apx11}, respectively. From the figures we see that
for $r_0=0.17\,R_*$ all the averaging kernels are indeed well behaved generally, so
we expect inferences for this location to be consistent. At $r_0=0.48$ 
(Figure \ref{fig:apx10}) the localization is acceptable as long as $N$ is low. 
Closer to the stellar surface, at $r_0=0.92$ (Fig.\,\ref{fig:apx11}),
the situation is similar although the RLS kernels degrade considerably already
when $N=28$. 

From these figures we conclude that consistent inferences are to be
found using 
either RLS or SOLA methods if the resolution is kept low, i.e., $N\lesssim 14$.
Such a resolution also implies a sensible model comparison through the AICc (see
Table\,\ref{aic}), while larger $N$  and small smoothing parameters would lead to an effective number of
degrees of freedom $\nu < 1$ and strongly negative AICc values, implying
overfitting from the viewpoint of model comparison.

\FloatBarrier

\section{Appendix D: Testing RLS inversions with a synthetic profile}
\label{apx_D}

 The following test is to check that the counter-rotation profiles are not 
  produced by some undesired property of the RLS inversion methods. We take the
  optimum two-zone model from Section\,5 and smooth it using the `low
  pass' filter with correlation length $\lambda=0.3$ described in
  Section\,7.  This is taken as the `actual' rotational profile, which does not
  exhibit counter-rotation. Subsequently, we calculate the associated exact rotational
  splittings via Eq.\,(\ref{eq:rotsplit}). To each of these 19 splittings we add
  random noise sampled from a Gaussian distribution with zero mean and
  the same standard deviation as the actual measurement errors (Error
  Set\,1). We set $N=8,\,\mu_{\rm RLS}=10^{-5}$ as used for the RLS inversions
  in the main text and proceed to calculate the inversion profile. 

\begin{figure*}[h!]
\centering
\includegraphics[width=0.9\linewidth]{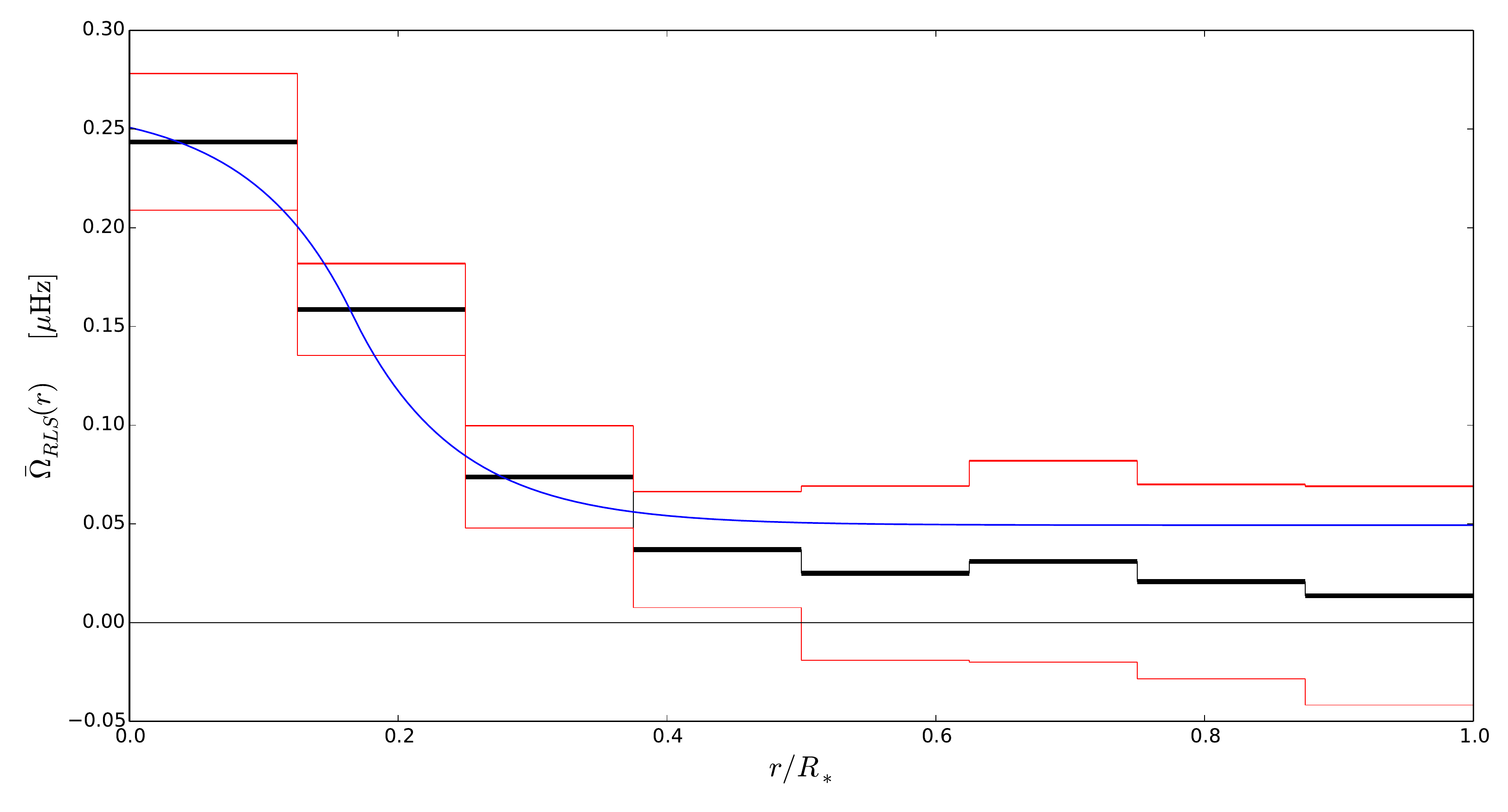}
\caption{The smoothed-out two-zone model (blue, correlation length
    $\lambda=0.3$) and the recovered profile from RLS inversion (black,
    $\mu_{\rm RLS}=10^{-5}$). Error bounds are in red.}
\label{fig:rls_test}
\end{figure*}

  At each radial bin we compare the inversion value with the integral average of
  the `actual' profile over the same radial bin. This gives us a direct estimate
  of the inversion error. By repeating this process a large number of times we
  can obtain well defined statistics (we used $10^7$
  iterations). Figure\,\ref{fig:rls_test} shows the `actual' profile in blue,
  the recovered profile in black and the estimated $1\sigma$ uncertainty range
  in red. The RLS method does a good job in recovering the actual profile, which
  always occurs within the errors. Some inversion profiles must counter-rotate
  mildly since the error region extends below zero in the outer half of the
  star. However, given the errors, we do not find a counter-rotating
  profile in this case.

\FloatBarrier

\newpage

\bibliographystyle{apj}
\bibliography{inversion}

\begin{thebibliography}{}
\expandafter\ifx\csname natexlab\endcsname\relax\def\natexlab#1{#1}\fi

\bibitem[{{Aerts} {et~al.}(2010){Aerts}, {Christensen-Dalsgaard}, \&
  {Kurtz}}]{Aerts2010asteroseismology}
{Aerts}, C., {Christensen-Dalsgaard}, J., \& {Kurtz}, D.~W. 2010,
  Asteroseismology, Astronomy and Astrophysics Library (Springer Heidelberg)

\bibitem[{{Aerts} {et~al.}(2003){Aerts}, {Thoul}, {Daszy{\'n}ska}, {Scuflaire},
  {Waelkens}, {Dupret}, {Niemczura}, \& {Noels}}]{Aerts2003}
{Aerts}, C., {Thoul}, A., {Daszy{\'n}ska}, J., {et~al.} 2003, Science, 300,
  1926

\bibitem[{{Beck} {et~al.}(2012){Beck}, {Montalban}, {Kallinger}, {De Ridder},
  {Aerts}, {Garc{\'{\i}}a}, {Hekker}, {Dupret}, {Mosser}, {Eggenberger},
  {Stello}, {Elsworth}, {Frandsen}, {Carrier}, {Hillen}, {Gruberbauer},
  {Christensen-Dalsgaard}, {Miglio}, {Valentini}, {Bedding}, {Kjeldsen},
  {Girouard}, {Hall}, \& {Ibrahim}}]{Beck2012}
{Beck}, P.~G., {Montalban}, J., {Kallinger}, T., {et~al.} 2012, \nat, 481, 55

\bibitem[{{Beck} {et~al.}(2014){Beck}, {Hambleton}, {Vos}, {Kallinger},
  {Bloemen}, {Tkachenko}, {Garc{\'{\i}}a}, {{\O}stensen}, {Aerts}, {Kurtz}, {De
  Ridder}, {Hekker}, {Pavlovski}, {Mathur}, {De Smedt}, {Derekas}, {Corsaro},
  {Mosser}, {Van Winckel}, {Huber}, {Degroote}, {Davies}, {Pr{\v s}a},
  {Debosscher}, {Elsworth}, {Nemeth}, {Siess}, {Schmid}, {P{\'a}pics}, {de
  Vries}, {van Marle}, {Marcos-Arenal}, \& {Lobel}}]{Beck2014}
{Beck}, P.~G., {Hambleton}, K., {Vos}, J., {et~al.} 2014, \aap, 564, A36

\bibitem[{Boyd \& Vandenberghe(2004)}]{boyd2004}
Boyd, S., \& Vandenberghe, L. 2004, Convex Optimization (Cambridge University
  Press)

\bibitem[{{Briquet} {et~al.}(2007){Briquet}, {Morel}, {Thoul}, {Scuflaire},
  {Miglio}, {Montalb{\'a}n}, {Dupret}, \& {Aerts}}]{Briquet2007}
{Briquet}, M., {Morel}, T., {Thoul}, A., {et~al.} 2007, \mnras, 381, 1482

\bibitem[{Burnham \& Anderson(2002)}]{burnham2002model}
Burnham, K., \& Anderson, D. 2002, Model Selection and Multimodel Inference: A
  Practical Information-Theoretic Approach (Springer)

\bibitem[{{Cantiello} {et~al.}(2014){Cantiello}, {Mankovich}, {Bildsten},
  {Christensen-Dalsgaard}, \& {Paxton}}]{Cantiello2014}
{Cantiello}, M., {Mankovich}, C., {Bildsten}, L., {Christensen-Dalsgaard}, J.,
  \& {Paxton}, B. 2014, \apj, 788, 93

\bibitem[{{Charpinet} {et~al.}(2009){Charpinet}, {Fontaine}, \&
  {Brassard}}]{Charpinet2009}
{Charpinet}, S., {Fontaine}, G., \& {Brassard}, P. 2009, \nat, 461, 501

\bibitem[{{Christensen-Dalsgaard}(2002)}]{JCD2002}
{Christensen-Dalsgaard}, J. 2002, Reviews of Modern Physics, 74, 1073

\bibitem[{{C{\'o}rsico} {et~al.}(2012){C{\'o}rsico}, {Althaus}, {Kawaler},
  {Miller Bertolami}, \& {Garc{\'{\i}}a-Berro}}]{Corsico2012}
{C{\'o}rsico}, A.~H., {Althaus}, L.~G., {Kawaler}, S.~D., {Miller Bertolami},
  M.~M., \& {Garc{\'{\i}}a-Berro}, E. 2012, in Astronomical Society of the
  Pacific Conference Series, Vol. 462, Progress in Solar/Stellar Physics with
  Helio- and Asteroseismology, ed. H.~{Shibahashi}, M.~{Takata}, \& A.~E.
  {Lynas-Gray}, 176

\bibitem[{Craig \& Brown(1986)}]{craig1986}
Craig, I., \& Brown, J. 1986, Inverse problems in astronomy: a guide to
  inversion strategies for remotely sensed data (A. Hilger)

\bibitem[{{Degroote} {et~al.}(2009){Degroote}, {Briquet}, {Catala},
  {Uytterhoeven}, {Lefever}, {Morel}, {Aerts}, {Carrier}, {Auvergne}, {Baglin},
  \& {Michel}}]{degroote2009}
{Degroote}, P., {Briquet}, M., {Catala}, C., {et~al.} 2009, \aap, 506, 111

\bibitem[{{Degroote} {et~al.}(2010){Degroote}, {Aerts}, {Baglin}, {Miglio},
  {Briquet}, {Noels}, {Niemczura}, {Montalban}, {Bloemen}, {Oreiro}, {Vu{\v
  c}kovi{\'c}}, {Smolders}, {Auvergne}, {Baudin}, {Catala}, \&
  {Michel}}]{Degroote2010}
{Degroote}, P., {Aerts}, C., {Baglin}, A., {et~al.} 2010, \nat, 464, 259

\bibitem[{{Deheuvels} {et~al.}(2012){Deheuvels}, {Garc{\'{\i}}a}, {Chaplin},
  {Basu}, {Antia}, {Appourchaux}, {Benomar}, {Davies}, {Elsworth}, {Gizon},
  {Goupil}, {Reese}, {Regulo}, {Schou}, {Stahn}, {Casagrande},
  {Christensen-Dalsgaard}, {Fischer}, {Hekker}, {Kjeldsen}, {Mathur}, {Mosser},
  {Pinsonneault}, {Valenti}, {Christiansen}, {Kinemuchi}, \&
  {Mullally}}]{Deheuvels2012}
{Deheuvels}, S., {Garc{\'{\i}}a}, R.~A., {Chaplin}, W.~J., {et~al.} 2012, \apj,
  756, 19

\bibitem[{{Deheuvels} {et~al.}(2014){Deheuvels}, {Doğan, G.}, {Goupil, M.~J.},
  {Appourchaux, T.}, {Benomar, O.}, {Bruntt, H.}, {Campante, T.~L.},
  {Casagrande, L.}, {Ceillier, T.}, {Davies, G.~R.}, {De Cat, P.}, {Fu, J.~N.},
  {García, R.~A.}, {Lobel, A.}, {Mosser, B.}, {Reese, D.~R.}, {Regulo, C.},
  {Schou, J.}, {Stahn, T.}, {Thygesen, A.~O.}, {Yang, X.~H.}, {Chaplin, W.~J.},
  {Christensen-Dalsgaard, J.}, {Eggenberger, P.}, {Gizon, L.}, {Mathis, S.},
  {Molenda-Żakowicz, J.}, \& {Pinsonneault, M.}}]{Deheuvels2014}
{Deheuvels}, S., {Doğan, G.}, {Goupil, M.~J.}, {et~al.} 2014, \aap, 564, A27

\bibitem[{{Eggenberger} {et~al.}(2012){Eggenberger}, {Montalb{\'a}n}, \&
  {Miglio}}]{Eggenberger2012}
{Eggenberger}, P., {Montalb{\'a}n}, J., \& {Miglio}, A. 2012, \aap, 544, L4

\bibitem[{{Gough}(1985)}]{gough1985}
{Gough}, D. 1985, \solphys, 100, 65

\bibitem[{{Hasan} {et~al.}(2005){Hasan}, {Zahn}, \&
  {Christensen-Dalsgaard}}]{hasan2005}
{Hasan}, S.~S., {Zahn}, J.-P., \& {Christensen-Dalsgaard}, J. 2005, \aap, 444,
  L29

\bibitem[{Hastie {et~al.}(2009)Hastie, Tibshirani, \&
  Friedman}]{hastie2009elements}
Hastie, T., Tibshirani, R., \& Friedman, J. 2009, The Elements of Statistical
  Learning: Data Mining, Inference, and Prediction, Springer series in
  statistics (Springer)

\bibitem[{Hurvich \& Tsai(1989)}]{hurvich1989}
Hurvich, C.~M., \& Tsai, C.-L. 1989, Biometrika, 76, 297

\bibitem[{{Kawaler} \& {Bradley}(1994)}]{kawaler1994}
{Kawaler}, S.~D., \& {Bradley}, P.~A. 1994, \apj, 427, 415

\bibitem[{{Kawaler} {et~al.}(1999){Kawaler}, {Sekii}, \& {Gough}}]{kawaler1999}
{Kawaler}, S.~D., {Sekii}, T., \& {Gough}, D. 1999, \apj, 516, 349

\bibitem[{Kuhn \& Tucker(1951)}]{kuhn1951}
Kuhn, H.~W., \& Tucker, A.~W. 1951, in Proceedings of the Second Berkeley
  Symposium on Mathematical Statistics and Probability (Berkeley, Calif.:
  University of California Press), 481--492

\bibitem[{{Kurtz} {et~al.}(2014){Kurtz}, {Saio}, {Takata}, {Shibahashi},
  {Murphy}, \& {Sekii}}]{Kurtz2014}
{Kurtz}, D.~W., {Saio}, H., {Takata}, M., {et~al.} 2014, \mnras, 444, 102

\bibitem[{{Ledoux}(1951)}]{ledoux1951nonradial}
{Ledoux}, P. 1951, \apj, 114, 373

\bibitem[{{Maeder}(2009)}]{Maeder2009}
{Maeder}, A. 2009, Physics, Formation and Evolution of Rotating Stars,
  Astronomy and Astrophysics Library (Springer Heidelberg)

\bibitem[{{Moravveji} {et~al.}(2015){Moravveji}, {Aerts}, {Papics}, {Andres
  Triana}, \& {Vandoren}}]{moravveji2015}
{Moravveji}, E., {Aerts}, C., {Papics}, P.~I., {Andres Triana}, S., \&
  {Vandoren}, B. 2015, \aap, in press (arXiv:1505.06902)

\bibitem[{{Mosser} {et~al.}(2012){Mosser}, {Goupil}, {Belkacem}, {Marques},
  {Beck}, {Bloemen}, {De Ridder}, {Barban}, {Deheuvels}, {Elsworth}, {Hekker},
  {Kallinger}, {Ouazzani}, {Pinsonneault}, {Samadi}, {Stello}, {Garc{\'{\i}}a},
  {Klaus}, {Li}, {Mathur}, \& {Morris}}]{Mosser2012}
{Mosser}, B., {Goupil}, M.~J., {Belkacem}, K., {et~al.} 2012, \aap, 548, A10

\bibitem[{{Pamyatnykh} {et~al.}(2004){Pamyatnykh}, {Handler}, \&
  {Dziembowski}}]{Pamyatnykh2004}
{Pamyatnykh}, A.~A., {Handler}, G., \& {Dziembowski}, W.~A. 2004, \mnras, 350,
  1022

\bibitem[{{P{\'a}pics} {et~al.}(2014){P{\'a}pics}, {Moravveji}, {Aerts},
  {Tkachenko}, {Triana}, {Bloemen}, \& {Southworth}}]{papics2014}
{P{\'a}pics}, P.~I., {Moravveji}, E., {Aerts}, C., {et~al.} 2014, \aap, 570, A8

\bibitem[{{Paxton} {et~al.}(2011){Paxton}, {Bildsten}, {Dotter}, {Herwig},
  {Lesaffre}, \& {Timmes}}]{Paxton2011}
{Paxton}, B., {Bildsten}, L., {Dotter}, A., {et~al.} 2011, \apjs, 192, 3

\bibitem[{{Paxton} {et~al.}(2013){Paxton}, {Cantiello}, {Arras}, {Bildsten},
  {Brown}, {Dotter}, {Mankovich}, {Montgomery}, {Stello}, {Timmes}, \&
  {Townsend}}]{Paxton2013}
{Paxton}, B., {Cantiello}, M., {Arras}, P., {et~al.} 2013, \apjs, 208, 4

\bibitem[{{Pijpers} \& {Thompson}(1994)}]{pijpers1994}
{Pijpers}, F.~P., \& {Thompson}, M.~J. 1994, \aap, 281, 231

\bibitem[{{Rogers} {et~al.}(2013){Rogers}, {Lin}, {McElwaine}, \&
  {Lau}}]{Rogers2013}
{Rogers}, T.~M., {Lin}, D.~N.~C., {McElwaine}, J.~N., \& {Lau}, H.~H.~B. 2013,
  \apj, 772, 21

\bibitem[{{Saio} {et~al.}(2015){Saio}, {Kurtz}, {Takata}, {Shibahashi},
  {Murphy}, {Sekii}, \& {Bedding}}]{Saio2015}
{Saio}, H., {Kurtz}, D.~W., {Takata}, M., {et~al.} 2015, \mnras, 447, 3264

\bibitem[{{Thompson} {et~al.}(2003){Thompson}, {Christensen-Dalsgaard},
  {Miesch}, \& {Toomre}}]{Thompson2003}
{Thompson}, M.~J., {Christensen-Dalsgaard}, J., {Miesch}, M.~S., \& {Toomre},
  J. 2003, \araa, 41, 599

\bibitem[{{Townsend} \& {Teitler}(2013)}]{Townsend2013}
{Townsend}, R.~H.~D., \& {Teitler}, S.~A. 2013, \mnras, 435, 3406

\bibitem[{{Triana} {et~al.}(2014){Triana}, {Zimmerman}, {Nataf}, {Thorette},
  {Lekic}, \& {Lathrop}}]{triana2014}
{Triana}, S.~A., {Zimmerman}, D.~S., {Nataf}, H.-C., {et~al.} 2014, New Journal
  of Physics, 16, 113005 (arXiv:1410.3641)

\bibitem[{{van Saders} \& {Pinsonneault}(2013)}]{vanSanders2013}
{van Saders}, J.~L., \& {Pinsonneault}, M.~H. 2013, \apj, 776, 67

\bibitem[{{Zwintz} {et~al.}(2014){Zwintz}, {Fossati}, {Ryabchikova},
  {Guenther}, {Aerts}, {Barnes}, {Theme{\ss}l}, {Lorenz}, {Cameron},
  {Kuschnig}, {Pollack-Drs}, {Moravveji}, {Baglin}, {Matthews}, {Moffat},
  {Poretti}, {Rainer}, {Rucinski}, {Sasselov}, \& {Weiss}}]{Zwintz2014b}
{Zwintz}, K., {Fossati}, L., {Ryabchikova}, T., {et~al.} 2014, Science, 345,
  550

\end{thebibliography}

\end{document}